\documentclass{ieeeaccessyd}

\newtheorem{theorem}{Theorem}
\newtheorem{proof}{Proof}

\newtheorem{definition}{Definition}

\newcommand{\qed}{\nobreak \ifvmode \relax \else
      \ifdim\lastskip<1.5em \hskip-\lastskip
      \hskip1.5em plus0em minus0.5em \fi \nobreak
      \vrule height0.75em width0.5em depth0.25em\fi}

\usepackage{cite}
\usepackage{amsmath,amssymb,amsfonts}
\usepackage{algorithmic}
\usepackage{graphicx}
\usepackage{textcomp}
\usepackage{mathtools}

\usepackage[english]{algorithm2e}

\def\BibTeX{{\rm B\kern-.05em{\sc i\kern-.025em b}\kern-.08em
    T\kern-.1667em\lower.7ex\hbox{E}\kern-.125emX}}

\begin{document}

\newcommand{\yqubitonesevspace}
{$\Yqubitonesevspace$}
\newcommand{\Yqubitonesevspace}
{{\cal E}}
%
%
%
%
\newcommand{\yqubitoneindexstd}
{$\Yqubitoneindexstd$}
\newcommand{\Yqubitoneindexstd}
{j}
%
%
%
%
%
\newcommand{\yqubitonespaceindexone}
{$\Yqubitonespaceindexone$}
\newcommand{\Yqubitonespaceindexone}
{\Yqubitonesevspace_{1}}
%
\newcommand{\yqubitonespaceindextwo}
{$\Yqubitonespaceindextwo$}
\newcommand{\Yqubitonespaceindextwo}
{\Yqubitonesevspace_{2}}
%
\newcommand{\yqubitonespaceindexstd}
{$\Yqubitonespaceindexstd$}
\newcommand{\Yqubitonespaceindexstd}
{\Yqubitonesevspace_{\Yqubitoneindexstd}}
%
%
%
\newcommand{\yqubitonetimeinit}
{$\Yqubitonetimeinit$}
\newcommand{\Yqubitonetimeinit}
{t_0}
%
%
%
%
\newcommand{\yqubitonetimefinal}
{$\Yqubitonetimefinal$}
\newcommand{\Yqubitonetimefinal}
{t}
%
%
%
%
%
%
\newcommand{\yqubitonetimeinitstateindexone}
{$\Yqubitonetimeinitstateindexone$}
\newcommand{\Yqubitonetimeinitstateindexone}
{| \psi_1 ( \Yqubitonetimeinit ) \rangle}
%
\newcommand{\yqubitonetimeinitstateindextwo}
{$\Yqubitonetimeinitstateindextwo$}
\newcommand{\Yqubitonetimeinitstateindextwo}
{| \psi_2 ( \Yqubitonetimeinit ) \rangle}
%
%
%
\newcommand{\yqubitonetimeinitveccomp}
{$\Yqubitonetimeinitveccomp$}
\newcommand{\Yqubitonetimeinitveccomp}
{C_{\Yqubitoneindexstd}(\Yqubitonetimeinit)}
%
\newcommand{\yqubitonetimeanyveccomp}
{$\Yqubitonetimeanyveccomp$}
\newcommand{\Yqubitonetimeanyveccomp}
{C_{\Yqubitoneindexstd}(t)}
%
%
%
\newcommand{\yqubitonetimeinitanyhamilton}
{$\Yqubitonetimeinitanyhamilton$}
\newcommand{\Yqubitonetimeinitanyhamilton}
{
\Yopmix
_{\Yqubitoneindexstd}}
%
%
%
%
\newcommand{\yqubitonetimenonestateindexqubitoneindexstd}
{$\Yqubitonetimenonestateindexqubitoneindexstd$}
\newcommand{\Yqubitonetimenonestateindexqubitoneindexstd}
{| \psi_{\Yqubitoneindexstd} \rangle}
%
%
%
%
%
%
%
%
\newcommand{\yqubitbothspace}
{$\Yqubitbothspace$}
\newcommand{\Yqubitbothspace}
{\Yqubitonesevspace}
%
%
%
%
%
%
%
%
%
%
\newcommand{\ymixsyststateinitial}
{$\Ymixsyststateinitial$}
\newcommand{\Ymixsyststateinitial}
{| \psi 
( \Yqubitonetimeinit ) \rangle
}
%
%
%
%
%
%
%
%
%
%
%
\newcommand{\ytwoqubitwritereadtimeinterval}
{$\Ytwoqubitwritereadtimeinterval$}
\newcommand{\Ytwoqubitwritereadtimeinterval}
{\tau}
%
%
%
%
%
\newcommand{\ytwoqubitwritereadtimeintervalindexone}
{$\Ytwoqubitwritereadtimeintervalindexone$}
\newcommand{\Ytwoqubitwritereadtimeintervalindexone}
{\Ytwoqubitwritereadtimeinterval_{1}}
%
\newcommand{\ytwoqubitwritereadtimeintervalindextwo}
{$\Ytwoqubitwritereadtimeintervalindextwo$}
\newcommand{\Ytwoqubitwritereadtimeintervalindextwo}
{\Ytwoqubitwritereadtimeinterval_{2}}
%
%
\newcommand{\ytwoqubitwritereadtimeintervalindexthree}
{$\Ytwoqubitwritereadtimeintervalindexthree$}
\newcommand{\Ytwoqubitwritereadtimeintervalindexthree}
{\Ytwoqubitwritereadtimeinterval_{3}}
%
%
%
%
%
%
%
%
%
%
%
%
\newcommand{\yqubitbothtimeinitstatecoefnot}
{$\Yqubitbothtimeinitstatecoefnot$}
\newcommand{\Yqubitbothtimeinitstatecoefnot}
{c}
%
\newcommand{\yqubitbothtimeinitstatecoefplusplus}
{$\Yqubitbothtimeinitstatecoefplusplus$}
\newcommand{\Yqubitbothtimeinitstatecoefplusplus}
{\Yqubitbothtimeinitstatecoefnot_{
1
}( \Yqubitonetimeinit )}
%
\newcommand{\yqubitbothtimeinitstatecoefplusminus}
{$\Yqubitbothtimeinitstatecoefplusminus$}
\newcommand{\Yqubitbothtimeinitstatecoefplusminus}
{\Yqubitbothtimeinitstatecoefnot_{
2
}( \Yqubitonetimeinit )}
%
\newcommand{\yqubitbothtimeinitstatecoefminusplus}
{$\Yqubitbothtimeinitstatecoefminusplus$}
\newcommand{\Yqubitbothtimeinitstatecoefminusplus}
{\Yqubitbothtimeinitstatecoefnot_{
3
}( \Yqubitonetimeinit )}
%
\newcommand{\yqubitbothtimeinitstatecoefminusminus}
{$\Yqubitbothtimeinitstatecoefminusminus$}
\newcommand{\Yqubitbothtimeinitstatecoefminusminus}
{\Yqubitbothtimeinitstatecoefnot_{
4
}( \Yqubitonetimeinit )}
%
%
%
%
\newcommand{\yqubitbothtimeinitstatecoefindexequalstd}
{$\Yqubitbothtimeinitstatecoefindexequalstd$}
\newcommand{\Yqubitbothtimeinitstatecoefindexequalstd}
{k}
%
\newcommand{\yqubitbothtimeinitstatecoefwithindexstd}
{$\Yqubitbothtimeinitstatecoefwithindexstd$}
\newcommand{\Yqubitbothtimeinitstatecoefwithindexstd}
{\Yqubitbothtimeinitstatecoefnot
_{\Yqubitbothtimeinitstatecoefindexequalstd}( \Yqubitonetimeinit )}
%
%
%
%
%
%
%
%
%
%
%
\newcommand{\yqubitbothtimefinalstatecoefplusplus}
{$\Yqubitbothtimefinalstatecoefplusplus$}
\newcommand{\Yqubitbothtimefinalstatecoefplusplus}
{\Yqubitbothtimeinitstatecoefnot_{1}( \Yqubitonetimefinal )}
%
\newcommand{\yqubitbothtimefinalstatecoefplusminus}
{$\Yqubitbothtimefinalstatecoefplusminus$}
\newcommand{\Yqubitbothtimefinalstatecoefplusminus}
{\Yqubitbothtimeinitstatecoefnot_{2}( \Yqubitonetimefinal )}
%
\newcommand{\yqubitbothtimefinalstatecoefminusplus}
{$\Yqubitbothtimefinalstatecoefminusplus$}
\newcommand{\Yqubitbothtimefinalstatecoefminusplus}
{\Yqubitbothtimeinitstatecoefnot_{3}( \Yqubitonetimefinal )}
%
\newcommand{\yqubitbothtimefinalstatecoefminusminus}
{$\Yqubitbothtimefinalstatecoefminusminus$}
\newcommand{\Yqubitbothtimefinalstatecoefminusminus}
{\Yqubitbothtimeinitstatecoefnot_{4}( \Yqubitonetimefinal )}
%
%
%
%
%

%
\newcommand{\yqubitbothtimefinalstatecoefindexindexstd}
{$\Yqubitbothtimefinalstatecoefindexindexstd$}
\newcommand{\Yqubitbothtimefinalstatecoefindexindexstd}
{k}
%
\newcommand{\yqubitbothtimefinalstatecoefindexstd}
{$\Yqubitbothtimefinalstatecoefindexstd$}
\newcommand{\Yqubitbothtimefinalstatecoefindexstd}
{\Yqubitbothtimeinitstatecoefnot
_{\Yqubitbothtimefinalstatecoefindexindexstd}( \Yqubitonetimefinal )}
%
\newcommand{\yqubitbothtimefinalstatecoefindexstdpolarmod}
{$\Yqubitbothtimefinalstatecoefindexstdpolarmod$}
\newcommand{\Yqubitbothtimefinalstatecoefindexstdpolarmod}
{\rho
_{\Yqubitbothtimefinalstatecoefindexindexstd}
}
%
\newcommand{\yqubitbothtimefinalstatecoefpolarphasenot}
{$\Yqubitbothtimefinalstatecoefpolarphasenot$}
\newcommand{\Yqubitbothtimefinalstatecoefpolarphasenot}
{\xi}
%
\newcommand{\yqubitbothtimefinalstatecoefindexstdpolarphase}
{$\Yqubitbothtimefinalstatecoefindexstdpolarphase$}
\newcommand{\Yqubitbothtimefinalstatecoefindexstdpolarphase}
{\Yqubitbothtimefinalstatecoefpolarphasenot
_{\Yqubitbothtimefinalstatecoefindexindexstd}
}
%
%
%
%
\newcommand{\yqubitbothtimefinalstatecoefpluspluspolarphase}
{$\Yqubitbothtimefinalstatecoefpluspluspolarphase$}
\newcommand{\Yqubitbothtimefinalstatecoefpluspluspolarphase}
{\Yqubitbothtimefinalstatecoefpolarphasenot
_{1}
}
%
\newcommand{\yqubitbothtimefinalstatecoefplusminuspolarphase}
{$\Yqubitbothtimefinalstatecoefplusminuspolarphase$}
\newcommand{\Yqubitbothtimefinalstatecoefplusminuspolarphase}
{\Yqubitbothtimefinalstatecoefpolarphasenot
_{2}
}
%
\newcommand{\yqubitbothtimefinalstatecoefminuspluspolarphase}
{$\Yqubitbothtimefinalstatecoefminuspluspolarphase$}
\newcommand{\Yqubitbothtimefinalstatecoefminuspluspolarphase}
{\Yqubitbothtimefinalstatecoefpolarphasenot
_{3}
}
%
\newcommand{\yqubitbothtimefinalstatecoefminusminuspolarphase}
{$\Yqubitbothtimefinalstatecoefminusminuspolarphase$}
\newcommand{\Yqubitbothtimefinalstatecoefminusminuspolarphase}
{\Yqubitbothtimefinalstatecoefpolarphasenot
_{4}
}
%
%
%
%
%
%
%
%
%
\newcommand{\yveccompsyststatetinit}
{$\Yveccompsyststatetinit$}
\newcommand{\Yveccompsyststatetinit}
{C_{+
} 
(
\Yqubitonetimeinit
)
}
%
%
%
%
\newcommand{\yveccompsyststatetfinal}
{$\Yveccompsyststatetfinal$}
\newcommand{\Yveccompsyststatetfinal}
{C_{+
} (t)}
%
%
%
%
%
%
%
\newcommand{\ymixsyststatefinal}
{$\Ymixsyststatefinal$}
\newcommand{\Ymixsyststatefinal}
{| \psi (t) 
\rangle
}
%
%
%
%
%
%
%
%
\newcommand{\yopmix}
{$\Yopmix$}
\newcommand{\Yopmix}
{M}
%
%
%
%
\newcommand{\yopmixestim}
{$\Yopmixestim$}
\newcommand{\Yopmixestim}
{\widehat{\Yopmix}}
%
%
%
%
%
%
%
%
%
%
\newcommand{\yopmixbases}
{$\Yopmixbases$}
\newcommand{\Yopmixbases}
{Q}
%
\newcommand{\yopmixbasesimagin}
{$\Yopmixbasesimagin$}
\newcommand{\Yopmixbasesimagin}
{\Yopmixbases_{\Ysqrtminusone}}
%
\newcommand{\yopmixdiag}
{$\Yopmixdiag$}
\newcommand{\Yopmixdiag}
{D}
%
%
%
%
\newcommand{\yopmixdiagestim}
{$\Yopmixdiagestim$}
\newcommand{\Yopmixdiagestim}
{\widehat{\Yopmixdiag}}
%
%
%
%
%
%
%
%
%
%
\newcommand{\yopsep}
{$\Yopsep$}
\newcommand{\Yopsep}
{U}
%
%
%
%
%
%
%
%
%
%
\newcommand{\ysepsyststateoutnnot}
{$\Ysepsyststateoutnot$}
\newcommand{\Ysepsyststateoutnot}
{\Phi}
%
%
%
\newcommand{\ysepsyststateout}
{$\Ysepsyststateout$}
\newcommand{\Ysepsyststateout}
{|
\Ysepsyststateoutnot
\rangle
%
}
%
%
%
%
%
%
%
%
%
%
%
\newcommand{\yveccompsepsyststateout}
{$\Yveccompsepsyststateout$}
\newcommand{\Yveccompsepsyststateout}
{C
%
}
%
%
%
%
\newcommand{\ysepsyststateoutcoefnot}
{$\Ysepsyststateoutcoefnot$}
\newcommand{\Ysepsyststateoutcoefnot}
{c}
%
%
%
%
\newcommand{\ysepsystdedicstateoutcoefnot}
{$\Ysepsystdedicstateoutcoefnot$}
\newcommand{\Ysepsystdedicstateoutcoefnot}
{c}
%
\newcommand{\ysepsystdedicstateoutcoefplusplus}
{$\Ysepsystdedicstateoutcoefplusplus$}
\newcommand{\Ysepsystdedicstateoutcoefplusplus}
{\Ysepsystdedicstateoutcoefnot_{
1
}}
%
\newcommand{\ysepsystdedicstateoutcoefplusminus}
{$\Ysepsystdedicstateoutcoefplusminus$}
\newcommand{\Ysepsystdedicstateoutcoefplusminus}
{\Ysepsystdedicstateoutcoefnot_{
2
}}
%
\newcommand{\ysepsystdedicstateoutcoefminusplus}
{$\Ysepsystdedicstateoutcoefminusplus$}
\newcommand{\Ysepsystdedicstateoutcoefminusplus}
{\Ysepsystdedicstateoutcoefnot_{
3
}}
%
\newcommand{\ysepsystdedicstateoutcoefminusminus}
{$\Ysepsystdedicstateoutcoefminusminus$}
\newcommand{\Ysepsystdedicstateoutcoefminusminus}
{\Ysepsystdedicstateoutcoefnot_{
4
}}
%
%
%
%
%
\newcommand{\ysepsystdedicstateoutcoefpart}
{$\Ysepsystdedicstateoutcoefpart$}
\newcommand{\Ysepsystdedicstateoutcoefpart}
{\Ysepsystdedicstateoutcoefnot_{5}}
%
%
%
%
%
\newcommand{\ysepsystdedicstateoutcoefpartphase}
{$\Ysepsystdedicstateoutcoefpartphase$}
\newcommand{\Ysepsystdedicstateoutcoefpartphase}
{\phi_{5}}
%
%
%
%
%
\newcommand{\ysepsystdedicstateoutcoefwithindexstd}
{$\Ysepsystdedicstateoutcoefwithindexstd$}
\newcommand{\Ysepsystdedicstateoutcoefwithindexstd}
{\Ysepsystdedicstateoutcoefnot_{\Yqubitbothtimeinitstatecoefindexequalstd}}
%
%
%
%
%
%
%
%
%
%
%
%

%
\newcommand{\ysepsystdedicstateoutcoefplusplusseqindex}
{$\Ysepsystdedicstateoutcoefplusplusseqindex$}
\newcommand{\Ysepsystdedicstateoutcoefplusplusseqindex}
{
\Ysepsystdedicstateoutcoefplusplus
(
\Ytwoqubitseqindex
)
}
%
\newcommand{\ysepsystdedicstateoutcoefplusminusseqindex}
{$\Ysepsystdedicstateoutcoefplusminusseqindex$}
\newcommand{\Ysepsystdedicstateoutcoefplusminusseqindex}
{
\Ysepsystdedicstateoutcoefplusminus
(
\Ytwoqubitseqindex
)
}
%
\newcommand{\ysepsystdedicstateoutcoefminusplusseqindex}
{$\Ysepsystdedicstateoutcoefminusplusseqindex$}
\newcommand{\Ysepsystdedicstateoutcoefminusplusseqindex}
{
\Ysepsystdedicstateoutcoefminusplus
(
\Ytwoqubitseqindex
)
}
%
\newcommand{\ysepsystdedicstateoutcoefminusminusseqindex}
{$\Ysepsystdedicstateoutcoefminusminusseqindex$}
\newcommand{\Ysepsystdedicstateoutcoefminusminusseqindex}
{
\Ysepsystdedicstateoutcoefminusminus
(
\Ytwoqubitseqindex
)
}
%
%
%
%
%
%
%
%
%
%
%
%
\newcommand{\ysepsystdedicstateoutcoefprobnot}
{$\Ysepsystdedicstateoutcoefprobnot$}
\newcommand{\Ysepsystdedicstateoutcoefprobnot}
{P}
%
\newcommand{\ysepsystdedicstateoutcoefprobplusplus}
{$\Ysepsystdedicstateoutcoefprobplusplus$}
\newcommand{\Ysepsystdedicstateoutcoefprobplusplus}
{\Ysepsystdedicstateoutcoefprobnot_{
1
%
%
%
z
}}
%
\newcommand{\ysepsystdedicstateoutcoefprobplusminus}
{$\Ysepsystdedicstateoutcoefprobplusminus$}
\newcommand{\Ysepsystdedicstateoutcoefprobplusminus}
{\Ysepsystdedicstateoutcoefprobnot_{
2
%
%
%
z
}}
%
\newcommand{\ysepsystdedicstateoutcoefprobminusplus}
{$\Ysepsystdedicstateoutcoefprobminusplus$}
\newcommand{\Ysepsystdedicstateoutcoefprobminusplus}
{\Ysepsystdedicstateoutcoefprobnot_{
3
%
%
%
z
}}
%
\newcommand{\ysepsystdedicstateoutcoefprobminusminus}
{$\Ysepsystdedicstateoutcoefprobminusminus$}
\newcommand{\Ysepsystdedicstateoutcoefprobminusminus}
{\Ysepsystdedicstateoutcoefprobnot_{
4
%
%
%
z
}}
%
%
%
%
%
\newcommand{\ysepsystdedicstateoutcoefprobplusplusseqindex}
{$\Ysepsystdedicstateoutcoefprobplusplusseqindex$}
\newcommand{\Ysepsystdedicstateoutcoefprobplusplusseqindex}
{
\Ysepsystdedicstateoutcoefprobplusplus
(
\Ytwoqubitseqindex
)
}
%
\newcommand{\ysepsystdedicstateoutcoefprobplusminusseqindex}
{$\Ysepsystdedicstateoutcoefprobplusminusseqindex$}
\newcommand{\Ysepsystdedicstateoutcoefprobplusminusseqindex}
{
\Ysepsystdedicstateoutcoefprobplusminus
(
\Ytwoqubitseqindex
)
}
%
\newcommand{\ysepsystdedicstateoutcoefprobminusplusseqindex}
{$\Ysepsystdedicstateoutcoefprobminusplusseqindex$}
\newcommand{\Ysepsystdedicstateoutcoefprobminusplusseqindex}
{
\Ysepsystdedicstateoutcoefprobminusplus
(
\Ytwoqubitseqindex
)
}
%
\newcommand{\ysepsystdedicstateoutcoefprobminusminusseqindex}
{$\Ysepsystdedicstateoutcoefprobminusminusseqindex$}
\newcommand{\Ysepsystdedicstateoutcoefprobminusminusseqindex}
{
\Ysepsystdedicstateoutcoefprobminusminus
(
\Ytwoqubitseqindex
)
}
%
%
%
%
%
%
%
%
\newcommand{\ysepsystdedicstateoutcoefprobnotox}
{$\Ysepsystdedicstateoutcoefprobnotox$}
\newcommand{\Ysepsystdedicstateoutcoefprobnotox}
{P}
%
\newcommand{\ysepsystdedicstateoutcoefprobplusplusox}
{$\Ysepsystdedicstateoutcoefprobplusplusox$}
\newcommand{\Ysepsystdedicstateoutcoefprobplusplusox}
{\Ysepsystdedicstateoutcoefprobnot_{1x}}
%
\newcommand{\ysepsystdedicstateoutcoefprobplusminusox}
{$\Ysepsystdedicstateoutcoefprobplusminusox$}
\newcommand{\Ysepsystdedicstateoutcoefprobplusminusox}
{\Ysepsystdedicstateoutcoefprobnot_{2x}}
%
\newcommand{\ysepsystdedicstateoutcoefprobminusplusox}
{$\Ysepsystdedicstateoutcoefprobminusplusox$}
\newcommand{\Ysepsystdedicstateoutcoefprobminusplusox}
{\Ysepsystdedicstateoutcoefprobnot_{3x}}
%
\newcommand{\ysepsystdedicstateoutcoefprobminusminusox}
{$\Ysepsystdedicstateoutcoefprobminusminusox$}
\newcommand{\Ysepsystdedicstateoutcoefprobminusminusox}
{\Ysepsystdedicstateoutcoefprobnot_{4x}}
%
%
%
%
\newcommand{\ysepsystdedicstateoutcoefprobplusplusoxseqindex}
{$\Ysepsystdedicstateoutcoefprobplusplusoxseqindex$}
\newcommand{\Ysepsystdedicstateoutcoefprobplusplusoxseqindex}
{
\Ysepsystdedicstateoutcoefprobplusplusox
(
\Ytwoqubitseqindex
)
}
%
\newcommand{\ysepsystdedicstateoutcoefprobplusminusoxseqindex}
{$\Ysepsystdedicstateoutcoefprobplusminusoxseqindex$}
\newcommand{\Ysepsystdedicstateoutcoefprobplusminusoxseqindex}
{
\Ysepsystdedicstateoutcoefprobplusminusox
(
\Ytwoqubitseqindex
)
}
%
\newcommand{\ysepsystdedicstateoutcoefprobminusplusoxseqindex}
{$\Ysepsystdedicstateoutcoefprobminusplusoxseqindex$}
\newcommand{\Ysepsystdedicstateoutcoefprobminusplusoxseqindex}
{
\Ysepsystdedicstateoutcoefprobminusplusox
(
\Ytwoqubitseqindex
)
}
%
\newcommand{\ysepsystdedicstateoutcoefprobminusminusoxseqindex}
{$\Ysepsystdedicstateoutcoefprobminusminusoxseqindex$}
\newcommand{\Ysepsystdedicstateoutcoefprobminusminusoxseqindex}
{
\Ysepsystdedicstateoutcoefprobminusminusox
(
\Ytwoqubitseqindex
)
}
%
%
%
%
%
%
%
%
%
\newcommand{\ysepsystdedicopmixbasesonestateoutcoefprobzznot}
{$\Ysepsystdedicopmixbasesonestateoutcoefprobzznot$}
\newcommand{\Ysepsystdedicopmixbasesonestateoutcoefprobzznot}
{\Ysepsystdedicstateoutcoefprobnot}
%
\newcommand{\ysepsystdedicopmixbasesonestateoutcoefprobzzplusplus}
{$\Ysepsystdedicopmixbasesonestateoutcoefprobzzplusplus$}
\newcommand{\Ysepsystdedicopmixbasesonestateoutcoefprobzzplusplus}
{\Ysepsystdedicopmixbasesonestateoutcoefprobzznot
_{1z}
( \Yopmixbases )
}
%
\newcommand{\ysepsystdedicopmixbasesonestateoutcoefprobzzplusminus}
{$\Ysepsystdedicopmixbasesonestateoutcoefprobzzplusminus$}
\newcommand{\Ysepsystdedicopmixbasesonestateoutcoefprobzzplusminus}
{\Ysepsystdedicopmixbasesonestateoutcoefprobzznot
_{2z}
( \Yopmixbases )
}
%
\newcommand{\ysepsystdedicopmixbasesonestateoutcoefprobzzminusplus}
{$\Ysepsystdedicopmixbasesonestateoutcoefprobzzminusplus$}
\newcommand{\Ysepsystdedicopmixbasesonestateoutcoefprobzzminusplus}
{\Ysepsystdedicopmixbasesonestateoutcoefprobzznot
_{3z}
( \Yopmixbases )
}
%
\newcommand{\ysepsystdedicopmixbasesonestateoutcoefprobzzminusminus}
{$\Ysepsystdedicopmixbasesonestateoutcoefprobzzminusminus$}
\newcommand{\Ysepsystdedicopmixbasesonestateoutcoefprobzzminusminus}
{\Ysepsystdedicopmixbasesonestateoutcoefprobzznot
_{4z}
( \Yopmixbases )
}
%
%
%
%
\newcommand{\ysepsystdedicopmixbasesimaginonestateoutcoefprobzznot}
{$\Ysepsystdedicopmixbasesimaginonestateoutcoefprobzznot$}
\newcommand{\Ysepsystdedicopmixbasesimaginonestateoutcoefprobzznot}
{\Ysepsystdedicstateoutcoefprobnot}
%
\newcommand{\ysepsystdedicopmixbasesimaginonestateoutcoefprobzzplusplus}
{$\Ysepsystdedicopmixbasesimaginonestateoutcoefprobzzplusplus$}
\newcommand{\Ysepsystdedicopmixbasesimaginonestateoutcoefprobzzplusplus}
{\Ysepsystdedicopmixbasesimaginonestateoutcoefprobzznot
_{1z}
( \Yopmixbasesimagin )
}
%
\newcommand{\ysepsystdedicopmixbasesimaginonestateoutcoefprobzzplusminus}
{$\Ysepsystdedicopmixbasesimaginonestateoutcoefprobzzplusminus$}
\newcommand{\Ysepsystdedicopmixbasesimaginonestateoutcoefprobzzplusminus}
{\Ysepsystdedicopmixbasesimaginonestateoutcoefprobzznot
_{2z}
( \Yopmixbasesimagin )
}
%
\newcommand{\ysepsystdedicopmixbasesimaginonestateoutcoefprobzzminusplus}
{$\Ysepsystdedicopmixbasesimaginonestateoutcoefprobzzminusplus$}
\newcommand{\Ysepsystdedicopmixbasesimaginonestateoutcoefprobzzminusplus}
{\Ysepsystdedicopmixbasesimaginonestateoutcoefprobzznot
_{3z}
( \Yopmixbasesimagin )
}
%
\newcommand{\ysepsystdedicopmixbasesimaginonestateoutcoefprobzzminusminus}
{$\Ysepsystdedicopmixbasesimaginonestateoutcoefprobzzminusminus$}
\newcommand{\Ysepsystdedicopmixbasesimaginonestateoutcoefprobzzminusminus}
{\Ysepsystdedicopmixbasesimaginonestateoutcoefprobzznot
_{4z}
( \Yopmixbasesimagin )
}
%
%
%
%
%
%
%
%
%
%
%
\newcommand{\ysepmixdiag}
{$\Ysepmixdiag$}
\newcommand{\Ysepmixdiag}
{\tilde{\Yopmixdiag}}
%
%
\newcommand{\ysepmixdiagelnot}
{$\Ysepmixdiagelnot$}
\newcommand{\Ysepmixdiagelnot}
{\gamma}
%
%
\newcommand{\ysepmixdiagelomegaoneone}
{$\Ysepmixdiagelomegaoneone$}
\newcommand{\Ysepmixdiagelomegaoneone}
{
\Ysepmixdiagelnot
_1}
%
\newcommand{\ysepmixdiagelomegaonezero}
{$\Ysepmixdiagelomegaonezero$}
\newcommand{\Ysepmixdiagelomegaonezero}
{
\Ysepmixdiagelnot
_2}
%
\newcommand{\ysepmixdiagelomegazerozero}
{$\Ysepmixdiagelomegazerozero$}
\newcommand{\Ysepmixdiagelomegazerozero}
{
\Ysepmixdiagelnot
_3}
%
\newcommand{\ysepmixdiagelomegaoneminusone}
{$\Ysepmixdiagelomegaoneminusone$}
\newcommand{\Ysepmixdiagelomegaoneminusone}
{
\Ysepmixdiagelnot
_4}
%
%
%
%
%
\newcommand{\yindexstdforsepmixdiagel}
{$\Yindexstdforsepmixdiagel$}
\newcommand{\Yindexstdforsepmixdiagel}
{k}
%
\newcommand{\ysepmixdiagelindexstd}
{$\Ysepmixdiagelindexstd$}
\newcommand{\Ysepmixdiagelindexstd}
{\gamma_{\Yindexstdforsepmixdiagel}}
%
\newcommand{\ysepmixdiagelphysnot}
{$\Ysepmixdiagelphysnot$}
\newcommand{\Ysepmixdiagelphysnot}
{V}
%
\newcommand{\ysepmixdiagelomegaoneonephys}
{$\Ysepmixdiagelomegaoneonephys$}
\newcommand{\Ysepmixdiagelomegaoneonephys}
{\Ysepmixdiagelphysnot_1}
%
\newcommand{\ysepmixdiagelomegaonezerophys}
{$\Ysepmixdiagelomegaonezerophys$}
\newcommand{\Ysepmixdiagelomegaonezerophys}
{\Ysepmixdiagelphysnot_2}
%
\newcommand{\ysepmixdiagelomegazerozerophys}
{$\Ysepmixdiagelomegazerozerophys$}
\newcommand{\Ysepmixdiagelomegazerozerophys}
{\Ysepmixdiagelphysnot_3}
%
\newcommand{\ysepmixdiagelomegaoneminusonephys}
{$\Ysepmixdiagelomegaoneminusonephys$}
\newcommand{\Ysepmixdiagelomegaoneminusonephys}
{\Ysepmixdiagelphysnot_4}
%
\newcommand{\ysepmixdiagelindexstdphys}
{$\Ysepmixdiagelindexstdphys$}
\newcommand{\Ysepmixdiagelindexstdphys}
{\Ysepmixdiagelphysnot_{\Yindexstdforsepmixdiagel}}
%
\newcommand{\ysepmixdiagelindexstdfuncphystoel}
{$\Ysepmixdiagelindexstdfuncphystoel$}
\newcommand{\Ysepmixdiagelindexstdfuncphystoel}
{g_{\Yindexstdforsepmixdiagel}}
%
%
%
%
%
%
\newcommand{\ysepmixdiagelomegaoneoneideal}
{$\Ysepmixdiagelomegaoneoneideal$}
\newcommand{\Ysepmixdiagelomegaoneoneideal}
{
\Ysepmixdiagelnot
_{1d}
}
%
\newcommand{\ysepmixdiagelomegaonezeroideal}
{$\Ysepmixdiagelomegaonezeroideal$}
\newcommand{\Ysepmixdiagelomegaonezeroideal}
{
\Ysepmixdiagelnot
_{2d}}
%
\newcommand{\ysepmixdiagelomegazerozeroideal}
{$\Ysepmixdiagelomegazerozeroideal$}
\newcommand{\Ysepmixdiagelomegazerozeroideal}
{
\Ysepmixdiagelnot
_{3d}}
%
\newcommand{\ysepmixdiagelomegaoneminusoneideal}
{$\Ysepmixdiagelomegaoneminusoneideal$}
\newcommand{\Ysepmixdiagelomegaoneminusoneideal}
{
\Ysepmixdiagelnot
_{4d}}
%
%
%
%
%
%
%
%
%
\newcommand{\ysepmixdiagelindexstdideal}
{$\Ysepmixdiagelindexstdideal$}
\newcommand{\Ysepmixdiagelindexstdideal}
{
\Ysepmixdiagelnot
_{\Yindexstdforsepmixdiagel d}}
%
%
%
%
%
%
%
%
%
%
%

%
%
\newcommand{\ysepmixdiagelnotestim}
{$\Ysepmixdiagelnotestim$}
\newcommand{\Ysepmixdiagelnotestim}
{\widehat{\Ysepmixdiagelnot}}
%
\newcommand{\ysepmixdiagelomegaoneoneestim}
{$\Ysepmixdiagelomegaoneoneestim$}
\newcommand{\Ysepmixdiagelomegaoneoneestim}
{\Ysepmixdiagelnotestim _1}
%
\newcommand{\ysepmixdiagelomegaonezeroestim}
{$\Ysepmixdiagelomegaonezeroestim$}
\newcommand{\Ysepmixdiagelomegaonezeroestim}
{\Ysepmixdiagelnotestim _2}
%
\newcommand{\ysepmixdiagelomegazerozeroestim}
{$\Ysepmixdiagelomegazerozeroestim$}
\newcommand{\Ysepmixdiagelomegazerozeroestim}
{\Ysepmixdiagelnotestim _3}
%
\newcommand{\ysepmixdiagelomegaoneminusoneestim}
{$\Ysepmixdiagelomegaoneminusoneestim$}
\newcommand{\Ysepmixdiagelomegaoneminusoneestim}
{\Ysepmixdiagelnotestim _4}
%
%
%
%
%
%
%
%
%
%
%
%
%
%
%
%
\newcommand{\yopglob}
{$\Yopglob$}
\newcommand{\Yopglob}
{G}
%
%
%
%
\newcommand{\yopglobdiag}
{$\Yopglobdiag$}
\newcommand{\Yopglobdiag}
{\Delta}
%
%
\newcommand{\yopglobdiagelnot}
{$\Yopglobdiagelnot$}
\newcommand{\Yopglobdiagelnot}
{\delta}
%
%
\newcommand{\yopglobdiagelomegaoneone}
{$\Yopglobdiagelomegaoneone$}
\newcommand{\Yopglobdiagelomegaoneone}
{
\Yopglobdiagelnot
_1}
%
\newcommand{\yopglobdiagelomegaonezero}
{$\Yopglobdiagelomegaonezero$}
\newcommand{\Yopglobdiagelomegaonezero}
{
\Yopglobdiagelnot
_2}
%
\newcommand{\yopglobdiagelomegazerozero}
{$\Yopglobdiagelomegazerozero$}
\newcommand{\Yopglobdiagelomegazerozero}
{
\Yopglobdiagelnot
_3}
%
\newcommand{\yopglobdiagelomegaoneminusone}
{$\Yopglobdiagelomegaoneminusone$}
\newcommand{\Yopglobdiagelomegaoneminusone}
{
\Yopglobdiagelnot
_4}
%
%
%
%
\newcommand{\yopglobdiagelindexstdforsepmixdiagel}
{$\Yopglobdiagelindexstdforsepmixdiagel$}
\newcommand{\Yopglobdiagelindexstdforsepmixdiagel}
{
\Yopglobdiagelnot_{\Yindexstdforsepmixdiagel}}
%
%
%
%
\newcommand{\yopglobdiagelcombinone}
{$\Yopglobdiagelcombinone$}
\newcommand{\Yopglobdiagelcombinone}
{
\Yopglobdiagelnot
_5}
%
%
%
%
\newcommand{\yalphaonebetatwomod}
{$\Yalphaonebetatwomod$}
\newcommand{\Yalphaonebetatwomod}
{A_1}
%
\newcommand{\yalphaonebetatwophase}
{$\Yalphaonebetatwophase$}
\newcommand{\Yalphaonebetatwophase}
{\xi_1}
%
%
%
\newcommand{\yalphatwobetaonemod}
{$\Yalphatwobetaonemod$}
\newcommand{\Yalphatwobetaonemod}
{A_2}
%
\newcommand{\yalphatwobetaonephase}
{$\Yalphatwobetaonephase$}
\newcommand{\Yalphatwobetaonephase}
{\xi_2}
%
%
%
%
%
%
%
%
%
%
%
\newcommand{\ynonentangcondmodulusquantityone}
{$\Ynonentangcondmodulusquantityone$}
\newcommand{\Ynonentangcondmodulusquantityone}
{
B
_1}
%
\newcommand{\ynonentangcondmodulusquantitytwo}
{$\Ynonentangcondmodulusquantitytwo$}
\newcommand{\Ynonentangcondmodulusquantitytwo}
{
B
_2}
%
%
%
%
%
%
%
%
%
%
%
\newcommand{\ytwoqubitsbasisplusplus}
{$\Ytwoqubitsbasisplusplus$}
\newcommand{\Ytwoqubitsbasisplusplus}
{{\cal B} _+}
%
%
%
%
\newcommand{\ytwoqubitsbasisplusxplusx}
{$\Ytwoqubitsbasisplusxplusx$}
\newcommand{\Ytwoqubitsbasisplusxplusx}
{{\cal B} _{+x}}
%
%
%
%
%
%
%
%
%
%
%
%
%
%
%
%
\newcommand{\ysepsystoutnonentangcondmodulusmultiple}
{$\Ysepsystoutnonentangcondmodulusmultiple$}
\newcommand{\Ysepsystoutnonentangcondmodulusmultiple}
{m}
%
%
%
%
%
%
%
%
%
%
\newcommand{\ysepsystoutnonentangcondmodulusmultipleestim}
{$\Ysepsystoutnonentangcondmodulusmultipleestim$}
\newcommand{\Ysepsystoutnonentangcondmodulusmultipleestim}
{\widehat{\Ysepsystoutnonentangcondmodulusmultiple}}
%
\newcommand{\ysepsystoutnonentangcondmodulusmultipleestimminusactual}
{$\Ysepsystoutnonentangcondmodulusmultipleestimminusactual$}
\newcommand{\Ysepsystoutnonentangcondmodulusmultipleestimminusactual}
{\Delta_{\Ysepsystoutnonentangcondmodulusmultiple}}
%
%
%
%
%
%
%
%
%
%
%
%
%
%
%
%
\newcommand{\yqubitnbarb}
{$\Yqubitnbarb$}
\newcommand{\Yqubitnbarb}
{Q}
%
%
%
%
%
%
\newcommand{\yqubitindexstd}
{$\Yqubitindexstd$}
\newcommand{\Yqubitindexstd}
{j}
%
%
%
%
%
%
%
%
\newcommand{\yqubitindexstdtwo}
          {$\Yqubitindexstdtwo$}
\newcommand{\Yqubitindexstdtwo}
           {q}                 
%
%
%
\newcommand{\yqubitonespaceindexqubitnbarb}
{$\Yqubitonespaceindexqubitnbarb$}
\newcommand{\Yqubitonespaceindexqubitnbarb}
{\Yqubitonesevspace_{\Yqubitnbarb}}
%
%
%
\newcommand{\yqubitnbarbtimenonestatenot}
{$\Yqubitnbarbtimenonestatenot$}
\newcommand{\Yqubitnbarbtimenonestatenot}
{\psi}
%
%
%
\newcommand{\yqubitnbarbtimenonestate}
{$\Yqubitnbarbtimenonestate$}
\newcommand{\Yqubitnbarbtimenonestate}
{| \Yqubitnbarbtimenonestatenot \rangle}
%
%
%
%
\newcommand{\yqubitnbarbtimeinitialstate}
{$\Yqubitnbarbtimeinitialstate$}
\newcommand{\Yqubitnbarbtimeinitialstate}
{| \Yqubitnbarbtimenonestatenot 
%
( \Yqubitonetimeinit )
\rangle}
%
\newcommand{\yqubitnbarbtimefinalstate}
{$\Yqubitnbarbtimefinalstate$}
\newcommand{\Yqubitnbarbtimefinalstate}
{| \Yqubitnbarbtimenonestatenot 
%
( \Yqubitonetimefinal )
\rangle}
%
%
%
%
\newcommand{\yqubitnbarbtimenonestateseqindex}
{$\Yqubitnbarbtimenonestateseqindex$}
\newcommand{\Yqubitnbarbtimenonestateseqindex}
{| \Yqubitnbarbtimenonestatenot
(
\Ytwoqubitseqindex
)
\rangle}
%
%
%
%
\newcommand{\yqubitnbarbstatecoefindexequalstd}
{$\Yqubitnbarbstatecoefindexequalstd$}
\newcommand{\Yqubitnbarbstatecoefindexequalstd}
{k}
%
%
%
\newcommand{\yqubitnbarbspaceoverallvecbasiscoefindexequalstd}
{$\Yqubitnbarbspaceoverallvecbasiscoefindexequalstd$}
\newcommand{\Yqubitnbarbspaceoverallvecbasiscoefindexequalstd}
{
|
%
{\Yqubitnbarbstatecoefindexequalstd}
\rangle
}
%
%
%
\newcommand{\yqubitnbarbspaceoverallcoefindexequalstdwithvecbasis}
{$\Yqubitnbarbspaceoverallcoefindexequalstdwithvecbasis$}
\newcommand{\Yqubitnbarbspaceoverallcoefindexequalstdwithvecbasis}
{
c
_
{\Yqubitnbarbstatecoefindexequalstd}
}
%
\newcommand{\yqubitnbarbspaceoverallcoefindexequalstdwithvecbasisseqindex}
{$\Yqubitnbarbspaceoverallcoefindexequalstdwithvecbasisseqindex$}
\newcommand{\Yqubitnbarbspaceoverallcoefindexequalstdwithvecbasisseqindex}
{
c
_
{\Yqubitnbarbstatecoefindexequalstd}
(
\Ytwoqubitseqindex
)
}
%
%
%
\newcommand{\yqubitnbarbspaceoveralleventindexequalstdwithvecbasis}
{$\Yqubitnbarbspaceoveralleventindexequalstdwithvecbasis$}
\newcommand{\Yqubitnbarbspaceoveralleventindexequalstdwithvecbasis}
{
A
_
{\Yqubitnbarbstatecoefindexequalstd}
}
%
%
%
\newcommand{\yqubitnbarbspaceoveralleventindexequalstdwithvecbasisprob}
{$\Yqubitnbarbspaceoveralleventindexequalstdwithvecbasisprob$}
\newcommand{\Yqubitnbarbspaceoveralleventindexequalstdwithvecbasisprob}
{
P
(
\Yqubitnbarbspaceoveralleventindexequalstdwithvecbasis
)
}
%
\newcommand{\yqubitnbarbspaceoveralleventindexequalstdwithvecbasisprobexpect}
{$\Yqubitnbarbspaceoveralleventindexequalstdwithvecbasisprobexpect$}
\newcommand{\Yqubitnbarbspaceoveralleventindexequalstdwithvecbasisprobexpect}
{
E
\{
\Yqubitnbarbspaceoveralleventindexequalstdwithvecbasisprob
\}
}
%
%
\newcommand{\yqubitnbarbspaceoveralleventindexequalstdwithvecbasisprobexpectapproxonetwoqubitseqnb}
{$\Yqubitnbarbspaceoveralleventindexequalstdwithvecbasisprobexpectapproxonetwoqubitseqnb$}
\newcommand{\Yqubitnbarbspaceoveralleventindexequalstdwithvecbasisprobexpectapproxonetwoqubitseqnb}
{
E
^{\prime}
\{
\Yqubitnbarbspaceoveralleventindexequalstdwithvecbasisprob
\}
}
%
%
\newcommand{\yqubitnbarbspaceoveralleventindexequalstdwithvecbasisprobexpectapproxtwotwoqubitseqnb}
{$\Yqubitnbarbspaceoveralleventindexequalstdwithvecbasisprobexpectapproxtwotwoqubitseqnb$}
\newcommand{\Yqubitnbarbspaceoveralleventindexequalstdwithvecbasisprobexpectapproxtwotwoqubitseqnb}
{
E
^{\prime \prime}
\{
\Yqubitnbarbspaceoveralleventindexequalstdwithvecbasisprob
\}
}
%
%
%
\newcommand{\yqubitnbarbspaceoveralleventindexequalstdwithvecbasisprobseqindex}
{$\Yqubitnbarbspaceoveralleventindexequalstdwithvecbasisprobseqindex$}
\newcommand{\Yqubitnbarbspaceoveralleventindexequalstdwithvecbasisprobseqindex}
{
P
(
\Yqubitnbarbspaceoveralleventindexequalstdwithvecbasis
,
\Ytwoqubitseqindex
)
}
%
%
%

\newcommand{\yqubitnbarbspaceoveralleventindexequalstdwithvecbasisprobstatewritereadseqindex}
{$\Yqubitnbarbspaceoveralleventindexequalstdwithvecbasisprobstatewritereadseqindex$}
\newcommand{\Yqubitnbarbspaceoveralleventindexequalstdwithvecbasisprobstatewritereadseqindex}
{
P
(
\Yqubitnbarbspaceoveralleventindexequalstdwithvecbasis
,
\Yqubitnbarbstatewritereadseqindex
)
}
%
%
%
\newcommand{\yqubitnbarbspaceoveralleventindexequalstdwithvecbasisprobseqindexapproxwritereadonestatenb}
{$\Yqubitnbarbspaceoveralleventindexequalstdwithvecbasisprobseqindexapproxwritereadonestatenb$}
\newcommand{\Yqubitnbarbspaceoveralleventindexequalstdwithvecbasisprobseqindexapproxwritereadonestatenb}
{
P
^{\prime}
(
\Yqubitnbarbspaceoveralleventindexequalstdwithvecbasis
,
\Ytwoqubitseqindex
,
\Ywritereadonestatenb
)
}
%
%
%
\newcommand{\yfuncnumberofofeventoccur}
{$\Yfuncnumberofofeventoccur$}
\newcommand{\Yfuncnumberofofeventoccur}
{{\cal N}}
%
\newcommand{\yqubitnbarbspaceoveralleventindexequalstdwithvecbasisnboccurseqindexwritereadonestatenb}
{$\Yqubitnbarbspaceoveralleventindexequalstdwithvecbasisnboccurseqindexwritereadonestatenb$}
\newcommand{\Yqubitnbarbspaceoveralleventindexequalstdwithvecbasisnboccurseqindexwritereadonestatenb}
{
\Yfuncnumberofofeventoccur
(
\Yqubitnbarbspaceoveralleventindexequalstdwithvecbasis
,
\Ytwoqubitseqindex
,
\Ywritereadonestatenb
)
}
%
\newcommand{\yqubitnbarbspaceoveralleventindexequalstdwithvecbasisnboccurqubitnbarbstatewritereadseqnb}
{$\Yqubitnbarbspaceoveralleventindexequalstdwithvecbasisnboccurqubitnbarbstatewritereadseqnb$}
\newcommand{\Yqubitnbarbspaceoveralleventindexequalstdwithvecbasisnboccurqubitnbarbstatewritereadseqnb}
{
\Yfuncnumberofofeventoccur
(
\Yqubitnbarbspaceoveralleventindexequalstdwithvecbasis
,
\Yqubitnbarbstatewritereadseqnb
)
}
%
\newcommand{\yqubitnbarbspaceoveralleventindexequalstdwithvecbasisindicfuncqubitnbarbstatewritereadseqindex}
{$\Yqubitnbarbspaceoveralleventindexequalstdwithvecbasisindicfuncqubitnbarbstatewritereadseqindex$}
\newcommand{\Yqubitnbarbspaceoveralleventindexequalstdwithvecbasisindicfuncqubitnbarbstatewritereadseqindex}
{
1\hspace{-.12cm}1
(
\Yqubitnbarbspaceoveralleventindexequalstdwithvecbasis
,
\Yqubitnbarbstatewritereadseqindex
)
}
%
%
%
%
%
%
%
%
%
%

\newcommand{\yketspacedim}{$\Yketspacedim$}
\newcommand{\Yketspacedim}{d}

\newcommand{\yindexstdforketbasis}{$\Yindexstdforketbasis$}
\newcommand{\Yindexstdforketbasis}{k}

\newcommand{\yketbasisindexstd}{$\Yketbasisindexstd$}
\newcommand{\Yketbasisindexstd}
{%
\lvert
\Yindexstdforketbasis
\rangle
}

%
%
%
%


\newcommand{\yketbasiscoefdetermnot}{$\Yketbasiscoefdetermnot$}
\newcommand{\Yketbasiscoefdetermnot}
{c}

\newcommand{\yketbasiscoefdetermindexstd}{$\Yketbasiscoefdetermindexstd$}
\newcommand{\Yketbasiscoefdetermindexstd}
{
\Yketbasiscoefdetermnot
_{
\Yindexstdforketbasis
}
}
%
%
%


\newcommand{\yketdetermnot}{$\Yketdetermnot$}
\newcommand{\Yketdetermnot}{\psi}

\newcommand{\yketdeterm}{$\Yketdeterm$}
\newcommand{\Yketdeterm}{%
\lvert
\Yketdetermnot \rangle}

\newcommand{\ybradeterm}{$\Ybradeterm$}
\newcommand{\Ybradeterm}{\langle \Yketdetermnot
\rvert
}

%
%
%
%
\newcommand{\ybranobardeterm}{$\Ybranobardeterm$}
\newcommand{\Ybranobardeterm}{\langle \Yketdetermnot}
%
%
%
%
%
%
%
%
%
%
\newcommand{\yregisterindexstdone}
          {$\Yregisterindexstdone$}
\newcommand{\Yregisterindexstdone}
           {r}           
%
%
%
\newcommand{\yqubitspaceregisterindexstdonequbitindexone}
          {$\Yqubitspaceregisterindexstdonequbitindexone$}
\newcommand{\Yqubitspaceregisterindexstdonequbitindexone}
           {\Yqubitonesevspace
           _{\Yregisterindexstdone 1}
           }                                 
%
%
%
\newcommand{\yqubitspaceregisterindexstdonequbitindexstdtwo}
          {$\Yqubitspaceregisterindexstdonequbitindexstdtwo$}
\newcommand{\Yqubitspaceregisterindexstdonequbitindexstdtwo}
           {\Yqubitonesevspace
           _{\Yregisterindexstdone \Yqubitindexstdtwo}
           }          
%
%
%
\newcommand{\yqubitspaceregisterindexstdonequbitindexqubitnbarb}
          {$\Yqubitspaceregisterindexstdonequbitindexqubitnbarb$}
\newcommand{\Yqubitspaceregisterindexstdonequbitindexqubitnbarb}
           {\Yqubitonesevspace
           _{\Yregisterindexstdone \Yqubitnbarb}
           } 																								
%
%
%
\newcommand{\yqubitspaceregisternoindexqubitnoindexbasisintstd}
          {$\Yqubitspaceregisternoindexqubitnoindexbasisintstd$}
\newcommand{\Yqubitspaceregisternoindexqubitnoindexbasisintstd}
           {k}
%
%
%
\newcommand{\yqubitspaceregisterindexstdonequbitonebasisintstd}
          {$\Yqubitspaceregisterindexstdonequbitonebasisintstd$}
\newcommand{\Yqubitspaceregisterindexstdonequbitonebasisintstd}
           {\Yqubitspaceregisternoindexqubitnoindexbasisintstd
           _{\Yregisterindexstdone 1}
           } 							
%
%
%
\newcommand{\yqubitspaceregisterindexstdonequbitindexstdtwobasisintstd}
          {$\Yqubitspaceregisterindexstdonequbitindexstdtwobasisintstd$}
\newcommand{\Yqubitspaceregisterindexstdonequbitindexstdtwobasisintstd}
           {\Yqubitspaceregisternoindexqubitnoindexbasisintstd
           _{\Yregisterindexstdone \Yqubitindexstdtwo}
           }                      
%
%
%
\newcommand{\yqubitspaceregisterindexstdonequbitlastbasisintstd}
          {$\Yqubitspaceregisterindexstdonequbitlastbasisintstd$}
\newcommand{\Yqubitspaceregisterindexstdonequbitlastbasisintstd}
           {\Yqubitspaceregisternoindexqubitnoindexbasisintstd
           _{\Yregisterindexstdone \Yqubitnbarb}
           }  
%
%
%
\newcommand{\yqubitspaceregisterindexstdonequbitanybasisintstd}
          {$\Yqubitspaceregisterindexstdonequbitanybasisintstd$}
\newcommand{\Yqubitspaceregisterindexstdonequbitanybasisintstd}
           {\Yqubitspaceregisternoindexqubitnoindexbasisintstd
           _{\Yregisterindexstdone \bullet}
           }						
%
%
%
\newcommand{\ysetnotnoarg}
          {$\Ysetnotnoarg$}
\newcommand{\Ysetnotnoarg}
           {{\cal S}}  																		
%
%
%
\newcommand{\ysetnotargqubitspaceregisterindexstdonequbitanybasisintstd}
          {$\Ysetnotargqubitspaceregisterindexstdonequbitanybasisintstd$}
\newcommand{\Ysetnotargqubitspaceregisterindexstdonequbitanybasisintstd}
           {\Ysetnotnoarg (
					 \Yqubitspaceregisterindexstdonequbitanybasisintstd )
					 } 																
%
%
%
\newcommand{\yqubitspaceregisterindexstdonequbitonebasisstateintstd}
          {$\Yqubitspaceregisterindexstdonequbitonebasisstateintstd$}
\newcommand{\Yqubitspaceregisterindexstdonequbitonebasisstateintstd}
           {| \Yqubitspaceregisterindexstdonequbitonebasisintstd
           \rangle
           _{\Yregisterindexstdone 1}
           }                	
%
%
%
\newcommand{\yqubitspaceregisterindexstdonequbitindexstdtwobasisstateintstd}
          {$\Yqubitspaceregisterindexstdonequbitindexstdtwobasisstateintstd$}
\newcommand{\Yqubitspaceregisterindexstdonequbitindexstdtwobasisstateintstd}
           {| \Yqubitspaceregisterindexstdonequbitindexstdtwobasisintstd
           \rangle
           _{\Yregisterindexstdone \Yqubitindexstdtwo}
           }     																										
%
%
%
\newcommand{\yqubitspaceregisterindexstdonequbitlastbasisstateintstd}
          {$\Yqubitspaceregisterindexstdonequbitlastbasisstateintstd$}
\newcommand{\Yqubitspaceregisterindexstdonequbitlastbasisstateintstd}
           {| \Yqubitspaceregisterindexstdonequbitlastbasisintstd
           \rangle
           _{\Yregisterindexstdone \Yqubitnbarb}
           } 																																			
%
%
%
\newcommand{\yqubitspaceregisterindexnoqubitsindexnobasisstatecoefnot}
          {$\Yqubitspaceregisterindexnoqubitsindexnobasisstatecoefnot$}
\newcommand{\Yqubitspaceregisterindexnoqubitsindexnobasisstatecoefnot}
           {c} 																																				
%
%
%
\newcommand{\yregisteronequbitspaceregisterindexstdonequbitsonetolastbasisstatecoefstd}
          {$\Yregisteronequbitspaceregisterindexstdonequbitsonetolastbasisstatecoefstd$}
\newcommand{\Yregisteronequbitspaceregisterindexstdonequbitsonetolastbasisstatecoefstd}
           {\Yqubitspaceregisterindexnoqubitsindexnobasisstatecoefnot
           _{1
					   \Yqubitspaceregisterindexstdonequbitonebasisintstd
					   \dots
						 \Yqubitspaceregisterindexstdonequbitlastbasisintstd}
           } 
%
%
%
\newcommand{\yregistertwoqubitspaceregisterindexstdonequbitsonetolastbasisstatecoefstd}
          {$\Yregistertwoqubitspaceregisterindexstdonequbitsonetolastbasisstatecoefstd$}
\newcommand{\Yregistertwoqubitspaceregisterindexstdonequbitsonetolastbasisstatecoefstd}
           {\Yqubitspaceregisterindexnoqubitsindexnobasisstatecoefnot
           _{2
					   \Yqubitspaceregisterindexstdonequbitonebasisintstd
					   \dots
						 \Yqubitspaceregisterindexstdonequbitlastbasisintstd}
           } 																		
%
%
%
\newcommand{\yregisterstdonequbitspaceregisterindexstdonequbitsonetolastbasisstatecoefstd}
          {$\Yregisterstdonequbitspaceregisterindexstdonequbitsonetolastbasisstatecoefstd$}
\newcommand{\Yregisterstdonequbitspaceregisterindexstdonequbitsonetolastbasisstatecoefstd}
           {\Yqubitspaceregisterindexnoqubitsindexnobasisstatecoefnot
           _{\Yregisterindexstdone
					   \Yqubitspaceregisterindexstdonequbitonebasisintstd
					   \dots
						 \Yqubitspaceregisterindexstdonequbitlastbasisintstd}
           } 
%
%
%
\newcommand{\ystateregisternot}
          {$\Ystateregisternot$}
\newcommand{\Ystateregisternot}
           {\psi}

%
%
%
\newcommand{\ystateregisterindexone}
          {$\Ystateregisterindexone$}
\newcommand{\Ystateregisterindexone}
           {| \Ystateregisternot
					_{1}
					  \rangle
           }
%
%
%
\newcommand{\ystateregisterindextwo}
          {$\Ystateregisterindextwo$}
\newcommand{\Ystateregisterindextwo}
           {| \Ystateregisternot
					_{2}
					  \rangle
           }
%
%
%
\newcommand{\ystateregisterindexstdone}
          {$\Ystateregisterindexstdone$}
\newcommand{\Ystateregisterindexstdone}
           {| \Ystateregisternot
					_{\Yregisterindexstdone}
					  \rangle
           }
%
%
%
%
%
%
%
%
%
%
%
%
%
%
\newcommand{\ytwoqubitseqindex}
{$\Ytwoqubitseqindex$}
\newcommand{\Ytwoqubitseqindex}
{n}
%
%
%
%
\newcommand{\ytwoqubitseqnb}
{$\Ytwoqubitseqnb$}
\newcommand{\Ytwoqubitseqnb}
{N}
%
%
%
\newcommand{\ytwoqubitseqnbox}
{$\Ytwoqubitseqnbox$}
\newcommand{\Ytwoqubitseqnbox}
{\Ytwoqubitseqnb _x}
%
%
%
\newcommand{\ytwoqubitseqnboz}
{$\Ytwoqubitseqnboz$}
\newcommand{\Ytwoqubitseqnboz}
{\Ytwoqubitseqnb _z}
%
%
%
%
%
\newcommand{\yqubitnbarbstatewritereadseqindex}
{$\Yqubitnbarbstatewritereadseqindex$}
\newcommand{\Yqubitnbarbstatewritereadseqindex}
{\ell}
%
%
%
%
\newcommand{\yqubitnbarbstatewritereadseqnb}
{$\Yqubitnbarbstatewritereadseqnb$}
\newcommand{\Yqubitnbarbstatewritereadseqnb}
{L}
%
%
%
%
%
\newcommand{\yalphaoneseqindex}
{$\Yalphaoneseqindex$}
\newcommand{\Yalphaoneseqindex}
{
\alpha _1
(
\Ytwoqubitseqindex
)
}
%
\newcommand{\ybetaoneseqindex}
{$\Ybetaoneseqindex$}
\newcommand{\Ybetaoneseqindex}
{
\beta _1
(
\Ytwoqubitseqindex
)
}
%
\newcommand{\yalphatwoseqindex}
{$\Yalphatwoseqindex$}
\newcommand{\Yalphatwoseqindex}
{
\alpha _2
(
\Ytwoqubitseqindex
)
}
%
\newcommand{\ybetatwoseqindex}
{$\Ybetatwoseqindex$}
\newcommand{\Ybetatwoseqindex}
{
\beta _2
(
\Ytwoqubitseqindex
)
}
%
%
%
%
%
%
%
%
%
\newcommand{\ycostfuncozone}
{$\Ycostfuncozone$}
\newcommand{\Ycostfuncozone}
{F_z}
%
\newcommand{\ycostfuncozoneeqindex}
{$\Ycostfuncozoneeqindex$}
\newcommand{\Ycostfuncozoneeqindex}
{f_z
(
\Ytwoqubitseqindex
)
}
%
\newcommand{\ycostfuncozoneeqindexcoefnot}
{$\Ycostfuncozoneeqindexcoefnot$}
\newcommand{\Ycostfuncozoneeqindexcoefnot}
{w}
%
\newcommand{\ycostfuncozoneeqindexcoefcos}
{$\Ycostfuncozoneeqindexcoefcos$}
\newcommand{\Ycostfuncozoneeqindexcoefcos}
{\Ycostfuncozoneeqindexcoefnot_1
(
\Ytwoqubitseqindex
)
}
%
\newcommand{\ycostfuncozoneeqindexcoefsin}
{$\Ycostfuncozoneeqindexcoefsin$}
\newcommand{\Ycostfuncozoneeqindexcoefsin}
{\Ycostfuncozoneeqindexcoefnot_2
(
\Ytwoqubitseqindex
)
}
%
\newcommand{\ycostfuncozoneeqindexcoefconst}
{$\Ycostfuncozoneeqindexcoefconst$}
\newcommand{\Ycostfuncozoneeqindexcoefconst}
{\Ycostfuncozoneeqindexcoefnot_3
(
\Ytwoqubitseqindex
)
}
%
%
%
\newcommand{\ycostfuncoxone}
{$\Ycostfuncoxone$}
\newcommand{\Ycostfuncoxone}
{F_x}
%
%
%
%
%
\newcommand{\ycostfuncoxoneeqindex}
{$\Ycostfuncoxoneeqindex$}
\newcommand{\Ycostfuncoxoneeqindex}
{f_x
(
\Ytwoqubitseqindex
)
}
%
%
%
%
%
%
%
%
\newcommand{\ypropertycomplexsingle}
{$\Ypropertycomplexsingle$}
\newcommand{\Ypropertycomplexsingle}
{d}
%
\newcommand{\ypropertycomplexseqindex}
{$\Ypropertycomplexseqindex$}
\newcommand{\Ypropertycomplexseqindex}
{\Ypropertycomplexsingle ( \Ytwoqubitseqindex )}
%
%
%
%
%
%
%
%
%
%
%
\newcommand{\ysepsystdensopoutqubitonesqrtracepartcondphasemultiple}
{$\Ysepsystdensopoutqubitonesqrtracepartcondphasemultiple$}
\newcommand{\Ysepsystdensopoutqubitonesqrtracepartcondphasemultiple}
{
k
}
%
%
%
%
%
%
%
%
%
%
%
%
%
%
\newcommand{\ysepsystdensopoutqubitonesqrtracepartcondphasemultipleestim}
{$\Ysepsystdensopoutqubitonesqrtracepartcondphasemultipleestim$}
\newcommand{\Ysepsystdensopoutqubitonesqrtracepartcondphasemultipleestim}
{\widehat{\Ysepsystdensopoutqubitonesqrtracepartcondphasemultiple}}
%
\newcommand{\ysepsystdensopoutqubitonesqrtracepartcondphasemultipleestimminusactual}
{$\Ysepsystdensopoutqubitonesqrtracepartcondphasemultipleestimminusactual$}
\newcommand{\Ysepsystdensopoutqubitonesqrtracepartcondphasemultipleestimminusactual}
{\Delta_{\Ysepsystdensopoutqubitonesqrtracepartcondphasemultiple}}
%
%
%
%
%
%
%
%
%
%
%


%
%
%
%
\newcommand{\ytwoqubitsprobaplusplus}
{$\Ytwoqubitsprobaplusplus$}
\newcommand{\Ytwoqubitsprobaplusplus}
{p_1}
%
\newcommand{\ytwoqubitsprobaplusminus}
{$\Ytwoqubitsprobaplusminus$}
\newcommand{\Ytwoqubitsprobaplusminus}
{p_2}
%
\newcommand{\ytwoqubitsprobaminusplus}
{$\Ytwoqubitsprobaminusplus$}
\newcommand{\Ytwoqubitsprobaminusplus}
{p_3}
%
\newcommand{\ytwoqubitsprobaminusminus}
{$\Ytwoqubitsprobaminusminus$}
\newcommand{\Ytwoqubitsprobaminusminus}
{p_4}
%
%
%
%
\newcommand{\ytwoqubitsprobaindexstd}
{$\Ytwoqubitsprobaindexstd$}
\newcommand{\Ytwoqubitsprobaindexstd}
{\Ytwoqubitsprobanot
_{\Yqubitbothtimefinalstatecoefindexindexstd}
}
%
%
%
%
\newcommand{\ytwoqubitsprobavec}
{$\Ytwoqubitsprobavec$}
\newcommand{\Ytwoqubitsprobavec}
{p}
%
%
%
%
%
%
\newcommand{\ytwoqubitsprobanot}
{$\Ytwoqubitsprobanot$}
\newcommand{\Ytwoqubitsprobanot}
{p}
%
\newcommand{\ytwoqubitsprobaplusplusdirzz}
{$\Ytwoqubitsprobaplusplusdirzz$}
\newcommand{\Ytwoqubitsprobaplusplusdirzz}
{\Ytwoqubitsprobanot
%
%
_{1zz}
}
%
\newcommand{\ytwoqubitsprobaplusminusdirzz}
{$\Ytwoqubitsprobaplusminusdirzz$}
\newcommand{\Ytwoqubitsprobaplusminusdirzz}
{\Ytwoqubitsprobanot
%
%
_{2zz}
}
%
\newcommand{\ytwoqubitsprobaminusplusdirzz}
{$\Ytwoqubitsprobaminusplusdirzz$}
\newcommand{\Ytwoqubitsprobaminusplusdirzz}
{\Ytwoqubitsprobanot
%
_{3zz}
}
%
\newcommand{\ytwoqubitsprobaminusminusdirzz}
{$\Ytwoqubitsprobaminusminusdirzz$}
\newcommand{\Ytwoqubitsprobaminusminusdirzz}
{\Ytwoqubitsprobanot
%
_{4zz}
}
%
%
%
%
\newcommand{\ytwoqubitsprobaindexstddirzz}
{$\Ytwoqubitsprobaindexstddirzz$}
\newcommand{\Ytwoqubitsprobaindexstddirzz}
{\Ytwoqubitsprobanot
_{\Yqubitbothtimefinalstatecoefindexindexstd zz}
}

%
%
\newcommand{\ytwoqubitsprobaplusplusdirxx}
{$\Ytwoqubitsprobaplusplusdirxx$}
\newcommand{\Ytwoqubitsprobaplusplusdirxx}
{\Ytwoqubitsprobanot 
_{1xx}
}
%
\newcommand{\ytwoqubitsprobaplusminusdirxx}
{$\Ytwoqubitsprobaplusminusdirxx$}
\newcommand{\Ytwoqubitsprobaplusminusdirxx}
{\Ytwoqubitsprobanot
_{2xx}
}
%
\newcommand{\ytwoqubitsprobaminusplusdirxx}
{$\Ytwoqubitsprobaminusplusdirxx$}
\newcommand{\Ytwoqubitsprobaminusplusdirxx}
{\Ytwoqubitsprobanot
_{3xx}
}
%
\newcommand{\ytwoqubitsprobaminusminusdirxx}
{$\Ytwoqubitsprobaminusminusdirxx$}
\newcommand{\Ytwoqubitsprobaminusminusdirxx}
{\Ytwoqubitsprobanot
_{4xx}
}
%
%
%
%
%
\newcommand{\ytwoqubitsprobaindexstddirxx}
{$\Ytwoqubitsprobaindexstddirxx$}
\newcommand{\Ytwoqubitsprobaindexstddirxx}
{\Ytwoqubitsprobanot
_{\Yqubitbothtimefinalstatecoefindexindexstd xx}
}
%
%
%
%
\newcommand{\ytwoqubitsprobaplusplusdirzx}
{$\Ytwoqubitsprobaplusplusdirzx$}
\newcommand{\Ytwoqubitsprobaplusplusdirzx}
{\Ytwoqubitsprobanot_{1zx}}
%
\newcommand{\ytwoqubitsprobaplusminusdirzx}
{$\Ytwoqubitsprobaplusminusdirzx$}
\newcommand{\Ytwoqubitsprobaplusminusdirzx}
{\Ytwoqubitsprobanot_{2zx}}
%
\newcommand{\ytwoqubitsprobaminusplusdirzx}
{$\Ytwoqubitsprobaminusplusdirzx$}
\newcommand{\Ytwoqubitsprobaminusplusdirzx}
{\Ytwoqubitsprobanot_{3zx}}
%
\newcommand{\ytwoqubitsprobaminusminusdirzx}
{$\Ytwoqubitsprobaminusminusdirzx$}
\newcommand{\Ytwoqubitsprobaminusminusdirzx}
{\Ytwoqubitsprobanot_{4zx}}
%
%
%
%
\newcommand{\ytwoqubitsprobaplusplusdirzy}
{$\Ytwoqubitsprobaplusplusdirzy$}
\newcommand{\Ytwoqubitsprobaplusplusdirzy}
{\Ytwoqubitsprobanot_{1zy}}
%
\newcommand{\ytwoqubitsprobaplusminusdirzy}
{$\Ytwoqubitsprobaplusminusdirzy$}
\newcommand{\Ytwoqubitsprobaplusminusdirzy}
{\Ytwoqubitsprobanot_{2zy}}
%
\newcommand{\ytwoqubitsprobaminusplusdirzy}
{$\Ytwoqubitsprobaminusplusdirzy$}
\newcommand{\Ytwoqubitsprobaminusplusdirzy}
{\Ytwoqubitsprobanot_{3zy}}
%
\newcommand{\ytwoqubitsprobaminusminusdirzy}
{$\Ytwoqubitsprobaminusminusdirzy$}
\newcommand{\Ytwoqubitsprobaminusminusdirzy}
{\Ytwoqubitsprobanot_{4zy}}
%
%
%
%
\newcommand{\ytwoqubitsprobaplusplusdirxz}
{$\Ytwoqubitsprobaplusplusdirxz$}
\newcommand{\Ytwoqubitsprobaplusplusdirxz}
{\Ytwoqubitsprobanot_{1xz}}
%
\newcommand{\ytwoqubitsprobaplusminusdirxz}
{$\Ytwoqubitsprobaplusminusdirxz$}
\newcommand{\Ytwoqubitsprobaplusminusdirxz}
{\Ytwoqubitsprobanot_{2xz}}
%
\newcommand{\ytwoqubitsprobaminusplusdirxz}
{$\Ytwoqubitsprobaminusplusdirxz$}
\newcommand{\Ytwoqubitsprobaminusplusdirxz}
{\Ytwoqubitsprobanot_{3xz}}
%
\newcommand{\ytwoqubitsprobaminusminusdirxz}
{$\Ytwoqubitsprobaminusminusdirxz$}
\newcommand{\Ytwoqubitsprobaminusminusdirxz}
{\Ytwoqubitsprobanot_{4xz}}
%
%
%
%
\newcommand{\ytwoqubitsprobaplusplusdiryz}
{$\Ytwoqubitsprobaplusplusdiryz$}
\newcommand{\Ytwoqubitsprobaplusplusdiryz}
{\Ytwoqubitsprobanot_{1yz}}
%
\newcommand{\ytwoqubitsprobaplusminusdiryz}
{$\Ytwoqubitsprobaplusminusdiryz$}
\newcommand{\Ytwoqubitsprobaplusminusdiryz}
{\Ytwoqubitsprobanot_{2yz}}
%
\newcommand{\ytwoqubitsprobaminusplusdiryz}
{$\Ytwoqubitsprobaminusplusdiryz$}
\newcommand{\Ytwoqubitsprobaminusplusdiryz}
{\Ytwoqubitsprobanot_{3yz}}
%
\newcommand{\ytwoqubitsprobaminusminusdiryz}
{$\Ytwoqubitsprobaminusminusdiryz$}
\newcommand{\Ytwoqubitsprobaminusminusdiryz}
{\Ytwoqubitsprobanot_{4yz}}
\newcommand{\ytwoqubitresultphaseinit}
{$\Ytwoqubitresultphaseinit$}
\newcommand{\Ytwoqubitresultphaseinit}
{\Delta _I}
%
%
%
\newcommand{\ytwoqubitresultphaseevol}
{$\Ytwoqubitresultphaseevol$}
\newcommand{\Ytwoqubitresultphaseevol}
{\Delta _E}
%
%
\newcommand{\ytwoqubitresultphaseevolindexd}
{$\Ytwoqubitresultphaseevolindexd$}
\newcommand{\Ytwoqubitresultphaseevolindexd}
{\Delta _{Ed}}
%
\newcommand{\ytwoqubitresultphaseevolindexdestim}
{$\Ytwoqubitresultphaseevolindexdestim$}
\newcommand{\Ytwoqubitresultphaseevolindexdestim}
{\widehat{\Delta} _{Ed}}
%
%
%
%
\newcommand{\ytwoqubitresultphaseevolsin}
{$\Ytwoqubitresultphaseevolsin$}
\newcommand{\Ytwoqubitresultphaseevolsin}
{v}
%
%
%
%
\newcommand{\ytwoqubitresultphaseevolsinindexone}
{$\Ytwoqubitresultphaseevolsinindexone$}
\newcommand{\Ytwoqubitresultphaseevolsinindexone}
{\Ytwoqubitresultphaseevolsin_{1}}
%
\newcommand{\ytwoqubitresultphaseevolsinindextwo}
{$\Ytwoqubitresultphaseevolsinindextwo$}
\newcommand{\Ytwoqubitresultphaseevolsinindextwo}
{\Ytwoqubitresultphaseevolsin_{2}}
\newcommand{\ytwoqubitresultphaseevolsinestim}
{$\Ytwoqubitresultphaseevolsinestim$}
\newcommand{\Ytwoqubitresultphaseevolsinestim}
{\overline{\Ytwoqubitresultphaseevolsin}}


%
%
\newcommand{\ytwoqubitresultphaseevolsinestimtwo}
{$\Ytwoqubitresultphaseevolsinestimtwo$}
\newcommand{\Ytwoqubitresultphaseevolsinestimtwo}
{\widehat{\Ytwoqubitresultphaseevolsin}}
%
%
\newcommand{\ytwoqubitresultphaseevolsinestimtwoindexone}
{$\Ytwoqubitresultphaseevolsinestimtwoindexone$}
\newcommand{\Ytwoqubitresultphaseevolsinestimtwoindexone}
{\Ytwoqubitresultphaseevolsinestimtwo_{1}}
%
%
\newcommand{\ytwoqubitresultphaseevolsinestimtwoindextwo}
{$\Ytwoqubitresultphaseevolsinestimtwoindextwo$}
\newcommand{\Ytwoqubitresultphaseevolsinestimtwoindextwo}
{\Ytwoqubitresultphaseevolsinestimtwo_{2}}
%


%
%
%
%
%
\newcommand{\yparamqubitbothstateplusmodulusnot}
{$\Yparamqubitbothstateplusmodulusnot$}
\newcommand{\Yparamqubitbothstateplusmodulusnot}
{r}
%
\newcommand{\yparamqubitonestateplusmodulus}
{$\Yparamqubitonestateplusmodulus$}
\newcommand{\Yparamqubitonestateplusmodulus}
{{\Yparamqubitbothstateplusmodulusnot}_1}
%
\newcommand{\yparamqubittwostateplusmodulus}
{$\Yparamqubittwostateplusmodulus$}
\newcommand{\Yparamqubittwostateplusmodulus}
{{\Yparamqubitbothstateplusmodulusnot}_2}
%
\newcommand{\yparamqubitindexstdstateplusmodulus}
{$\Yparamqubitindexstdstateplusmodulus$}
\newcommand{\Yparamqubitindexstdstateplusmodulus}
{{\Yparamqubitbothstateplusmodulusnot}_{\Yqubitindexstd}}
%
\newcommand{\yparamqubitbothstateminusmodulusnot}
{$\Yparamqubitbothstateminusmodulusnot$}
\newcommand{\Yparamqubitbothstateminusmodulusnot}
{q}
%
\newcommand{\yparamqubitonestateminusmodulus}
{$\Yparamqubitonestateminusmodulus$}
\newcommand{\Yparamqubitonestateminusmodulus}
{{\Yparamqubitbothstateminusmodulusnot}_1}
%
\newcommand{\yparamqubittwostateminusmodulus}
{$\Yparamqubittwostateminusmodulus$}
\newcommand{\Yparamqubittwostateminusmodulus}
{{\Yparamqubitbothstateminusmodulusnot}_2}
%
\newcommand{\yparamqubitindexstdstateminusmodulus}
{$\Yparamqubitindexstdstateminusmodulus$}
\newcommand{\Yparamqubitindexstdstateminusmodulus}
{{\Yparamqubitbothstateminusmodulusnot}_{\Yqubitindexstd}}
%
%
%
%
%
\newcommand{\yparamqubitbothstateplusphasenot}
{$\Yparamqubitbothstateplusphasenot$}
\newcommand{\Yparamqubitbothstateplusphasenot}
{\theta}
%
\newcommand{\yparamqubitonestateplusphase}
{$\Yparamqubitonestateplusphase$}
\newcommand{\Yparamqubitonestateplusphase}
{\Yparamqubitbothstateplusphasenot_1}
%
\newcommand{\yparamqubittwostateplusphase}
{$\Yparamqubittwostateplusphase$}
\newcommand{\Yparamqubittwostateplusphase}
{\Yparamqubitbothstateplusphasenot_2}
%
%
\newcommand{\yparamqubitindexstdstateplusphase}
{$\Yparamqubitindexstdstateplusphase$}
\newcommand{\Yparamqubitindexstdstateplusphase}
{{\Yparamqubitbothstateplusphasenot}_{\Yqubitindexstd}}
%
%
%
%
%
%
\newcommand{\yparamqubitbothstateminusphasenot}
{$\Yparamqubitbothstateminusphasenot$}
\newcommand{\Yparamqubitbothstateminusphasenot}
{\phi}
%
%
\newcommand{\yparamqubitonestateminusphase}
{$\Yparamqubitonestateminusphase$}
\newcommand{\Yparamqubitonestateminusphase}
{\Yparamqubitbothstateminusphasenot_1}
%
\newcommand{\yparamqubittwostateminusphase}
{$\Yparamqubittwostateminusphase$}
\newcommand{\Yparamqubittwostateminusphase}
{\Yparamqubitbothstateminusphasenot_2}
%
\newcommand{\yparamqubitindexstdstateminusphase}
{$\Yparamqubitindexstdstateminusphase$}
\newcommand{\Yparamqubitindexstdstateminusphase}
{{\Yparamqubitbothstateminusphasenot}_{\Yqubitindexstd}}
%
%
%
%
%
%
%
%
%
%
%
\newcommand{\ygennumberonerealpart}
{$\Ygennumberonerealpart$}
\newcommand{\Ygennumberonerealpart}
{a_1}
%
%
%
\newcommand{\ygennumbertworealpart}
{$\Ygennumbertworealpart$}
\newcommand{\Ygennumbertworealpart}
{a_2}
%
\newcommand{\ygennumbertwoimagpart}
{$\Ygennumbertwoimagpart$}
\newcommand{\Ygennumbertwoimagpart}
{b_2}
%
%
%
%
%
%
%
%
\newcommand{\ymagfieldnot}
{$\Ymagfieldnot$}
\newcommand{\Ymagfieldnot}
{B}
%
\newcommand{\ymagfieldvec}
{$\Ymagfieldvec$}
\newcommand{\Ymagfieldvec}
{\overrightarrow{\Ymagfieldnot}}
%
%
%
\newcommand{\yhamiltonfieldscale}
{$\Yhamiltonfieldscale$}
\newcommand{\Yhamiltonfieldscale}
{G}
%
%
%
%
%
%
%
%
%
%
%

\newcommand{\yexchangetensorppalvaluexy}
{$\Yexchangetensorppalvaluexy$}
\newcommand{\Yexchangetensorppalvaluexy}
{J_{xy}}

\newcommand{\yexchangetensorppalvaluexyestim}
{$\Yexchangetensorppalvaluexyestim$}
\newcommand{\Yexchangetensorppalvaluexyestim}
{\widehat{J}_{xy}}
%
%
%
%
\newcommand{\yexchangetensorppalvaluexyestimindexone}
{$\Yexchangetensorppalvaluexyestimindexone$}
\newcommand{\Yexchangetensorppalvaluexyestimindexone}
{\widehat{J}_{xy1}}
%
\newcommand{\yexchangetensorppalvaluexyestimindextwo}
{$\Yexchangetensorppalvaluexyestimindextwo$}
\newcommand{\Yexchangetensorppalvaluexyestimindextwo}
{\widehat{J}_{xy2}}
%

%
\newcommand{\yexchangetensorppalvaluexyestimargtwoqubitwritereadtimeintervalindexone}
{$\Yexchangetensorppalvaluexyestimargtwoqubitwritereadtimeintervalindexone$}
\newcommand{\Yexchangetensorppalvaluexyestimargtwoqubitwritereadtimeintervalindexone}
{\Yexchangetensorppalvaluexyestim
(
\Ytwoqubitwritereadtimeintervalindexone
)}
%

%
\newcommand{\yexchangetensorppalvaluexyshiftindetermint}
{$\Yexchangetensorppalvaluexyshiftindetermint$}
\newcommand{\Yexchangetensorppalvaluexyshiftindetermint}
{k_{xy}}
%
\newcommand{\yexchangetensorppalvaluexyshiftindetermintindexone}
{$\Yexchangetensorppalvaluexyshiftindetermintindexone$}
\newcommand{\Yexchangetensorppalvaluexyshiftindetermintindexone}
{k_{xy1}}
%
\newcommand{\yexchangetensorppalvaluexyshiftindetermintindextwo}
{$\Yexchangetensorppalvaluexyshiftindetermintindextwo$}
\newcommand{\Yexchangetensorppalvaluexyshiftindetermintindextwo}
{k_{xy2}}
%
%
%
\newcommand{\yexchangetensorppalvaluexyshiftindetermintestim}
{$\Yexchangetensorppalvaluexyshiftindetermintestim$}
\newcommand{\Yexchangetensorppalvaluexyshiftindetermintestim}
{\widehat{k}_{xy}}
%
\newcommand{\yexchangetensorppalvaluexyshiftindetermintestimindexone}
{$\Yexchangetensorppalvaluexyshiftindetermintestimindexone$}
\newcommand{\Yexchangetensorppalvaluexyshiftindetermintestimindexone}
{\widehat{k}_{xy1}}
%
\newcommand{\yexchangetensorppalvaluexyshiftindetermintestimindextwo}
{$\Yexchangetensorppalvaluexyshiftindetermintestimindextwo$}
\newcommand{\Yexchangetensorppalvaluexyshiftindetermintestimindextwo}
{\widehat{k}_{xy2}}
%
%
%
%
\newcommand{\yexchangetensorppalvaluexyshiftindetermintestimminusactual}
{$\Yexchangetensorppalvaluexyshiftindetermintestimminusactual$}
\newcommand{\Yexchangetensorppalvaluexyshiftindetermintestimminusactual}
{\Delta k _{xy}}
%
%
\newcommand{\yexchangetensorppalvaluexyshiftindetermintestimminusactualindexone}
{$\Yexchangetensorppalvaluexyshiftindetermintestimminusactualindexone$}
\newcommand{\Yexchangetensorppalvaluexyshiftindetermintestimminusactualindexone}
{\Delta k _{xy1}}
%
\newcommand{\yexchangetensorppalvaluexyshiftindetermintestimminusactualindextwo}
{$\Yexchangetensorppalvaluexyshiftindetermintestimminusactualindextwo$}
\newcommand{\Yexchangetensorppalvaluexyshiftindetermintestimminusactualindextwo}
{\Delta k _{xy2}}
%
%
%
\newcommand{\yexchangetensorppalvaluexyshiftindetermintestimmin}
{$\Yexchangetensorppalvaluexyshiftindetermintestimmin$}
\newcommand{\Yexchangetensorppalvaluexyshiftindetermintestimmin}
{\widehat{k}_{xy}^{min}}
%
\newcommand{\yexchangetensorppalvaluexyshiftindetermintestimminindexone}
{$\Yexchangetensorppalvaluexyshiftindetermintestimminindexone$}
\newcommand{\Yexchangetensorppalvaluexyshiftindetermintestimminindexone}
{\widehat{k}_{xy1}^{min}}
%
\newcommand{\yexchangetensorppalvaluexyshiftindetermintestimminindextwo}
{$\Yexchangetensorppalvaluexyshiftindetermintestimminindextwo$}
\newcommand{\Yexchangetensorppalvaluexyshiftindetermintestimminindextwo}
{\widehat{k}_{xy2}^{min}}
%
%
%
\newcommand{\yexchangetensorppalvaluexyshiftindetermintestimmax}
{$\Yexchangetensorppalvaluexyshiftindetermintestimmax$}
\newcommand{\Yexchangetensorppalvaluexyshiftindetermintestimmax}
{\widehat{k}_{xy}^{max}}
%
\newcommand{\yexchangetensorppalvaluexyshiftindetermintestimmaxindexone}
{$\Yexchangetensorppalvaluexyshiftindetermintestimmaxindexone$}
\newcommand{\Yexchangetensorppalvaluexyshiftindetermintestimmaxindexone}
{\widehat{k}_{xy1}^{max}}
%
\newcommand{\yexchangetensorppalvaluexyshiftindetermintestimmaxindextwo}
{$\Yexchangetensorppalvaluexyshiftindetermintestimmaxindextwo$}
\newcommand{\Yexchangetensorppalvaluexyshiftindetermintestimmaxindextwo}
{\widehat{k}_{xy2}^{max}}
%

\newcommand{\yexchangetensorppalvaluez}
{$\Yexchangetensorppalvaluez$}
\newcommand{\Yexchangetensorppalvaluez}
{J_z}

\newcommand{\yexchangetensorppalvaluezestim}
{$\Yexchangetensorppalvaluezestim$}
\newcommand{\Yexchangetensorppalvaluezestim}
{\widehat{J}_z}
%
%
%
%
%
\newcommand{\yexchangetensorppalvaluezestimindexone}
{$\Yexchangetensorppalvaluezestimindexone$}
\newcommand{\Yexchangetensorppalvaluezestimindexone}
{\widehat{J}_{z1}}
%
\newcommand{\yexchangetensorppalvaluezestimindextwo}
{$\Yexchangetensorppalvaluezestimindextwo$}
\newcommand{\Yexchangetensorppalvaluezestimindextwo}
{\widehat{J}_{z2}}
%

%
\newcommand{\yexchangetensorppalvaluezestimargtwoqubitwritereadtimeintervalindextwo}
{$\Yexchangetensorppalvaluezestimargtwoqubitwritereadtimeintervalindextwo$}
\newcommand{\Yexchangetensorppalvaluezestimargtwoqubitwritereadtimeintervalindextwo}
{\Yexchangetensorppalvaluezestim
(
\Ytwoqubitwritereadtimeintervalindextwo
)}
%
%
%
%
%
\newcommand{\yexchangetensorppalvaluezshiftindetermint}
{$\Yexchangetensorppalvaluezshiftindetermint$}
\newcommand{\Yexchangetensorppalvaluezshiftindetermint}
{k_{z}}
%
\newcommand{\yexchangetensorppalvaluezshiftindetermintindexone}
{$\Yexchangetensorppalvaluezshiftindetermintindexone$}
\newcommand{\Yexchangetensorppalvaluezshiftindetermintindexone}
{k_{z1}}
%
\newcommand{\yexchangetensorppalvaluezshiftindetermintindextwo}
{$\Yexchangetensorppalvaluezshiftindetermintindextwo$}
\newcommand{\Yexchangetensorppalvaluezshiftindetermintindextwo}
{k_{z2}}
%
%
%
\newcommand{\yexchangetensorppalvaluezshiftindetermintestim}
{$\Yexchangetensorppalvaluezshiftindetermintestim$}
\newcommand{\Yexchangetensorppalvaluezshiftindetermintestim}
{\widehat{k}_{z}}
%
\newcommand{\yexchangetensorppalvaluezshiftindetermintestimindexone}
{$\Yexchangetensorppalvaluezshiftindetermintestimindexone$}
\newcommand{\Yexchangetensorppalvaluezshiftindetermintestimindexone}
{\widehat{k}_{z1}}
%
\newcommand{\yexchangetensorppalvaluezshiftindetermintestimindextwo}
{$\Yexchangetensorppalvaluezshiftindetermintestimindextwo$}
\newcommand{\Yexchangetensorppalvaluezshiftindetermintestimindextwo}
{\widehat{k}_{z2}}
%
%
%
%
%
\newcommand{\yexchangetensorppalvaluezshiftindetermintestimminusactual}
{$\Yexchangetensorppalvaluezshiftindetermintestimminusactual$}
\newcommand{\Yexchangetensorppalvaluezshiftindetermintestimminusactual}
{\Delta k _{z}}
%
%
\newcommand{\yexchangetensorppalvaluezshiftindetermintestimminusactualindexone}
{$\Yexchangetensorppalvaluezshiftindetermintestimminusactualindexone$}
\newcommand{\Yexchangetensorppalvaluezshiftindetermintestimminusactualindexone}
{\Delta k _{z1}}
%
\newcommand{\yexchangetensorppalvaluezshiftindetermintestimminusactualindextwo}
{$\Yexchangetensorppalvaluezshiftindetermintestimminusactualindextwo$}
\newcommand{\Yexchangetensorppalvaluezshiftindetermintestimminusactualindextwo}
{\Delta k _{z2}}
%
%
%
\newcommand{\yexchangetensorppalvaluezshiftindetermintestimmin}
{$\Yexchangetensorppalvaluezshiftindetermintestimmin$}
\newcommand{\Yexchangetensorppalvaluezshiftindetermintestimmin}
{\widehat{k}_{z}^{min}}
%
\newcommand{\yexchangetensorppalvaluezshiftindetermintestimminindexone}
{$\Yexchangetensorppalvaluezshiftindetermintestimminindexone$}
\newcommand{\Yexchangetensorppalvaluezshiftindetermintestimminindexone}
{\widehat{k}_{z1}^{min}}
%
\newcommand{\yexchangetensorppalvaluezshiftindetermintestimminindextwo}
{$\Yexchangetensorppalvaluezshiftindetermintestimminindextwo$}
\newcommand{\Yexchangetensorppalvaluezshiftindetermintestimminindextwo}
{\widehat{k}_{z2}^{min}}
%
%
%
\newcommand{\yexchangetensorppalvaluezshiftindetermintestimmax}
{$\Yexchangetensorppalvaluezshiftindetermintestimmax$}
\newcommand{\Yexchangetensorppalvaluezshiftindetermintestimmax}
{\widehat{k}_{z}^{max}}
%
\newcommand{\yexchangetensorppalvaluezshiftindetermintestimmaxindexone}
{$\Yexchangetensorppalvaluezshiftindetermintestimmaxindexone$}
\newcommand{\Yexchangetensorppalvaluezshiftindetermintestimmaxindexone}
{\widehat{k}_{z1}^{max}}
%
\newcommand{\yexchangetensorppalvaluezshiftindetermintestimmaxindextwo}
{$\Yexchangetensorppalvaluezshiftindetermintestimmaxindextwo$}
\newcommand{\Yexchangetensorppalvaluezshiftindetermintestimmaxindextwo}
{\widehat{k}_{z2}^{max}}
%
%
%
%
%
%
%
%
%
%
%
%
%
%
%
%
%
%


%
%
%
\newcommand{\ytwoqubitsprobadirxxplusplussumminusminusphasediff}
{$\Ytwoqubitsprobadirxxplusplussumminusminusphasediff$}
\newcommand{\Ytwoqubitsprobadirxxplusplussumminusminusphasediff}
{\Delta \Phi_{1,-1}}
%
%
%
\newcommand{\ytwoqubitsprobadirxxplusplusdiffminusminusfactorreal}
{$\Ytwoqubitsprobadirxxplusplusdiffminusminusfactorreal$}
\newcommand{\Ytwoqubitsprobadirxxplusplusdiffminusminusfactorreal}
{R_{14}}
%
%
%
\newcommand{\ytwoqubitsprobadirxxplusplusdiffminusminusfactorimag}
{$\Ytwoqubitsprobadirxxplusplusdiffminusminusfactorimag$}
\newcommand{\Ytwoqubitsprobadirxxplusplusdiffminusminusfactorimag}
{I_{14}}
%
\newcommand{\ytwoqubitsprobadirxxplusplusdiffminusminusphasediff}
{$\Ytwoqubitsprobadirxxplusplusdiffminusminusphasediff$}
\newcommand{\Ytwoqubitsprobadirxxplusplusdiffminusminusphasediff}
{\Delta \Phi_{1,0}}
%
%
%
%
%
\newcommand{\ytwoqubitsprobadirxxplusplusdiffminusminusphasediffindexd}
{$\Ytwoqubitsprobadirxxplusplusdiffminusminusphasediffindexd$}
\newcommand{\Ytwoqubitsprobadirxxplusplusdiffminusminusphasediffindexd}
{\Delta \Phi_{1,0d}}
%
\newcommand{\ytwoqubitsprobadirxxplusplusdiffminusminusphasediffindexdestim}
{$\Ytwoqubitsprobadirxxplusplusdiffminusminusphasediffindexdestim$}
\newcommand{\Ytwoqubitsprobadirxxplusplusdiffminusminusphasediffindexdestim}
{\widehat{\Delta \Phi}_{1,0d}}
%
%
%
%
%
\newcommand{\ytwoqubitsprobadirxxplusplusdiffminusminusphasediffcosgen}
{$\Ytwoqubitsprobadirxxplusplusdiffminusminusphasediffcosgen$}
\newcommand{\Ytwoqubitsprobadirxxplusplusdiffminusminusphasediffcosgen}
{w_1}
%
%
\newcommand{\ytwoqubitsprobadirxxplusplusdiffminusminusphasediffcosgenestim}
{$\Ytwoqubitsprobadirxxplusplusdiffminusminusphasediffcosgenestim$}
\newcommand{\Ytwoqubitsprobadirxxplusplusdiffminusminusphasediffcosgenestim}
{\widehat{w}_1}
%
%
%
%
\newcommand{\ytwoqubitsprobadirxxplusplusdiffminusminusphasediffsingen}
{$\Ytwoqubitsprobadirxxplusplusdiffminusminusphasediffsingen$}
\newcommand{\Ytwoqubitsprobadirxxplusplusdiffminusminusphasediffsingen}
{w_2}
%
%
\newcommand{\ytwoqubitsprobadirxxplusplusdiffminusminusphasediffsingenestim}
{$\Ytwoqubitsprobadirxxplusplusdiffminusminusphasediffsingenestim$}
\newcommand{\Ytwoqubitsprobadirxxplusplusdiffminusminusphasediffsingenestim}
{\widehat{w}_2}
%
%
%
%



\newcommand{\yopdensity}
{$\Yopdensity$}
\newcommand{\Yopdensity}
{\rho}



\newcommand{\yopdensityindexstdforket}
{$\Yopdensityindexstdforket$}
\newcommand{\Yopdensityindexstdforket}
{\alpha}

\newcommand{\yopdensityketindexstd}
{$\Yopdensityketindexstd$}
\newcommand{\Yopdensityketindexstd}
{
|
\Yketdetermnot
_
{\Yopdensityindexstdforket}
\rangle
}

\newcommand{\yopdensitybraindexstd}
{$\Yopdensitybraindexstd$}
\newcommand{\Yopdensitybraindexstd}
{
\langle
\Yketdetermnot
_
{\Yopdensityindexstdforket}
|
}

%
\newcommand{\yopdensitybranobarindexstd}
{$\Yopdensitybranobarindexstd$}
\newcommand{\Yopdensitybranobarindexstd}
{
\langle
\Yketdetermnot
_
{\Yopdensityindexstdforket}
}

\newcommand{\yopdensityketprobindexstd}
{$\Yopdensityketprobindexstd$}
\newcommand{\Yopdensityketprobindexstd}
{
p
_
{\Yopdensityindexstdforket}
}




\newcommand{\yphysicalquantitystdone}
{$\Yphysicalquantitystdone$}
\newcommand{\Yphysicalquantitystdone}
{A}

\newcommand{\yobservableopstdone}
{$\Yobservableopstdone$}
\newcommand{\Yobservableopstdone}
{\hat{\Yphysicalquantitystdone}}

\newcommand{\yobservableopstdoneindexelone}
{$\Yobservableopstdoneindexelone$}
\newcommand{\Yobservableopstdoneindexelone}
{
k
}

\newcommand{\yobservableopstdoneindexeltwo}
{$\Yobservableopstdoneindexeltwo$}
\newcommand{\Yobservableopstdoneindexeltwo}
{
\ell
}

\newcommand{\yobservableopstdoneelnot}
{$\Yobservableopstdoneelnot$}
\newcommand{\Yobservableopstdoneelnot}
{a}

\newcommand{\yobservableopstdoneelindexstdoneindexstdtwo}
{$\Yobservableopstdoneelindexstdoneindexstdtwo$}
\newcommand{\Yobservableopstdoneelindexstdoneindexstdtwo}
{
\Yobservableopstdoneelnot
_{
\Yobservableopstdoneindexelone
\Yobservableopstdoneindexeltwo
}
}



\newcommand{\yobservableopstdoneeigenbasisindexeigenone}
{$\Yobservableopstdoneeigenbasisindexeigenone$}
\newcommand{\Yobservableopstdoneeigenbasisindexeigenone}
{m}

\newcommand{\yobservableopstdoneeigenbasiseigenketindexone}
{$\Yobservableopstdoneeigenbasiseigenketindexone$}
\newcommand{\Yobservableopstdoneeigenbasiseigenketindexone}
{| 
e_{\Yobservableopstdoneeigenbasisindexeigenone}
\rangle}

\newcommand{\yobservableopstdoneeigenbasiseigenbraindexone}
{$\Yobservableopstdoneeigenbasiseigenbraindexone$}
\newcommand{\Yobservableopstdoneeigenbasiseigenbraindexone}
{\langle
e_{\Yobservableopstdoneeigenbasisindexeigenone}
|}

\newcommand{\yobservableopstdoneeigenbasiseigenbranobarindexone}
{$\Yobservableopstdoneeigenbasiseigenbranobarindexone$}
\newcommand{\Yobservableopstdoneeigenbasiseigenbranobarindexone}
{\langle
e_{\Yobservableopstdoneeigenbasisindexeigenone}
}

\newcommand{\yobservableopstdoneeigenbasiseigenvalindexone}
{$\Yobservableopstdoneeigenbasiseigenvalindexone$}
\newcommand{\Yobservableopstdoneeigenbasiseigenvalindexone}
{
a
_{\Yobservableopstdoneeigenbasisindexeigenone}
}

\newcommand{\yobservableopstdoneeigenbasisketbasiscoefdetermindexstd}
{$\Yobservableopstdoneeigenbasisketbasiscoefdetermindexstd$}
\newcommand{\Yobservableopstdoneeigenbasisketbasiscoefdetermindexstd}
{
\Yketbasiscoefdetermnot
_{
\Yobservableopstdoneeigenbasisindexeigenone
}
^{\prime}
}

\newcommand{\yobservableopstdoneeigenbasisketbasiscoefdetermindexstdop}
{$\Yobservableopstdoneeigenbasisketbasiscoefdetermindexstdop$}
\newcommand{\Yobservableopstdoneeigenbasisketbasiscoefdetermindexstdop}
{
M
_{
\Yobservableopstdoneeigenbasisindexeigenone
}
(
\Yphysicalquantitystdone
)
}

%
%
%
\newcommand{\ysqrtminusone}
{$\Ysqrtminusone$}
\newcommand{\Ysqrtminusone}
{i}
\newcommand{\yboltzmannconst}
{$\Yboltzmannconst$}
\newcommand{\Yboltzmannconst}
{k_B}
%
%
%
%
\newcommand{\ytrace}
{$\Ytrace$}
\newcommand{\Ytrace}
{\mathrm{Tr}}

%
%
%
%
%
%
%
%
%
%
%
%
\newcommand{\ywritereadonestatenb}
{$\Ywritereadonestatenb$}
\newcommand{\Ywritereadonestatenb}
{K}
%
\newcommand{\ywritereadonestatenbmixident}
{$\Ywritereadonestatenbmixident$}
\newcommand{\Ywritereadonestatenbmixident}
{\Ywritereadonestatenb
}
%
%
%
%
%
%
%
%
%
%
%
\newcommand{\yphaseoptfitveccompsyststatetinitandveccompsepsyststateout}
{$\Yphaseoptfitveccompsyststatetinitandveccompsepsyststateout$}
\newcommand{\Yphaseoptfitveccompsyststatetinitandveccompsepsyststateout}
{\phi_{opt}}
%
\newcommand{\yphaseextfitveccompsyststatetinitandveccompsepsyststateout}
{$\Yphaseextfitveccompsyststatetinitandveccompsepsyststateout$}
\newcommand{\Yphaseextfitveccompsyststatetinitandveccompsepsyststateout}
{\phi_{ext}}
%
%
%
\newcommand{\yrmsemoduli}
{$\Yrmsemoduli$}
\newcommand{\Yrmsemoduli}
{RMSE_m}
%
\newcommand{\yrmsecoef}
{$\Yrmsecoef$}
\newcommand{\Yrmsecoef}
{RMSE_c}
%
%
%
%

\newcommand{\yprocessmat}
{$\Yprocessmat$}
\newcommand{\Yprocessmat}
{U}

\newcommand{\yprocessmattimesyprocessinmixedonedensitymatrixeigenunit}
{$\Yprocessmattimesyprocessinmixedonedensitymatrixeigenunit$}
\newcommand{\Yprocessmattimesyprocessinmixedonedensitymatrixeigenunit}
{\Yprocessmat_{g}}

\newcommand{\yprocessmatcolindexstd}
{$\Yprocessmatcolindexstd$}
\newcommand{\Yprocessmatcolindexstd}
{k}

\newcommand{\yprocessmatcolumnnot}
{$\Yprocessmatcolumnnot$}
\newcommand{\Yprocessmatcolumnnot}
{u}

\newcommand{\yprocessmatcolumnfirst}
{$\Yprocessmatcolumnfirst$}
\newcommand{\Yprocessmatcolumnfirst}
{\Yprocessmatcolumnnot
_{1}}

\newcommand{\yprocessmatcolumnlast}
{$\Yprocessmatcolumnlast$}
\newcommand{\Yprocessmatcolumnlast}
{\Yprocessmatcolumnnot
_{\Yketspacedim}}

\newcommand{\yprocessinmixedonedensitymatrix}
{$\Yprocessinmixedonedensitymatrix$}
\newcommand{\Yprocessinmixedonedensitymatrix}
{\Yopdensity_{1}}

\newcommand{\yprocessinmixedonedensitymatrixeigendiag}
{$\Yprocessinmixedonedensitymatrixeigendiag$}
\newcommand{\Yprocessinmixedonedensitymatrixeigendiag}
{\Yopdensity_{I1D}}

\newcommand{\yprocessinmixedonedensitymatrixeigenunit}
{$\Yprocessinmixedonedensitymatrixeigenunit$}
\newcommand{\Yprocessinmixedonedensitymatrixeigenunit}
{U_{I1}}

\newcommand{\yprocessoutmixedonedensitymatrix}
{$\Yprocessoutmixedonedensitymatrix$}
\newcommand{\Yprocessoutmixedonedensitymatrix}
{\Yopdensity_{2}}

\newcommand{\yketnotstdone}
{$\Yketnotstdone$}
\newcommand{\Yketnotstdone}
{\Psi}

\newcommand{\yprocessinpuretwoket}
{$\Yprocessinpuretwoket$}
\newcommand{\Yprocessinpuretwoket}
{\lvert
\Yketnotstdone_{1}
\rangle
}

\newcommand{\yprocessoutpuretwoket}
{$\Yprocessoutpuretwoket$}
\newcommand{\Yprocessoutpuretwoket}
{\lvert
\Yketnotstdone_{2}
\rangle
}

\newcommand{\yprocessinmixedtwodensitymatrix}
{$\Yprocessinmixedtwodensitymatrix$}
\newcommand{\Yprocessinmixedtwodensitymatrix}
{\Yopdensity_{5}}

\newcommand{\yprocessoutmixedtwodensitymatrix}
{$\Yprocessoutmixedtwodensitymatrix$}
\newcommand{\Yprocessoutmixedtwodensitymatrix}
{\Yopdensity_{6}}

\newcommand{\yprocessmatestimone}
{$\Yprocessmatestimone$}
\newcommand{\Yprocessmatestimone}
{\widehat{\Yprocessmat}_{1}}

\newcommand{\yprocessmatestimtwo}
{$\Yprocessmatestimtwo$}
\newcommand{\Yprocessmatestimtwo}
{\widehat{\Yprocessmat}_{2}}

\newcommand{\yprocessmatestimtwocolindexstd}
{$\Yprocessmatestimtwo$colindexstd}
\newcommand{\Yprocessmatestimtwocolindexstd}
{
\Yprocessmatcolindexstd
}

\newcommand{\yprocessmatestimtwophaseindetermnot}
{$\Yprocessmatestimtwophaseindetermnot$}
\newcommand{\Yprocessmatestimtwophaseindetermnot}
{\phi}

\newcommand{\yprocessmatestimtwophaseindetermfirst}
{$\Yprocessmatestimtwophaseindetermfirst$}
\newcommand{\Yprocessmatestimtwophaseindetermfirst}
{\Yprocessmatestimtwophaseindetermnot
_{1}}

\newcommand{\yprocessmatestimtwophaseindetermsecond}
{$\Yprocessmatestimtwophaseindetermsecond$}
\newcommand{\Yprocessmatestimtwophaseindetermsecond}
{\Yprocessmatestimtwophaseindetermnot
_{2}}

\newcommand{\yprocessmatestimtwophaseindetermindexstd}
{$\Yprocessmatestimtwophaseindetermindexstd$}
\newcommand{\Yprocessmatestimtwophaseindetermindexstd}
{\Yprocessmatestimtwophaseindetermnot
_{\Yprocessmatcolindexstd}}

\newcommand{\yprocessmatestimtwophaseindetermlast}
{$\Yprocessmatestimtwophaseindetermlast$}
\newcommand{\Yprocessmatestimtwophaseindetermlast}
{\Yprocessmatestimtwophaseindetermnot
_{\Yketspacedim}}

\newcommand{\yprocessmatestimtwophaseindetermmatrixdiag}
{$\Yprocessmatestimtwophaseindetermmatrixdiag$}
\newcommand{\Yprocessmatestimtwophaseindetermmatrixdiag}
{D
_{\Yprocessmatestimtwophaseindetermnot}}

\newcommand{\yprocessmatestimtwobis}
{$\Yprocessmatestimtwobis$}
\newcommand{\Yprocessmatestimtwobis}
{\widehat{\Yprocessmat}_{c}}

\newcommand{\yprocessinmixedonedensitymatrixestim}
{$\Yprocessinmixedonedensitymatrixestim$}
\newcommand{\Yprocessinmixedonedensitymatrixestim}
{\widehat{\Yopdensity}_{1}}

\newcommand{\yprocessoutmixedonedensitymatrixestim}
{$\Yprocessoutmixedonedensitymatrixestim$}
\newcommand{\Yprocessoutmixedonedensitymatrixestim}
{\widehat{\Yopdensity}_{2}}

\newcommand{\yprocessoutmixedonedensitymatrixestimherm}
{$\Yprocessoutmixedonedensitymatrixestimherm$}
\newcommand{\Yprocessoutmixedonedensitymatrixestimherm}
{\widehat{\Yopdensity}_{3}}

\newcommand{\yprocessoutmixedonedensitymatrixestimhermunittrace}
{$\Yprocessoutmixedonedensitymatrixestimhermunittrace$}
\newcommand{\Yprocessoutmixedonedensitymatrixestimhermunittrace}
{\widehat{\Yopdensity}_{4}}

\newcommand{\yprocessoutmixedonedensitymatrixestimhermunittraceeigenmatval}
{$\Yprocessoutmixedonedensitymatrixestimhermunittraceeigenmatval$}
\newcommand{\Yprocessoutmixedonedensitymatrixestimhermunittraceeigenmatval}
{D_{4}}

\newcommand{\yprocessoutmixedonedensitymatrixestimhermunittraceeigenmatvec}
{$\Yprocessoutmixedonedensitymatrixestimhermunittraceeigenmatvec$}
\newcommand{\Yprocessoutmixedonedensitymatrixestimhermunittraceeigenmatvec}
{V_{4}}

\newcommand{\yprocessmatestimthree}
{$\Yprocessmatestimthree$}
\newcommand{\Yprocessmatestimthree}
{\widehat{\Yprocessmat}_{5}}

\newcommand{\yprocessoutpuretwoketestim}
{$\Yprocessoutpuretwoketestim$}
\newcommand{\Yprocessoutpuretwoketestim}
{\lvert
\widehat{\Yketnotstdone}_{2}
\rangle
}

\newcommand{\yprocessoutpuretwoketestimphase}
{$\Yprocessoutpuretwoketestimphase$}
\newcommand{\Yprocessoutpuretwoketestimphase}
{\theta}

\newcommand{\yprocessmatestimthreeketinterm}
{$\Yprocessmatestimthreeketinterm$}
\newcommand{\Yprocessmatestimthreeketinterm}
{\lvert
\Yketnotstdone
_{3}
\rangle
}

\newcommand{\yprocessmatestimthreeketintermtwo}
{$\Yprocessmatestimthreeketintermtwo$}
\newcommand{\Yprocessmatestimthreeketintermtwo}
{\lvert
\Yketnotstdone
_{4}
\rangle
}

\newcommand{\yprocessoutmixedtwodensitymatrixestimhermunittrace}
{$\Yprocessoutmixedtwodensitymatrixestimhermunittrace$}
\newcommand{\Yprocessoutmixedtwodensitymatrixestimhermunittrace}
{\widehat{\Yopdensity}_{8}}

\newcommand{\yprocessmatestimthreedensitymatrixinterm}
{$\Yprocessmatestimthreedensitymatrixinterm$}
\newcommand{\Yprocessmatestimthreedensitymatrixinterm}
{\Yopdensity
_{9}
}

\newcommand{\yconjug}
{$\Yconjug$}
\newcommand{\Yconjug}
{^*}

\newcommand{\ytransconjug}
{$\Ytransconjug$}
\newcommand{\Ytransconjug}
{^{\dag}}

\newcommand{\ymathspace}
{$\Ymathspace$}
\newcommand{\Ymathspace}
{\hspace{1mm}}

\newcommand{\yeigenvalnbmax}
{$\Yeigenvalnbmax$}
\newcommand{\Yeigenvalnbmax}
{n_{max}}

\newcommand{\yprocessinmixedonedensitymatrixvalmultiplicity}
{$\Yprocessinmixedonedensitymatrixvalmultiplicity$}
\newcommand{\Yprocessinmixedonedensitymatrixvalmultiplicity}
{\Yketspacedim_{1}}

\newcommand{\yprocessinmixedonedensitymatrixvalnb}
{$\Yprocessinmixedonedensitymatrixvalnb$}
\newcommand{\Yprocessinmixedonedensitymatrixvalnb}
{\Yketspacedim_{2}}

\newcommand{\yprocessinmixedonedensitymatrixvaldiagnot}
{$\Yprocessinmixedonedensitymatrixvaldiagnot$}
\newcommand{\Yprocessinmixedonedensitymatrixvaldiagnot}
{r}

\newcommand{\yprocessinmixedonedensitymatrixvaldiagdifffirst}
{$\Yprocessinmixedonedensitymatrixvaldiagdifffirst$}
\newcommand{\Yprocessinmixedonedensitymatrixvaldiagdifffirst}
{\Yprocessinmixedonedensitymatrixvaldiagnot_{1}}

\newcommand{\yprocessinmixedonedensitymatrixvaldiagdiffsecond}
{$\Yprocessinmixedonedensitymatrixvaldiagdiffsecond$}
\newcommand{\Yprocessinmixedonedensitymatrixvaldiagdiffsecond}
{\Yprocessinmixedonedensitymatrixvaldiagnot_{2}}

\newcommand{\yprocessinmixedonedensitymatrixvaldiagdifflast}
{$\Yprocessinmixedonedensitymatrixvaldiagdifflast$}
\newcommand{\Yprocessinmixedonedensitymatrixvaldiagdifflast}
{\Yprocessinmixedonedensitymatrixvaldiagnot_{\Yprocessinmixedonedensitymatrixvalnb}}

\newcommand{\yprocessinmixedonedensitymatrixvaldiagdiffeigensubsetindexstd}
{$\Yprocessinmixedonedensitymatrixvaldiagdiffeigensubsetindexstd$}
\newcommand{\Yprocessinmixedonedensitymatrixvaldiagdiffeigensubsetindexstd}
{\Yprocessinmixedonedensitymatrixvaldiagnot
_{\Yprocessinmixedoneeigensubsetindexstd}}

\newcommand{\yprocessinmixedtwodensitymatrixvaldiagdiffeigensubsetindexstd}
{$\Yprocessinmixedtwodensitymatrixvaldiagdiffeigensubsetindexstd$}
\newcommand{\Yprocessinmixedtwodensitymatrixvaldiagdiffeigensubsetindexstd}
{\Yprocessinmixedonedensitymatrixvaldiagnot
_{\Yprocessinmixedtwoeigensubsetindexstd}}

\newcommand{\yprocessinmixedoneeigensubsetindexstd}
{$\Yprocessinmixedoneeigensubsetindexstd$}
\newcommand{\Yprocessinmixedoneeigensubsetindexstd}
{m_
1}

\newcommand{\yprocessinmixedtwoeigensubsetindexstd}
{$\Yprocessinmixedtwoeigensubsetindexstd$}
\newcommand{\Yprocessinmixedtwoeigensubsetindexstd}
{m_
2}

\newcommand{\yprocessinmixedoneeigensubsetindexstdvalindexstd}
{$\Yprocessinmixedoneeigensubsetindexstdvalindexstd$}
\newcommand{\Yprocessinmixedoneeigensubsetindexstdvalindexstd}
{n_
1}

\newcommand{\yprocessinmixedtwoeigensubsetindexstdvalindexstd}
{$\Yprocessinmixedtwoeigensubsetindexstdvalindexstd$}
\newcommand{\Yprocessinmixedtwoeigensubsetindexstdvalindexstd}
{n_
2}

\newcommand{\yprocessinmixedoneeigensubsetindexstdspace}
{$\Yprocessinmixedoneeigensubsetindexstdspace$}
\newcommand{\Yprocessinmixedoneeigensubsetindexstdspace}
{{\cal S}_{%
1,
\Yprocessinmixedoneeigensubsetindexstd}}

\newcommand{\yprocessinmixedtwoeigensubsetindexstdspace}
{$\Yprocessinmixedtwoeigensubsetindexstdspace$}
\newcommand{\Yprocessinmixedtwoeigensubsetindexstdspace}
{{\cal S}_{%
2,
\Yprocessinmixedtwoeigensubsetindexstd}}

\newcommand{\ytwostageprocessinmixedtwodensitymatrix}
{$\Ytwostageprocessinmixedtwodensitymatrix$}
\newcommand{\Ytwostageprocessinmixedtwodensitymatrix}
{\Yopdensity_{5}}

\newcommand{\ytwostageprocessoutmixedtwodensitymatrix}
{$\Ytwostageprocessoutmixedtwodensitymatrix$}
\newcommand{\Ytwostageprocessoutmixedtwodensitymatrix}
{\Yopdensity_{6}}

\newcommand{\ytwostageprocessmatestimthree}
{$\Ytwostageprocessmatestimthree$}
\newcommand{\Ytwostageprocessmatestimthree}
{\widehat{\Yprocessmat}_{3}}

\newcommand{\yprocessoutmixedtwodensitymatrixestim}
{$\Yprocessoutmixedtwodensitymatrixestim$}
\newcommand{\Yprocessoutmixedtwodensitymatrixestim}
{\widehat{\Yopdensity}_{6}}

\newcommand{\ytwostageprocessoutmixedtwodensitymatrixestimherm}
{$\Ytwostageprocessoutmixedtwodensitymatrixestimherm$}
\newcommand{\Ytwostageprocessoutmixedtwodensitymatrixestimherm}
{\widehat{\Yopdensity}_{7}}

\newcommand{\ytwostageprocessoutmixedtwodensitymatrixestimhermunittrace}
{$\Ytwostageprocessoutmixedtwodensitymatrixestimhermunittrace$}
\newcommand{\Ytwostageprocessoutmixedtwodensitymatrixestimhermunittrace}
{\widehat{\Yopdensity}_{8}}

\newcommand{\ytwostageprocessoutmixedtwodensitymatrixestimhermunittraceeigenmatval}
{$\Ytwostageprocessoutmixedtwodensitymatrixestimhermunittraceeigenmatval$}
\newcommand{\Ytwostageprocessoutmixedtwodensitymatrixestimhermunittraceeigenmatval}
{D_{8}}

\newcommand{\ytwostageprocessoutmixedtwodensitymatrixestimhermunittraceeigenmatvec}
{$\Ytwostageprocessoutmixedtwodensitymatrixestimhermunittraceeigenmatvec$}
\newcommand{\Ytwostageprocessoutmixedtwodensitymatrixestimhermunittraceeigenmatvec}
{V_{8}}

\newcommand{\ytwostageprocessmatestimfour}
{$\Ytwostageprocessmatestimfour$}
\newcommand{\Ytwostageprocessmatestimfour}
{\widehat{\Yprocessmat}_{4}}

\newcommand{\ymultistageyprocessinmixeddensitymatrixblocknb}
{$\Ymultistageyprocessinmixeddensitymatrixblocknb$}
\newcommand{\Ymultistageyprocessinmixeddensitymatrixblocknb}
{b_{n}}

\newcommand{\ymultistageyprocessinmixeddensitymatrixblocknblogbasetwo}
{$\Ymultistageyprocessinmixeddensitymatrixblocknblogbasetwo$}
\newcommand{\Ymultistageyprocessinmixeddensitymatrixblocknblogbasetwo}
{{b_{\ell}}}

\newcommand{\ymultistageyprocessinmixeddensitymatrixblocknblogbasetwomax}
{$\Ymultistageyprocessinmixeddensitymatrixblocknblogbasetwomax$}
\newcommand{\Ymultistageyprocessinmixeddensitymatrixblocknblogbasetwomax}
{{b_{\ell m}}}
\newcommand{\ymultistageyprocessinmixeddensitymatrixblocknblogbasetworevers}
{$\Ymultistageyprocessinmixeddensitymatrixblocknblogbasetworevers$}
\newcommand{\Ymultistageyprocessinmixeddensitymatrixblocknblogbasetworevers}
{{b_{r}}}

\newcommand{\ymultistageyprocessinmixeddensitymatrixblocksize}
{$\Ymultistageyprocessinmixeddensitymatrixblocksize$}
\newcommand{\Ymultistageyprocessinmixeddensitymatrixblocksize}
{b_{s}}

\newcommand{\ymultistageyprocessinmixeddensitymatrixblockindex}
{$\Ymultistageyprocessinmixeddensitymatrixblockindex$}
\newcommand{\Ymultistageyprocessinmixeddensitymatrixblockindex}
{b_{i}}

\newcommand{\ymultistageblocknblogbasetwoeigensubsetindexzero}
{$\Ymultistageblocknblogbasetwoeigensubsetindexzero$}
\newcommand{\Ymultistageblocknblogbasetwoeigensubsetindexzero}
{m_{0}}

\newcommand{\ymultistageblocknblogbasetwoeigensubsetindexzerospace}
{$\Ymultistageblocknblogbasetwoeigensubsetindexzerospace$}
\newcommand{\Ymultistageblocknblogbasetwoeigensubsetindexzerospace}
{{\cal S}_{
0
,
\Ymultistageblocknblogbasetwoeigensubsetindexzero}}

\newcommand{\ymultistageblocknblogbasetwoeigensubsetindexone}
{$\Ymultistageblocknblogbasetwoeigensubsetindexone$}
\newcommand{\Ymultistageblocknblogbasetwoeigensubsetindexone}
{m_{1}}

\newcommand{\ymultistageblocknblogbasetwoeigensubsetindexonespace}
{$\Ymultistageblocknblogbasetwoeigensubsetindexonespace$}
\newcommand{\Ymultistageblocknblogbasetwoeigensubsetindexonespace}
{{\cal S}_{
1
,
\Ymultistageblocknblogbasetwoeigensubsetindexone}}

\newcommand{\ymultistageblocknblogbasetwoeigensubsetindextwo}
{$\Ymultistageblocknblogbasetwoeigensubsetindextwo$}
\newcommand{\Ymultistageblocknblogbasetwoeigensubsetindextwo}
{m_{2}}

\newcommand{\ymultistageblocknblogbasetwoeigensubsetindextwospace}
{$\Ymultistageblocknblogbasetwoeigensubsetindextwospace$}
\newcommand{\Ymultistageblocknblogbasetwoeigensubsetindextwospace}
{{\cal S}_{
2
,
\Ymultistageblocknblogbasetwoeigensubsetindextwo}}

\newcommand{\ymultistageblocknblogbasetwoeigensubsetindexstd}
{$\Ymultistageblocknblogbasetwoeigensubsetindexstd$}
\newcommand{\Ymultistageblocknblogbasetwoeigensubsetindexstd}
{m_{\Ymultistageyprocessinmixeddensitymatrixblocknblogbasetwo}}

\newcommand{\ymultistageblocknblogbasetwoeigensubsetindexshiftedstd}
{$\Ymultistageblocknblogbasetwoeigensubsetindexshiftedstd$}
\newcommand{\Ymultistageblocknblogbasetwoeigensubsetindexshiftedstd}
{\mu_{\Ymultistageyprocessinmixeddensitymatrixblocknblogbasetwo}}

\newcommand{\ymultistageblocknblogbasetwoeigensubsetindexstdspace}
{$\Ymultistageblocknblogbasetwoeigensubsetindexstdspace$}
\newcommand{\Ymultistageblocknblogbasetwoeigensubsetindexstdspace}
{{\cal S}_{
\Ymultistageyprocessinmixeddensitymatrixblocknblogbasetwo
,
\Ymultistageblocknblogbasetwoeigensubsetindexstd}}

\newcommand{\ymultistageblocknblogbasetwoeigensubsetindexstdinvaldiagdiffeigensubsetindexstd}
{$\Ymultistageblocknblogbasetwoeigensubsetindexstdinvaldiagdiffeigensubsetindexstd$}
\newcommand{\Ymultistageblocknblogbasetwoeigensubsetindexstdinvaldiagdiffeigensubsetindexstd}
{\Yprocessinmixedonedensitymatrixvaldiagnot
_{\Ymultistageblocknblogbasetwoeigensubsetindexstd}}

\newcommand{\ymultistageblocknblogbasetwoeigensubsetindexshiftedanyspacenot}
{$\Ymultistageblocknblogbasetwoeigensubsetindexshiftedanyspacenot$}
\newcommand{\Ymultistageblocknblogbasetwoeigensubsetindexshiftedanyspacenot}
{{\cal T}}

\newcommand{\ymultistageblocknblogbasetwoeigensubsetindexshiftedstdspace}
{$\Ymultistageblocknblogbasetwoeigensubsetindexshiftedstdspace$}
\newcommand{\Ymultistageblocknblogbasetwoeigensubsetindexshiftedstdspace}
{
\Ymultistageblocknblogbasetwoeigensubsetindexshiftedanyspacenot
_{
\Ymultistageyprocessinmixeddensitymatrixblocknblogbasetwo
,
\Ymultistageblocknblogbasetwoeigensubsetindexshiftedstd}}

\newcommand{\yketestimflucwidth}
{$\Yketestimflucwidth$}
\newcommand{\Yketestimflucwidth}
{w}

\newcommand{\yqubitnb}
{$\Yqubitnb$}
\newcommand{\Yqubitnb}
{q}

\newcommand{\ysubspaceintersecrecurs}
{$\Ysubspaceintersecrecurs$}
\newcommand{\Ysubspaceintersecrecurs}
{{\cal I}}

\newcommand{\yfaire}[1]
           {~\\
            . \hrulefill \framebox{DEBUT : A FAIRE} \hrulefill .
            \\
            ~\\
            {
            \scriptsize
            #1
            }
            ~\\
            . \hrulefill \framebox{\footnotesize Fin : \`a faire} \hrulefill .
            ~\\
            }

\newcommand{\faire}[1]{}

\newcommand{\yverifierjustesse}[1]
{{\color{red}#1}}

\newcommand{\ytextartitionehundredninetyfourvonemodifstepone}[1]
{#1}

\newcommand{\ytextartitionehundredninetyfourvonemodifsteptwo}[1]
{#1}

\newcommand{\ytextartitionehundredninetyfourvonemodifstepthree}[1]
{#1}

\newcommand{\ytextartitionehundredninetyfourvonemodifstepfour}[1]
{#1}

\newcommand{\ytextartitionehundredninetyfourvonemodifstepfive}[1]
{#1}

\newcommand{\ytextartitionehundredninetyfourvthreemodifstepone}[1]
{#1}

\newcommand{\ytextartitionehundredninetyfourvfourmodifstepone}[1]
{#1}

\newcommand{\ytextartitionehundredninetyfourvfivemodifstepone}[1]
{#1}

\history{}
\doi{: None}

\title{%
Multi-stage
tomography
based on eigenanalysis
for high-dimensional 
dense
unitary
\ytextartitionehundredninetyfourvfourmodifstepone{processes in
gate-based quantum computers}%
}
\author{\uppercase{Yannick Deville}\authorrefmark{1}, \IEEEmembership{%
Member,
IEEE},
and
\uppercase{Alain Deville\authorrefmark{2}%
}%
}
\address[1]{%
Universit\'e de Toulouse%
,
UPS, CNRS, CNES,
OMP,
IRAP
(Institut de Recherche en Astrophysique et Plan\'etologie),
F-31400
Toulouse,
France
(email: yannick.deville@irap.omp.eu)%
}
\address[2]{%
Aix-Marseille Universit\'e,
CNRS%
,
IM2NP UMR 7334%
,
F-%
13397
Marseille, France
(email: alain.deville@univ-amu.fr)%
}

\markboth
{\ytextartitionehundredninetyfourvfourmodifstepone{Y. Deville, A. Deville:
Tomography
of high-dimensional 
dense
unitary
quantum processes}}
{\ytextartitionehundredninetyfourvfourmodifstepone{Y. Deville, A. Deville:
Tomography
of high-dimensional 
dense
unitary
quantum processes}}

\corresp{Corresponding author: 
Yannick Deville (email: yannick.deville@irap.omp.eu).}

\begin{abstract}
Quantum Process Tomography (QPT) methods aim at
identifying, i.e. estimating,
a 
quantum process.
QPT 
is a major quantum information processing 
tool, since it especially allows one to 
experimentally
characterize
the actual behavior of
quantum gates,
that may be used as
the building blocks
of 
quantum computers.
We here 
consider 
unitary,
possibly
dense (i.e. without sparsity 
{constraints)}
processes%
{,
which corresponds to 
{isolated}
systems.
Moreover,}
we 
\ytextartitionehundredninetyfourvfourmodifstepone{develop}
QPT methods that are applicable to
a 
significant
number of qubits 
and hence to a high state space dimension%
{, which 
allows one to tackle more complex problems.}
Using the unitarity of the process allows us
to develop 
methods that first achieve part of QPT by 
performing an
eigenanalysis
of the estimated density matrix
of a process output.
Building upon this idea, we first develop a class
of complete
algorithms that are single-stage,
\ytextartitionehundredninetyfourvfivemodifstepone{in the
sense that they}
use only one eigendecomposition.
We then extend them to multiple-stage
algorithms
(i.e. with several eigendecompositions),
in order to address high-dimensional
state spaces 
{while
being less limited}
by the
estimation errors made when using an
arbitrary given Quantum State Tomography 
{(QST)}
algorithm
as a building block of our overall methods.
We 
{first
propose}
two-stage methods 
{and we then extend them to dichotomic methods, whose number of
stages increases with the considered state space
dimension.}
The relevance of 
our methods
is validated 
\ytextartitionehundredninetyfourvfourmodifstepone{with}
simulations.
\ytextartitionehundredninetyfourvfourmodifstepone{Single-stage
and two-stage methods 
efficiently apply
up to 13 qubits 
on a standard PC
\ytextartitionehundredninetyfourvfivemodifstepone{(with 16 GB
of RAM)}%
.
Multi-stage 
methods 
yield an even
higher accuracy
(Normalized Mean Square Error
down to $
10^{-3}
$).}
\end{abstract}

\begin{keywords}
\ytextartitionehundredninetyfourvfourmodifstepone{%
eigenanalysis of density matrix,
high-dimensional quantum state space,
quantum computing,
quantum gate characterization,
quantum machine learning
(supervised / semi-supervised / unsupervised),
quantum process tomography,
quantum state tomography,
unitary quantum gate}%
\end{keywords}

\titlepgskip=-15pt

\maketitle

\section{Introduction}
\label{sec-intro}
Quantum process tomography (QPT) 
\ytextartitionehundredninetyfourvthreemodifstepone{may be 
considered as the quantum counterpart of
classical system identification
\cite{booknielsen}, 
\ytextartitionehundredninetyfourvfivemodifstepone{that is,}
system parameter estimation. It
was
\ytextartitionehundredninetyfourvfivemodifstepone{mainly}%
\footnote{%
See also
\cite{booknielsen}
p. 398 for the other earliest references.%
}
introduced in 1997 in 
\cite{amq30official}
and
then extended 
(together with quantum process learning)
e.g.
in
\ytextartitionehundredninetyfourvfivemodifstepone{%
\cite{booknielsen,
amq-baldwin-physreva-2014,
amq75,
amq45official,
amq50-physical-review,
amq59,
amq48,
amq52-physical-review,
amq56,
amq41,
amq182,
amq183,
PhysRevA.63.020101,
Sakhouf_2023,
10383498,
StanchevVitanovSciRep2024,
xiao2024twostagesolutionquantumprocess-official-10814064,
PRXQuantum.5.010325,
dobrynin2024compressedsensinglindbladianquantumtomography,
PhysRevA.108.032419,
amoi6-172,
amq158,amq159,amq154officiel,EscandonMonardes2024estimationofhigh,PRXQuantum.5.040306}.%
}
QPT}
is a major 
quantum data processing tool, since it especially
allows one to
characterize the actual behavior of quantum gates, that
may
be
used as the building blocks of quantum computers.
\ytextartitionehundredninetyfourvthreemodifstepone{This
gate-based (or circuit-based) approach 
builds upon the unitary transforms performed by
quantum gates 
(see Chapter 4 of 
\cite{booknielsen}). It
is currently
one of the most prominent
quantum computing paradigms
and various commercial products based on this
paradigm are available.
This e.g. includes IBM's Qiskit environment
(see
\cite{IBM-webpage-quantum,IBM-webpage-quantum-qiskit,IBM-webpage-quantum-circuit-library}
and especially
\cite{IBM-webpage-quantum-circuit-library-standard-gates}
for the use of reversible unitary gates),
Amazon's Braket tool
(see
\cite{Amazon-webpage-braket-terms}
especially for 
reversible transformations),
and Mathworks's software package
(see
\cite{Mathworks-quantum-intro}
e.g. for their statement
``Quantum gates represent reversible operations that transform the
quantum state according to unitary matrices'')%
.}
\ytextartitionehundredninetyfourvthreemodifstepone{%
Due to the importance of the above-mentioned
{\it
unitary} processes, various methods
dedicated to such processes have been proposed,
for performing
unitary
quantum process estimation/learning
(see e.g.
\cite{amq-baldwin-physreva-2014,amoi6-172,amq183,amq158,amq159,amq154officiel,EscandonMonardes2024estimationofhigh,PRXQuantum.5.040306})
and e.g. for the
related topic of the
inversion
and
transposition
of
unknown
unitary
operations
\cite{odake2024analyticallowerboundnumber}.}

QPT is a ```data-driven'' approach, and can thus be considered as a 
type of quantum
machine learning (QML) problem
(see
\cite{amoi-submitted-moi-m2garss2024}
for a taxonomy of QML problems),
because
the input-output transform performed by the
considered process is 
generally
inferred 
from 
the values of
input quantum states 
and/or
from measurement results for output quantum states of that process.
More precisely, 
usual QPT methods require one to 
fix a priori 
and hence
know 
the above
input 
\ytextartitionehundredninetyfourvonemodifsteptwo{``values''
(i.e. states)}
of the process
and 
\ytextartitionehundredninetyfourvonemodifsteptwo{to use
information about the corresponding output ``values''
(i.e. states),
that is derived from measurements performed at the
output of the considered process.
This}
corresponds to the
\ytextartitionehundredninetyfourvfivemodifstepone{supervised 
version 
of 
machine learning,
also called the
non-blind version 
of 
machine learning}
\cite{amoi6-148}.
In contrast,
the other extreme case is the one which requires the most
limited knowledge about the process input
whereas output measurements are still used, 
and this
corresponds to the 
unsupervised, 
\ytextartitionehundredninetyfourvfivemodifstepone{or}
blind, QPT (BQPT) configuration
\cite{amoi6-148}.
BQPT methods were
introduced in \cite{amoi6-46}
and then especially developed in
\cite{amoi6-79,amoi6-118,amoi6-148}.
In that case, the individual input state values are 
completely unknown,
that is,
neither fixed a priori
nor estimated by
means of measurements:
in
\cite{amoi6-46,amoi6-79},
the input consists of
unknown 
unentangled
deterministic-coefficients pure states
(DCPS);
in
\cite{amoi6-118,amoi6-148},
it
consists of
random-coefficient pure states (RCPS)
and
the 
information available about these RCPS
is restricted to
some of their global, i.e. statistical,
properties.
DCPS are the usual form of pure states of quantum physics,
whereas
RCPS were
introduced in
\cite{amoi5-31}
and then discussed in various related papers,
especially including 
\cite{amoi-moi-journalqip-2023}.

Between the above two extreme cases, 
a recently
introduced approach 
(see
\cite{amoi6-172}
and references therein)
was claimed to be semi-blind (i.e. semi-supervised), 
because
it does not require the 
values of
the input quantum states to be fixed a priori,
but it estimates them, by also performing
measurements at the input of the process
for
part of the available copies of these
input
states.
This
method thus allows one to
use 
almost
any 
values for the input states,
whereas various supervised QPT methods are more
constraining because they are only applicable
to a specific, predefined, set of input values.

As detailed below,
the different parts and variants of the QPT algorithms proposed in the
present paper cover all above three types of learning.
Some of them require their input states to be known,
so that these states may be fixed a priori (supervised learning) or
estimated from measurements
(semi-supervised learning).
Other parts and variants of our algorithms use input states whose
values are not known, but that are required to meet a given
condition (unsupervised learning).
Besides, part of the proposed algorithms use
``pure states''%
\ytextartitionehundredninetyfourvfivemodifstepone{,
where
we employ the term ``pure states'' in the
above-defined usual sense,
that is, DCPS. In contrast, other 
proposed algorithms}
use
statistical mixtures.

Within the above framework, we focus on an
original case which combines 
\ytextartitionehundredninetyfourvonemodifsteptwo{
the following
two features.
First,
the considered process is constrained 
to
be unitary, 
which corresponds to addressing
a system that is
\ytextartitionehundredninetyfourvonemodifstepfour{isolated}
from its
environment.
Second, 
we are mainly interested in
the case when 
the process
has a high dimension,
\ytextartitionehundredninetyfourvfivemodifstepone{because}
it 
involves a 
significant
number of qubits%
\ytextartitionehundredninetyfourvfivemodifstepone{.
This}
allows one to tackle more complex problems.}
We stress that we do not set any other 
\ytextartitionehundredninetyfourvonemodifsteptwo{conditions}
on
the considered process.
In particular, 
\ytextartitionehundredninetyfourvonemodifsteptwo{we allow 
(but do not request)
the matrix
that represents that process in the considered basis to be a so-called
dense, i.e. non-sparse, matrix
(we here use the standard 
\ytextartitionehundredninetyfourvfivemodifstepone{terminology:}
a sparse matrix
is a matrix
that has a large percentage of
\ytextartitionehundredninetyfourvfivemodifstepone{zeros;}
see
e.g.
\cite{mathworks-sparse-matrix}).}
\ytextartitionehundredninetyfourvonemodifsteptwo{In contrast,}
some approaches
\ytextartitionehundredninetyfourvonemodifsteptwo{from the
literature}
for QPT and the related topic of Hamiltonian estimation
\ytextartitionehundredninetyfourvonemodifsteptwo{request
sparsity}
(see e.g.
\cite{amq181,amq182})%
\ytextartitionehundredninetyfourvonemodifsteptwo{, but this}
was considered to be less relevant than unitarity in
\cite{amoi6-172}
and
\ytextartitionehundredninetyfourvonemodifsteptwo{we explained above 
when the unitarity assumption is
relevant.}

Concerning
the ``mathematical principles''
of the proposed QPT methods, 
in this paper we
show
that
one of the key ingredients of
our approach is the eigendecomposition
of the (estimated)
density matrix of a mixed output
state of the considered process.
Before moving to technical details and a
variety of required extensions of this
idea 
in the subsequent sections, 
this major
idea 
can here be outlined
as follows.
The 
first
output density
matrix of the considered process
is here denoted as
\yprocessoutmixedonedensitymatrix .
As discussed in Section
\ref{sec-single-stage},
this
density matrix
reads
\begin{equation}
\label{eq-processoutmixedonedensitymatrixvsinmixedonedensitymatrix}
\Yprocessoutmixedonedensitymatrix
=
\Yprocessmat
\Ymathspace
\Yprocessinmixedonedensitymatrix
\Ymathspace
\Yprocessmat
\Ytransconjug
\end{equation}
where
\yprocessmat\
is the unitary matrix that represents
the process to be estimated in the considered
basis,
$.\Ytransconjug$
is the transconjugate (i.e. Hermitian transpose)
and
\yprocessinmixedonedensitymatrix\
is the density matrix of the 
first
input mixed state
that yields the 
output state
\yprocessoutmixedonedensitymatrix .
The
idea is then that,
when
\ytextartitionehundredninetyfourvfourmodifstepone{the density
matrix}
\yprocessinmixedonedensitymatrix\
is 
selected 
to be
diagonal
\ytextartitionehundredninetyfourvfourmodifstepone{in the
considered and fixed basis (computational basis), Eq.}
(\ref{eq-processoutmixedonedensitymatrixvsinmixedonedensitymatrix})
is an eigendecomposition of
\yprocessoutmixedonedensitymatrix .
Therefore, in practice, starting from an estimate of
\yprocessoutmixedonedensitymatrix\ derived from measurements,
its eigendecomposition may be hoped to yield 
a result directly related to
an estimate of
\yprocessmat .
This idea will then have to be refined further in this paper
because: 1) eigendecomposition yields indeterminacies
(defined below)
and 2) fluctuations in output density matrix
estimation limit the performance
of this preliminary approach and then 
lead
us to
extend it in this paper.

The above preliminary description
shows that the type of approaches proposed in this paper
\ytextartitionehundredninetyfourvfivemodifstepone{has 
some connections with}
works that were previously reported
both in the classical and quantum domains.
\ytextartitionehundredninetyfourvfivemodifstepone{The connections
with classical approaches are discussed in Appendix
\ref{sec-appendix-connections-clasical-processing},
whereas those with the quantum domain may be defined as follows.
A few recent papers dealing with quantum data}
also consider
(\ref{eq-processoutmixedonedensitymatrixvsinmixedonedensitymatrix})
in the framework of QPT. 
However, they only have the following limited relationships
with our investigation.
Ref.
\cite{amq153}
aims at deriving
the 
number (and values) of quantum states that should be
applied to the input of a quantum process to be able to
identify that process. However, that paper does not focus
on the same input values as in the present paper and,
as also noted in
\cite{amoi6-172},
it
does
not provide an explicit QPT algorithm.
Ref.
\cite{amq-baldwin-physreva-2014}
provides such 
algorithms, that are different from
ours
and that are tested only 
for a Hilbert space dimension
$
\Yketspacedim
=
5
$, whereas we consider dimensions up to
$
\Yketspacedim
=
2
^{13}
\simeq
8000
$.

QPT algorithms may also be split in two classes,
from the following point of view.
Part of them
first use the results of measurements at the
output of the system by explicitly
performing quantum state 
tomography
(QST)%
\ytextartitionehundredninetyfourvfivemodifstepone{. This means that
they estimate}
the ket (for a pure state)
or the density matrix (for a statistical
mixture)
that represents that ouput state.
They then use these estimated output states 
(and the
corresponding input states, their estimates
or their properties)
so
as to infer the 
considered process.
Other QPT methods
instead directly use the results of the
considered types of measurements.
In the present paper, we consider the
first class of QPT methods and we focus
on the part dedicated to QPT itself, as
opposed to QST,
as detailed
in Section
\ref{sec-test-results}.

The remainder of this paper is 
organized as
follows.
Section
\ref{sec-single-stage}
is devoted to the introduction of 
several variants of our basic type of QPT
methods, that is, single-stage methods.
These methods have a limitation,
that is defined
in Section
\ref{sec-two-stage}.
Therefore,
we first
introduce 
the
two-stage
extensions
of the above methods
in Section
\ref{sec-two-stage},
and then their general
multi-stage 
form
in Section
\ref{sec-multi-stage}.
Numerical results illustrate the performance of the proposed methods in
Section
\ref{sec-test-results}
and conclusions are drawn from this investigation in Section
\ref{sec-concl}.
\section{Proposed single-stage QPT methods}
\label{sec-single-stage}
As stated above, in all the configurations considered in this paper, the
quantum system is assumed to be
\ytextartitionehundredninetyfourvonemodifstepfour{isolated}
from its
environment
(the proposed approaches may therefore also be used
as an approximation for systems that have a limited interaction
with their environment 
and that 
therefore 
lead to near-unitary
processes, as in
\cite{amq-baldwin-physreva-2014}).
It is well-known 
that, in that case, the 
temporal
evolution
of the 
state of 
that system 
between given initial and final times
is governed by a 
unitary 
\ytextartitionehundredninetyfourvonemodifstepfour{operator}
(see
\cite{booknielsen,livremessiahtomeun};
see also \cite{livrefeynmanstatisticalmeca,articlefano1957}
for the relationship with the system's Hamiltonian).
More precisely, 
using the above-defined notations,
we
consider
a
(first)
mixed state defined by a
density matrix
\yprocessinmixedonedensitymatrix\
at an initial time.
As explained
e.g. in
\cite{booknielsen,livremessiahtomeun},
the state of that system at a subsequent, final, time
is then defined by a
density matrix
\yprocessoutmixedonedensitymatrix\
that may be expressed as in
(\ref{eq-processoutmixedonedensitymatrixvsinmixedonedensitymatrix}),
where
\yprocessmat\
is a unitary matrix, that depends on the dynamics of the
considered system.

The algorithms proposed in this section to estimate
\yprocessmat\
then consist of two parts. Each part involves a given
type of input quantum state
and an
algorithm
for exploiting 
the associated output quantum state
(and possibly the input
state).
Each of the following subsections describes one of
these two parts.
\subsection{%
First part
of the methods}
\label{sec-single-stage-part-one}
We here consider the situation when the first state
applied to the analyzed process is a given
statistical mixture, 
represented by the density matrix
\yprocessinmixedonedensitymatrix .
The process output is then also a mixed state
and its density matrix
\yprocessoutmixedonedensitymatrix\
may be expressed as
(\ref{eq-processoutmixedonedensitymatrixvsinmixedonedensitymatrix}).
More specifically, we here introduce a first variant,
in which the matrix 
\yprocessinmixedonedensitymatrix\
is diagonal
\ytextartitionehundredninetyfourvfivemodifstepone{(see
Appendix
\ref{sec-appendix-diag-density-matrix}
about how to create it)}%
,
whereas 
Appendix \ref{appendix-sec-single-stage-part-one}
addresses
the more general case when this constraint is 
\ytextartitionehundredninetyfourvfivemodifstepone{removed.}

The matrix
\yprocessoutmixedonedensitymatrix\
is known to be Hermitian, since it is a density matrix
(see e.g. 
\cite{livremessiahtomeun}
p. 283).
Therefore, it is
normal
(see \cite{book-horn-johnson} p. 100%
\ytextartitionehundredninetyfourvonemodifsteptwo{;
\footnote{\ytextartitionehundredninetyfourvonemodifsteptwo{Throughout
this paper, the reader may refer to
\cite{book-horn-johnson-2013},
instead of
\cite{book-horn-johnson}.}%
}%
}%
)
and hence
unitarily diagonalizable:
see Theorem
\ref{theorem-normalmatrix-eigendecomposition}
and comments in
its proof
(all theorems cited in the core of this paper are provided in
Appendix
\ref{sec-math-appendix}).
The first part of the QPT method proposed below
is based on the eigendecomposition of
\yprocessoutmixedonedensitymatrix\
(because 
(\ref{eq-processoutmixedonedensitymatrixvsinmixedonedensitymatrix})
is such an eigendecomposition, as detailed below).
More precisely,
in practice, what is initially available for performing an
eigendecomposition is not the theoretical matrix
\yprocessoutmixedonedensitymatrix\
but an estimate 
\yprocessoutmixedonedensitymatrixestim\
of
\yprocessoutmixedonedensitymatrix ,
provided by a QST algorithm
\ytextartitionehundredninetyfourvthreemodifstepone{(we here use
standard notations, e.g. as in
\cite{icabook-oja,book-papoulis,book-scharf}:
an
estimate of an unknown, deterministic, scalar, vector or
matrix
$
\theta
$
is denoted as
$
\widehat{\theta}
$).}
\ytextartitionehundredninetyfourvfivemodifstepone{Therefore,
our QPT method 
preferably
starts with two
preprocessing steps, that consist of
first deriving a Hermitian estimate
\yprocessoutmixedonedensitymatrixestimherm\
from
\yprocessoutmixedonedensitymatrixestim\
and then a unit-trace estimate
\yprocessoutmixedonedensitymatrixestimhermunittrace\
from
\yprocessoutmixedonedensitymatrixestimherm\
(see details
in Appendix \ref{sec-appendix-single-stage-algo}), unless 
the considered QST algorithm already
guarantees that 
\yprocessoutmixedonedensitymatrixestim\
meets these two properties.

Then, the core of our QPT algorithm performs}
an
eigendecomposition
of
the Hermitian matrix
\yprocessoutmixedonedensitymatrixestimhermunittrace .
This eigendecomposition problem has several solutions,
corresponding to the indeterminacies of
eigendecomposition 
detailed below.
Any of these solutions is defined by two matrices,
hereafter denoted as
\yprocessoutmixedonedensitymatrixestimhermunittraceeigenmatval\
and
\yprocessoutmixedonedensitymatrixestimhermunittraceeigenmatvec .
\yprocessoutmixedonedensitymatrixestimhermunittraceeigenmatval\
is a diagonal matrix that contains the eigenvalues of
\yprocessoutmixedonedensitymatrixestimhermunittrace\
in a arbitrary order
\ytextartitionehundredninetyfourvfivemodifstepone{(%
see e.g. 
\cite{book-horn-johnson} p. 102).}
Moreover,
each column of
\yprocessoutmixedonedensitymatrixestimhermunittraceeigenmatvec\
is an eigenvector of
\yprocessoutmixedonedensitymatrixestimhermunittrace\
(see 
(\ref{eq-processoutmixedonedensitymatrixestimhermunittrace-eigendecompos-with-inverse-itself})
and
Theorem \ref{theorem-eigendecomposition-eigenvectors-eigenvalues})
and
\yprocessoutmixedonedensitymatrixestimhermunittraceeigenmatvec\
is here unitary
(because
\yprocessoutmixedonedensitymatrixestimhermunittrace\
is unitarily diagonalizable),
hence
$
\Yprocessoutmixedonedensitymatrixestimhermunittraceeigenmatvec
^{-1}
=
\Yprocessoutmixedonedensitymatrixestimhermunittraceeigenmatvec
\Ytransconjug
$.
These results
\yprocessoutmixedonedensitymatrixestimhermunittraceeigenmatvec\
and
\yprocessoutmixedonedensitymatrixestimhermunittraceeigenmatval\
of the eigendecomposition are linked
to the initial matrix
\yprocessoutmixedonedensitymatrixestimhermunittrace\
as follows:
\begin{eqnarray}
\Yprocessoutmixedonedensitymatrixestimhermunittrace
&
=
&
\Yprocessoutmixedonedensitymatrixestimhermunittraceeigenmatvec
\Ymathspace
\Yprocessoutmixedonedensitymatrixestimhermunittraceeigenmatval
\Ymathspace
\Yprocessoutmixedonedensitymatrixestimhermunittraceeigenmatvec
^{-1}
\label{eq-processoutmixedonedensitymatrixestimhermunittrace-eigendecompos-with-inverse-itself}
\\
&
=
&
\Yprocessoutmixedonedensitymatrixestimhermunittraceeigenmatvec
\Ymathspace
\Yprocessoutmixedonedensitymatrixestimhermunittraceeigenmatval
\Ymathspace
\Yprocessoutmixedonedensitymatrixestimhermunittraceeigenmatvec
\Ytransconjug
.
\label{eq-processoutmixedonedensitymatrixestimhermunittrace-eigendecompos-with-inverse-itself-transconj}
\end{eqnarray}

Comparing
(\ref{eq-processoutmixedonedensitymatrixestimhermunittrace-eigendecompos-with-inverse-itself-transconj})
to
(\ref{eq-processoutmixedonedensitymatrixvsinmixedonedensitymatrix}),
with
\yprocessoutmixedonedensitymatrixestimhermunittrace\
an
estimate of
\yprocessoutmixedonedensitymatrix\
and with
\yprocessinmixedonedensitymatrix\
a diagonal matrix,
shows that
(\ref{eq-processoutmixedonedensitymatrixvsinmixedonedensitymatrix})
is one of the possible eigendecompositions of
\yprocessoutmixedonedensitymatrixestimhermunittrace\
up to estimation errors.
Therefore,
the matrices 
\yprocessoutmixedonedensitymatrixestimhermunittraceeigenmatval\
and
\yprocessoutmixedonedensitymatrixestimhermunittraceeigenmatvec\
obtained in practice with our algorithm
are respectively equal to
\yprocessinmixedonedensitymatrix\
and
\yprocessmat\
up to the indeterminacies of
the eigendecomposition process
(and up to estimation errors).
In particular,
\yprocessoutmixedonedensitymatrixestimhermunittraceeigenmatvec\
is the first obtained
quantity related to the estimation of
\yprocessmat\
(it could therefore be denoted as
\yprocessmatestimone ).
We now define its
indeterminacies in more detail,
and we show how to solve them.

In our first type 
of QPT methods described here,
we
constrain
all diagonal values of the 
user-defined matrix
\yprocessinmixedonedensitymatrix\
to be different.
This
means that all eigenvalues in the eigendecompositions
of
\yprocessoutmixedonedensitymatrix\
and hence
\yprocessoutmixedonedensitymatrixestimhermunittrace\
(ignoring estimation errors)
are different.
Therefore, each of them is associated with a different
one-dimensional eigenspace%
\ytextartitionehundredninetyfourvfivemodifstepone{. In other 
words,}
all eigenvectors corresponding to a given eigenvalue are 
along the same direction
and the directions
associated with two different eigenvalues are
different
(this is true for any complex square matrix:
see
\cite{book-horn-johnson} p. 47)
and
even orthogonal:
see
Theorem
\ref{theorem-normal-matrix-eigenvectors-orthogonal}.
Each eigenvalue is thus coupled with a single eigenvector 
in any
considered eigendecomposition.
The first indeterminacy of the eigendecomposition
of
\yprocessoutmixedonedensitymatrixestimhermunittrace\
is 
then
a
possible 
{\it
coupled}
permutation 
of 
1) the columns of the matrix 
\yprocessoutmixedonedensitymatrixestimhermunittraceeigenmatvec\
obtained
as a first quantity towards the estimation of
\yprocessmat\
in this eigendecomposition
and 2) the
values 
along the diagonal of
the matrix 
\yprocessoutmixedonedensitymatrixestimhermunittraceeigenmatval,
due to the above-mentioned arbitrary order of the
latter diagonal values.
This 
permutation
problem may be solved as
follows. 
As stated above,
we 
constrain
all diagonal values of 
\yprocessinmixedonedensitymatrix\
to be different 
(at this stage of the discussion, 
we restrict ourselves to a theoretical approach,
where we only need the eigenvalues to be different;
further in this paper,
we will discuss the gap required between them, when considering
practical implementations of our approach%
\ytextartitionehundredninetyfourvfivemodifstepone{:
see Appendix
\ref{sec-appendix-input-density-matrix}
for the method considered in the present section}%
).
Moreover, we
constrain
all 
diagonal values of
\yprocessinmixedonedensitymatrix\
to be placed on
that diagonal
according to a given order.
Hereafter, we will consider the case
when they are in decreasing order.
With respect to the above-defined 
machine learning
terminology,
this part of our
algorithm is thus unsupervised:
it does not require 
the exact value of
the input state, i.e.
the input density matrix 
\yprocessinmixedonedensitymatrix,
to be known
but it sets limited conditions on some of 
its
\ytextartitionehundredninetyfourvonemodifsteptwo{properties.}

The above known order of the diagonal
values
of
\yprocessinmixedonedensitymatrix\
is 
used 
to postprocess
\yprocessoutmixedonedensitymatrixestimhermunittraceeigenmatval,
by permuting
the values
obtained
on its diagonal
so that they are in the same order as
the order that we impose on
\yprocessinmixedonedensitymatrix .
More precisely, in the case considered
hereafter, we permute the values
obtained
on the diagonal of 
\yprocessoutmixedonedensitymatrixestimhermunittraceeigenmatval\
so that they are in decreasing order
(if some eigenvalues obtained in practice are nonreal, they are
ordered according to their moduli).
Since we now know which permutation is
required to this end,
we then use that specific permutation to
reorder the columns of
\yprocessoutmixedonedensitymatrixestimhermunittraceeigenmatvec\
in the same way.
The matrix with reordered columns thus obtained is denoted as
\yprocessmatestimtwo,
since it is the second quantity that we build for estimating
\yprocessmat .

Thanks to the above approach,
the only indeterminacy that remains in
\yprocessmatestimtwo\
with respect to
\yprocessmat\
is an unknown complex scale factor in each of
its columns,
since any such column belongs to the
adequate one-dimensional eigenspace
and is therefore
defined up to a complex scale factor.
Moreover, we create
\yprocessmatestimtwo\
so that each of its columns has unit 
\ytextartitionehundredninetyfourvfivemodifstepone{norm. To
this end,}
if the considered eigendecomposition
algorithm does not directly guarantee that
this 
\ytextartitionehundredninetyfourvfivemodifstepone{unit-norm}
condition is 
\ytextartitionehundredninetyfourvfivemodifstepone{met,}
we postprocess the value of
\yprocessmatestimtwo\
that it yields, by dividing each column of
\yprocessmatestimtwo\
by its norm.
After this postprocessing,
the scale factor of each column of
\yprocessmatestimtwo,
with respect to the column of
\yprocessmat\
that has the same index, reduces to a
unit-modulus factor, i.e. a phase
factor,
because
\yprocessmat\
is unitary and therefore all its columns
have unit norm, as those of
\ytextartitionehundredninetyfourvfivemodifstepone{%
\yprocessmatestimtwo .}
\yprocessmatestimtwo\
then reads
\begin{equation}
\Yprocessmatestimtwo
=
\left[
e^
{
\Ysqrtminusone
\Yprocessmatestimtwophaseindetermfirst
}
\Yprocessmatcolumnfirst
,
\dots
,
e^
{
\Ysqrtminusone
\Yprocessmatestimtwophaseindetermlast
}
\Yprocessmatcolumnlast
\right]
\label{eq-processmatestimtwo-vs-processmat-and-phases}
\end{equation}
where
$
\Yprocessmatcolumnfirst
,
\dots
,
\Yprocessmatcolumnlast
$
are the column vectors of the actual matrix
\yprocessmat\
and
$
\Yprocessmatestimtwophaseindetermfirst
,
\dots
,
\Yprocessmatestimtwophaseindetermlast
$
are the phase indeterminacies 
with respect to the above vectors
that remain at this stage.
Denoting
\yprocessmatestimtwophaseindetermmatrixdiag\
the diagonal matrix that contains the
unknown phase factors
$
e^
{
\Ysqrtminusone
\Yprocessmatestimtwophaseindetermfirst
}
,
\dots
,
e^
{
\Ysqrtminusone
\Yprocessmatestimtwophaseindetermlast
}
$
on its main diagonal,
(\ref{eq-processmatestimtwo-vs-processmat-and-phases})
reads
\begin{equation}
\Yprocessmatestimtwo
=
\Yprocessmat
\Yprocessmatestimtwophaseindetermmatrixdiag
.
\label{eq-processmatestimtwo-vs-processmatestimtwophaseindetermmatrixdiag}
\end{equation}

The second
part of our algorithm,
presented below in Section
\ref{sec-single-stage-part-two},
aims at 
determining 
all
phase factors 
$
e^
{
\Ysqrtminusone
\Yprocessmatestimtwophaseindetermfirst
}
,
\dots
,
e^
{
\Ysqrtminusone
\Yprocessmatestimtwophaseindetermlast
}
$,
up to a global phase factor
\ytextartitionehundredninetyfourvfivemodifstepone{(a process matrix
\yprocessmat\
can only be defined up to a single, i.e. global,
phase factor).}

\subsection{%
Second part
of the methods}
\label{sec-single-stage-part-two}
Thanks to the limited number of unknowns 
(namely
the phase factors 
$
e^
{
\Ysqrtminusone
\Yprocessmatestimtwophaseindetermfirst
}
,
\dots
,
e^
{
\Ysqrtminusone
\Yprocessmatestimtwophaseindetermlast
}
$)
that remain in the matrix
\yprocessmatestimtwo\
of
(\ref{eq-processmatestimtwo-vs-processmat-and-phases})
obtained in Section
\ref{sec-single-stage-part-one},
a single and pure state is then sufficient for 
estimating
these unknowns.
This yields the first variant of the second part
of our methods, presented here.
However, one may instead use a mixed state for estimating these
unknowns, as shown in Appendix
\ref{appendix-sec-single-stage-part-two}.

We therefore here
apply a known pure state, defined by a ket
\yprocessinpuretwoket,
at the input of the process
(this part of the algorithm is therefore supervised,
or semi-supervised if
\yprocessinpuretwoket\
is initially unknown but estimated
by performing measurements
for part of the available copies of that state%
\ytextartitionehundredninetyfourvfivemodifstepone{;
for the supervised case, Appendix
\ref{sec-appendix-single-stage-algo-input-ket}
explains how
\yprocessinpuretwoket\
is chosen}%
).
The corresponding
pure state 
at the output of that process
is defined by the ket
\begin{equation}
\Yprocessoutpuretwoket
=
\Yprocessmat
\Yprocessinpuretwoket
.
\label{eq-processoutpuretwoket}
\end{equation}

What is available
in practice
is
a
ket 
\yprocessoutpuretwoketestim,
obtained by
performing a
QST
at the output of the process,
and
equal to
\yprocessoutpuretwoket\ up to an unknown phase factor 
and
to estimation errors
\ytextartitionehundredninetyfourvfivemodifstepone{(see again
Appendix
\ref{sec-appendix-single-stage-algo}
concerning a possible normalization of the
QST output used here).}
When neglecting the above estimation
errors,
\begin{eqnarray}
\Yprocessoutpuretwoketestim
&
=
&
e
^{
\Ysqrtminusone
\Yprocessoutpuretwoketestimphase
}
\Yprocessoutpuretwoket
\\
&
=
&
e
^{
\Ysqrtminusone
\Yprocessoutpuretwoketestimphase
}
\Yprocessmat
\Yprocessinpuretwoket
.
\label{eq-processoutpuretwoketestim-vs-processinpuretwoket}
\end{eqnarray}

Our algorithm 
computes 
the ket
\begin{equation}
\Yprocessmatestimthreeketinterm
=
\Yprocessmatestimtwo
\Ytransconjug
\Yprocessoutpuretwoketestim
,
\label{eq-def-processmatestimthreeketinterm}
\end{equation}
based on the following motivation:
combining
(\ref{eq-def-processmatestimthreeketinterm}),
(\ref{eq-processmatestimtwo-vs-processmatestimtwophaseindetermmatrixdiag}),
(\ref{eq-processoutpuretwoketestim-vs-processinpuretwoket})
and considering that
\yprocessmat\ is unitary
(hence
$
\Yprocessmat
\Ytransconjug
\Yprocessmat
=
I
$)
yields
\begin{equation}
\Yprocessmatestimthreeketinterm
=
e
^{
\Ysqrtminusone
\Yprocessoutpuretwoketestimphase
}
\Yprocessmatestimtwophaseindetermmatrixdiag
\Yconjug
\Yprocessinpuretwoket
\label{eq-processmatestimthreeketinterm-vs-processinpuretwoket}
\end{equation}
where
\yconjug\
stands for conjugation.
This equation is very useful because, apart from
the global phase
\yprocessoutpuretwoketestimphase ,
its only unknown is
\yprocessmatestimtwophaseindetermmatrixdiag ,
i.e. the set of 
phase factors
$
e^
{
\Ysqrtminusone
\Yprocessmatestimtwophaseindetermfirst
}
,
\dots
,
e^
{
\Ysqrtminusone
\Yprocessmatestimtwophaseindetermlast
}
$
to be determined.
This problem is easily solved by considering each
component
\yprocessmatcolindexstd ,
denoted as
$
\lvert
.
\rangle
_{\Yprocessmatcolindexstd}
$,
of the above kets.
Using a ket
\yprocessinpuretwoket\
with nonzero components,
(\ref{eq-processmatestimthreeketinterm-vs-processinpuretwoket})
yields
\begin{equation}
\frac
{
\Yprocessmatestimthreeketinterm
_{\Yprocessmatcolindexstd}
}
{
\Yprocessinpuretwoket
_{\Yprocessmatcolindexstd}
}
=
e
^{
\Ysqrtminusone
(
\Yprocessoutpuretwoketestimphase
-
\Yprocessmatestimtwophaseindetermindexstd
)
}
\hspace{10mm}
\forall
\Yprocessmatcolindexstd
\in
\{
1,
\dots,
\Yketspacedim
\}
.
\label{eq-def-one-component}
\end{equation}
\ytextartitionehundredninetyfourvfivemodifstepone{The}
associated matrix form reads
\begin{equation}
\mathrm{diag}
(
\Yprocessmatestimthreeketinterm
\oslash
\Yprocessinpuretwoket
)
=
e
^{
\Ysqrtminusone
\Yprocessoutpuretwoketestimphase
}
\Yprocessmatestimtwophaseindetermmatrixdiag
\Yconjug
\label{eq-def-matrix-diag}
\end{equation}
where
$
\oslash
$
is the element-wise division for vectors
\ytextartitionehundredninetyfourvonemodifstepfour{(as defined in
Appendix
\ref{sec-math-appendix})}
and
$
\mathrm{diag}
(
v
)
$
is the diagonal matrix that contains all elements of a vector
$
v
$
on its main diagonal.

Based on the above properties, the final step of our algorithm creates,
as follows,
a new estimate of
\yprocessmat\
denoted as
\yprocessmatestimthree\
(the intermediate notations
\ytwostageprocessmatestimthree\
and
\ytwostageprocessmatestimfour\
are not used here but are required for the two-stage methods of
Section
\ref{sec-two-stage}):
\begin{equation}
\Yprocessmatestimthree
=
\Yprocessmatestimtwo
\Ymathspace
\mathrm{diag}
(
\Yprocessmatestimthreeketinterm
\oslash
\Yprocessinpuretwoket
).
\label{eq-def-processmatestimthree}
\end{equation}
This approach is used
because, due to
(\ref{eq-def-matrix-diag})
and
(\ref{eq-processmatestimtwo-vs-processmatestimtwophaseindetermmatrixdiag}),
Eq.
(\ref{eq-def-processmatestimthree})
yields
\begin{equation}
\Yprocessmatestimthree
=
e
^{
\Ysqrtminusone
\Yprocessoutpuretwoketestimphase
}
\Yprocessmat
\label{eq-processmatestimthree-vs-processmat}
\end{equation}
\ytextartitionehundredninetyfourvfivemodifstepone{which
means that}
\yprocessmatestimthree\
succeeds in restoring
\yprocessmat ,
up to a global phase 
\ytextartitionehundredninetyfourvfivemodifstepone{factor}
(and again up to the estimation errors that were mentioned above but implicit
in the above equations).

A pseudo-code of the version of our 
algorithm corresponding to all this
Section
\ref{sec-single-stage}
is provided in
Algorithm
\ref{label-algodef-eqptone}.
This algorithm is called EQPT1, because it is 
our first proposed
Eigenanalysis-based QPT
algorithm.

\begin{algorithm}[t]
{%
\SetKwInOut{Input}{Input}\SetKwInOut{Output}{Output}
\Input{%
a) 
Estimate 
\ytextartitionehundredninetyfourvfivemodifstepone{%
\yprocessoutmixedonedensitymatrixestimhermunittrace}
of output density matrix
\yprocessoutmixedonedensitymatrix\
(provided by Quantum State Tomography (QST)
and obtained for
input density matrix
(\ref{eq-opdensity-diag-val-uniform})).
b) 
Estimate 
\yprocessoutpuretwoketestim\
of output ket
\yprocessoutpuretwoket\
(provided by QST
and obtained for all components 
of
input ket
\yprocessinpuretwoket\
equal to
$
1
/
\sqrt{
\Yketspacedim
}
$).%
}
\Output{%
Estimate
\yprocessmatestimthree\
of 
quantum process matrix
\yprocessmat .}
\BlankLine
\Begin{
\tcc{\ytextartitionehundredninetyfourvfivemodifstepone{Exploit}
\yprocessoutmixedonedensitymatrixestimhermunittrace :}
Eigendecomposition:
derive 
1) a diagonal matrix
\yprocessoutmixedonedensitymatrixestimhermunittraceeigenmatval\
that contains the eigenvalues of
\yprocessoutmixedonedensitymatrixestimhermunittrace\
in an arbitrary order
and 2) a matrix
\yprocessoutmixedonedensitymatrixestimhermunittraceeigenmatvec\
whose columns are 
eigenvectors of
\yprocessoutmixedonedensitymatrixestimhermunittrace\
in the same order as eigenvalues\;
Reorder the eigenvalues in
\yprocessoutmixedonedensitymatrixestimhermunittraceeigenmatval\
in decreasing order
and 
apply the same permutation to the columns of
\yprocessoutmixedonedensitymatrixestimhermunittraceeigenmatvec\
to create the matrix
\yprocessmatestimtwo\;
\tcc{%
If the above eigendecomposition
algorithm does not
yield \emph{unit-norm}
eigenvectors, then create them as follows:}
Divide each column of
\yprocessmatestimtwo\
by its norm\;
\tcc{\ytextartitionehundredninetyfourvfivemodifstepone{Exploit}
\yprocessoutpuretwoketestim :}
\ytextartitionehundredninetyfourvfivemodifstepone{$
\Yprocessmatestimthreeketinterm
=
\Yprocessmatestimtwo
\Ytransconjug
\Yprocessoutpuretwoketestim
$\;}
\ytextartitionehundredninetyfourvfivemodifstepone{$
\Yprocessmatestimthree
=
\Yprocessmatestimtwo
\Ymathspace
\mathrm{diag}
(
\Yprocessmatestimthreeketinterm
\oslash
\Yprocessinpuretwoket
)$\;}
}
\caption{EQPT1: first Eigenanalysis-based 
Quantum Process Tomography
algorithm (composed of one stage).
\ytextartitionehundredninetyfourvfivemodifstepone{This version
is for the case when the QST algorithm provides
estimates that meet the known properties of the
actual quantities. Otherwise, see Appendix
\ref{sec-appendix-single-stage-algo}.}%
}
}
\label{label-algodef-eqptone}
\end{algorithm}
\section{Proposed two-stage QPT methods}
\label{sec-two-stage}
\subsection{Motivation}
\label{sec-two-stage-motivation}
The dimension \yketspacedim\
of the 
state
space
(and hence the dimension 
$
\Yketspacedim
\times
\Yketspacedim
$
of the 
considered
process matrix
\yprocessmat )
that can be addressed by the type of QPT methods proposed in Section
\ref{sec-single-stage} is limited by the following phenomenon.
As explained in that section,
these methods require all eigenvalues of the input density
matrix
\yprocessinmixedonedensitymatrix\
of the 
process
to be different.
Moreover, they should be
separated by a minimum gap
to be distinguishable,
so that their estimates are not permuted in practice
(in order to reorder them 
and the associated eigenvectors
correctly).
Besides,
all of them range from 0 to 1,
because they are nonnegative and their sum is equal to one,
as stated above.
All these conditions
yield an upper 
bound,
denoted as
\yeigenvalnbmax ,
for the acceptable
number of eigenvalues
of
\yprocessinmixedonedensitymatrix .
Since
the number of (different) eigenvalues of
\yprocessinmixedonedensitymatrix\
is always equal to
\yketspacedim\
for the
methods of Section
\ref{sec-single-stage},
this yields an upper bound for the
state
space dimension 
\yketspacedim\
that can be
handled by these methods,
and this bound on 
\yketspacedim\
is also equal to
\yeigenvalnbmax :
\begin{equation}
\Yketspacedim
\leq
\Yeigenvalnbmax
.
\label{eq-ketspacedim-bound-method-one-stage}
\end{equation}

In the present section, we aim at addressing higher values of
\yketspacedim .
To this end, we move to the case when some eigenvalues of
the input density matrix
\yprocessinmixedonedensitymatrix\
are equal
and we therefore introduce extensions
of the 
QPT methods of Section
\ref{sec-single-stage}
to be able to handle that more complex case.
\subsection{%
First part
of the methods}
\label{sec-two-stage-part-one}

As in Section
\ref{sec-single-stage-part-one},
the method proposed here starts by considering a 
given
mixed state that has
a
diagonal
density matrix
\yprocessinmixedonedensitymatrix\
and that is
applied to the input of the considered unitary process
defined by a matrix
\yprocessmat .
However, we here accept to have identical values on the main diagonal
of
\yprocessinmixedonedensitymatrix ,
at the expense of then having to handle more complex 
eigendecomposition properties for the process output,
as shown below.
More precisely,
each
value that appears on the main diagonal of
\yprocessinmixedonedensitymatrix\
appears
\yprocessinmixedonedensitymatrixvalmultiplicity\
times,
on adjacent rows,
and
that diagonal contains
\yprocessinmixedonedensitymatrixvalnb\
different values.
The number of rows of
\yprocessinmixedonedensitymatrix ,
equal to the dimension of the considered 
state
space,
is thus equal to
\begin{equation}
\Yketspacedim
=
\Yprocessinmixedonedensitymatrixvalmultiplicity
\Yprocessinmixedonedensitymatrixvalnb
\label{eq-def-ketspacedim-two-stage}
\end{equation}
(in practice, we aim at considering the case
when
\yprocessinmixedonedensitymatrixvalmultiplicity,
\yprocessinmixedonedensitymatrixvalnb,
and hence
\yketspacedim\
are powers of 2,
with an application to qubits).
If
requesting the same minimum gap between
different eigenvalues as in Section
\ref{sec-single-stage-part-one},
\ytextartitionehundredninetyfourvfivemodifstepone{or
in other words, if}
replacing the bound 
(\ref{eq-ketspacedim-bound-method-one-stage})
on
\yketspacedim\
in 
Section
\ref{sec-single-stage}
by the same bound value for
\yprocessinmixedonedensitymatrixvalnb\
instead of
\yketspacedim\
here,
\ytextartitionehundredninetyfourvfivemodifstepone{then}
the maximum 
value of
the
state
space dimension 
(\ref{eq-def-ketspacedim-two-stage})
that can be handled here
is 
\yprocessinmixedonedensitymatrixvalmultiplicity\
times
higher than the maximum dimension that can be addressed by the
methods of Section
\ref{sec-single-stage},
which is our goal, ideally.
However, it is not guaranteed that the methods proposed hereafter
can handle the same gap between
different eigenvalues as in Section
\ref{sec-single-stage-part-one}
with the same accuracy, especially due to the more complex
structure of these extended methods.
The practical accuracy of all proposed methods
is therefore investigated numerically further in this paper.

Denoting
$
\Yprocessinmixedonedensitymatrixvaldiagdifffirst
,
\dots
,
\Yprocessinmixedonedensitymatrixvaldiagdifflast
$
the 
different 
eigenvalues of
\yprocessinmixedonedensitymatrix
,
that 
\ytextartitionehundredninetyfourvonemodifsteptwo{chosen}
density matrix reads
\begin{equation}
\Yprocessinmixedonedensitymatrix
=
\left[
\begin{tabular}{llllllll}
\yprocessinmixedonedensitymatrixvaldiagdifffirst
&
0
&
$
\dots
$
&
&
&
&
&
0
\\
&
$
\ddots
$
\\
&
&
\yprocessinmixedonedensitymatrixvaldiagdifffirst
\\
&
&
&
\yprocessinmixedonedensitymatrixvaldiagdiffsecond
\\
&
&
&
&
$
\ddots
$
\\
&
&
&
&
&
\yprocessinmixedonedensitymatrixvaldiagdiffsecond
\\
&
&
&
&
&
&
$
\ddots
$
\\
0
&
$
\dots
$
&
&
&
&
&
0
&
\yprocessinmixedonedensitymatrixvaldiagdifflast
\end{tabular}
\right]
\label{eq-def-processinmixedonedensitymatrix-method-two-stage}
\end{equation}
with each diagonal value repeated
\yprocessinmixedonedensitymatrixvalmultiplicity\
times.
\ytextartitionehundredninetyfourvfivemodifstepone{This
may also be expressed as
\begin{equation}
\Yprocessinmixedonedensitymatrix
= 
\mathrm{diag}
(
r
)
\otimes
I
_{\Yprocessinmixedonedensitymatrixvalmultiplicity}
\label{eq-def-processinmixedonedensitymatrix-method-two-stage-compact}
\end{equation}
where
$
r
$
is the vector that contains the values
$
\Yprocessinmixedonedensitymatrixvaldiagdifffirst
,
\dots
,
\Yprocessinmixedonedensitymatrixvaldiagdifflast
$,
whereas
$
I
_{\Yprocessinmixedonedensitymatrixvalmultiplicity}
$
is the
$
\Yprocessinmixedonedensitymatrixvalmultiplicity
\times
\Yprocessinmixedonedensitymatrixvalmultiplicity
$
identity matrix
and
$
\otimes
$
stands for the Kronecker product
\cite{book-multi-way-smiled-bro-geladi}.}
Without loss of generality,
we here again request the values on the diagonal of
\yprocessinmixedonedensitymatrix\
to be in 
nonincreasing
order%
\ytextartitionehundredninetyfourvfivemodifstepone{:}
$
\Yprocessinmixedonedensitymatrixvaldiagdifffirst
>
\Yprocessinmixedonedensitymatrixvaldiagdiffsecond
>
\dots
>
\Yprocessinmixedonedensitymatrixvaldiagdifflast
$.
Moreover, 
\ytextartitionehundredninetyfourvfivemodifstepone{we
preferably use 
values that are uniformly
distributed
(that is,
multiples of the
same step, with zero excluded),
as in the method of Section
\ref{sec-single-stage}
(see details for the latter method in Appendix
\ref{sec-appendix-input-density-matrix}).
In the configuration of the present section,
(\ref{eq-opdensity-diag-val-uniform})
is thus}
replaced by:
\begin{equation}
\Yprocessinmixedonedensitymatrixvaldiagnot
_{
\Yprocessmatcolindexstd
}
=
\frac{
2
(
\Yprocessinmixedonedensitymatrixvalnb
-
\Yprocessmatcolindexstd
+
1
)
}
{
\Yketspacedim
(
\Yprocessinmixedonedensitymatrixvalnb
+
1
)
}
\hspace{5mm}
\Yprocessmatcolindexstd
\in
\{
1
,
\dots
,
\Yprocessinmixedonedensitymatrixvalnb
\}
.
\label{eq-opdensity-diag-val-uniform-two-stage}
\end{equation}

Here again, the process output corresponding to
\yprocessinmixedonedensitymatrix\
is a mixed state
and its density matrix
\yprocessoutmixedonedensitymatrix\
may be expressed as
(\ref{eq-processoutmixedonedensitymatrixvsinmixedonedensitymatrix}).
Moreover, as in Section
\ref{sec-single-stage-part-one},
\yprocessoutmixedonedensitymatrix\
is Hermitian and hence unitarily diagonalizable,
and it has unit trace.
Besides,
in practice,
an estimate 
\yprocessoutmixedonedensitymatrixestim\
of
\yprocessoutmixedonedensitymatrix\
is derived by a QST algorithm and,
if required by the considered QST algorithm,
we preprocess
this estimate of
\yprocessoutmixedonedensitymatrix\
in order to derive its Hermitian and unit-trace version,
denoted as
\yprocessoutmixedonedensitymatrixestimhermunittrace .
We then perform an eigendecomposition of
\yprocessoutmixedonedensitymatrixestimhermunittrace .
The analysis of the properties of this
eigendecomposition here requires additional care,
because the considered matrix has various
{\it
identical} eigenvalues.

The properties of Section
\ref{sec-single-stage-part-one}
that remain true here are as follows.
The eigendecomposition is not unique,
Eq. 
(\ref{eq-processoutmixedonedensitymatrixestimhermunittrace-eigendecompos-with-inverse-itself})-%
(\ref{eq-processoutmixedonedensitymatrixestimhermunittrace-eigendecompos-with-inverse-itself-transconj})
still apply,
\yprocessoutmixedonedensitymatrixestimhermunittraceeigenmatvec\
and
\yprocessoutmixedonedensitymatrixestimhermunittraceeigenmatval\
are respectively a unitary matrix and a diagonal matrix,
the
$k$th
column of
\yprocessoutmixedonedensitymatrixestimhermunittraceeigenmatvec\
is an eigenvector of
\yprocessoutmixedonedensitymatrixestimhermunittrace\
and the associated eigenvalue
is
the
$k$th
value on the diagonal of
\yprocessoutmixedonedensitymatrixestimhermunittraceeigenmatval\
(see again
Theorem \ref{theorem-eigendecomposition-eigenvectors-eigenvalues}
for the latter property).

In contrast, what is different here as compared with
Section
\ref{sec-single-stage-part-one}
is that the same eigenvalue here appears
\yprocessinmixedonedensitymatrixvalmultiplicity\
time and is thus associated with
\yprocessinmixedonedensitymatrixvalmultiplicity\
eigenvectors,
instead of only one in Section
\ref{sec-single-stage-part-one}.
\ytextartitionehundredninetyfourvfivemodifstepone{Eigendecomposition
properties are more complex than above in this general
framework but, fortunately, their complexity is here
limited by}
the fact that the
matrix 
\yprocessoutmixedonedensitymatrixestimhermunittrace\
to be decomposed is Hermitian.
For this class of
matrices
(and for all normal matrices), the 
\ytextartitionehundredninetyfourvfivemodifstepone{so-called}
geometric multiplicity of
each eigenvalue is guaranteed to be equal to its
\ytextartitionehundredninetyfourvfivemodifstepone{so-called}
algebraic multiplicity,
i.e. these matrices are nondefective
(see 
\ytextartitionehundredninetyfourvfivemodifstepone{details in}
\cite{book-horn-johnson}
p. 
\ytextartitionehundredninetyfourvfivemodifstepone{57, 58,}
103).
\ytextartitionehundredninetyfourvfivemodifstepone{More explicitly,
this means that}
each eigenvalue of
\yprocessoutmixedonedensitymatrixestimhermunittrace\
is here known to appear
\yprocessinmixedonedensitymatrixvalmultiplicity\
times 
at arbitrary locations
on the diagonal of
\yprocessoutmixedonedensitymatrixestimhermunittraceeigenmatval\
(algebraic multiplicity),
since these eigenvalues are the same, in a different order,
as in another eigendecomposition of the same matrix
\yprocessoutmixedonedensitymatrixestimhermunittrace\
(neglecting
estimation errors), namely the eigendecomposition
(\ref{eq-processoutmixedonedensitymatrixvsinmixedonedensitymatrix}).
Therefore, the dimension of the eigenspace associated
with any eigenvalue of
\yprocessoutmixedonedensitymatrixestimhermunittrace\
is also equal to
\yprocessinmixedonedensitymatrixvalmultiplicity .
So, when performing the eigendecomposition of
\yprocessoutmixedonedensitymatrixestimhermunittrace ,
the considered decomposition algorithm should provide
\yprocessinmixedonedensitymatrixvalmultiplicity\
linearly independent eigenvectors
for each eigenvalue,
and we hereafter assume that 
practical 
eigendecomposition algorithms
succeed in finding
them
\ytextartitionehundredninetyfourvonemodifsteptwo{(that was indeed the case in
all the tests reported further in this paper).}
As an overall result, when performing the eigendecomposition
of
\yprocessoutmixedonedensitymatrixestimhermunittrace ,
 1) we get a diagonal matrix
\yprocessoutmixedonedensitymatrixestimhermunittraceeigenmatval\
with
$
\Yketspacedim
=
\Yprocessinmixedonedensitymatrixvalmultiplicity
\Yprocessinmixedonedensitymatrixvalnb
$
eigenvalues that may be in an arbitrary order,
and with each encountered value
repeated
\yprocessinmixedonedensitymatrixvalmultiplicity\
times
and
2) we get a
unitary matrix
\yprocessoutmixedonedensitymatrixestimhermunittraceeigenmatvec\
whose columns are 
eigenvectors 
that have
the above-defined
properties.

As in Section
\ref{sec-single-stage-part-one},
we then apply a joint permutation to the diagonal
elements of
\yprocessoutmixedonedensitymatrixestimhermunittraceeigenmatval\
and to the columns of
\yprocessoutmixedonedensitymatrixestimhermunittraceeigenmatvec,
so that the values corresponding to the
diagonal
elements of
\yprocessoutmixedonedensitymatrixestimhermunittraceeigenmatval\
become in 
nonincreasing
order.
Considering
the resulting complete set
of values 
from the top to the bottom of the diagonal
of the reordered version of
\yprocessoutmixedonedensitymatrixestimhermunittraceeigenmatval,
we thus get
\yprocessinmixedonedensitymatrixvalnb\
successive
subsets,
with the same value repeated
\yprocessinmixedonedensitymatrixvalmultiplicity\
times
in each subset
(in practice, these 
\yprocessinmixedonedensitymatrixvalmultiplicity\
values may be slightly different,
due to estimation errors),
and
with 
different
and decreasing values
from one subset to the next one.
In other words, up to estimation errors,
the reordered version of
\yprocessoutmixedonedensitymatrixestimhermunittraceeigenmatval\
is nothing but the matrix
\yprocessinmixedonedensitymatrix\
of
(\ref{eq-def-processinmixedonedensitymatrix-method-two-stage}).
Each subset of identical values
along the diagonal of the obtained matrix 
(values equal to
eigenvalue
\yprocessinmixedonedensitymatrixvaldiagdiffeigensubsetindexstd )
has an index
\yprocessinmixedoneeigensubsetindexstd,
with
$
\Yprocessinmixedoneeigensubsetindexstd
\in
\{
1,
\dots
,
\Yprocessinmixedonedensitymatrixvalnb
\}
$,
and
consists of elements with indices%
\ytextartitionehundredninetyfourvonemodifstepone{%
\footnote{%
\ytextartitionehundredninetyfourvonemodifstepone{In 
\ytextartitionehundredninetyfourvonemodifsteptwo{this whole}
paper
except Appendix
\ref{sec-appendix-relevance-multi-stage},
the index of the first row and column of a matrix is equal to 1.}}}
\begin{equation}
(
\Yprocessinmixedoneeigensubsetindexstd
-
1
)
\Yprocessinmixedonedensitymatrixvalmultiplicity
+
\Yprocessinmixedoneeigensubsetindexstdvalindexstd,
\hspace{5mm}
\mathrm{with}
\hspace{5mm}
\Yprocessinmixedoneeigensubsetindexstdvalindexstd
\in
\{
1,
\dots
,
\Yprocessinmixedonedensitymatrixvalmultiplicity
\}
.
\label{eq-def-processinmixedoneeigensubsetindexstdvalindexall}
\end{equation}
Moreover,
the corresponding
matrix,
derived from
\yprocessoutmixedonedensitymatrixestimhermunittraceeigenmatvec,
with reordered 
eigenvectors in its columns
is here again denoted as
\yprocessmatestimtwo.
Its subset
of columns
that also have the indices defined by
(\ref{eq-def-processinmixedoneeigensubsetindexstdvalindexall})
is a set of 
\yprocessinmixedonedensitymatrixvalmultiplicity\
eigenvectors that 
form a basis of the
\yprocessinmixedonedensitymatrixvalmultiplicity -dimensional
eigenspace of
\yprocessoutmixedonedensitymatrixestimhermunittrace\
associated with its eigenvalue
\yprocessinmixedonedensitymatrixvaldiagdiffeigensubsetindexstd .
This eigenspace, obtained in the first part of this method,
is hereafter denoted as
\yprocessinmixedoneeigensubsetindexstdspace .

Importantly,
\yprocessinmixedoneeigensubsetindexstdspace\
also has the following relationship with
the actual matrix
\yprocessmat\
to be identified.
Eigenvectors corresponding to distinct eigenvalues are orthogonal
for a normal matrix
(see
Theorem
\ref{theorem-normal-matrix-eigenvectors-orthogonal})
and hence especially
for a Hermitian matrix
(see
also
\cite{book-horn-johnson} p. 171).
This has the following consequence,
considering that
(\ref{eq-processoutmixedonedensitymatrixvsinmixedonedensitymatrix})
and
(\ref{eq-def-processinmixedonedensitymatrix-method-two-stage})
define an eigendecomposition of
\yprocessoutmixedonedensitymatrix :
each
\ytextartitionehundredninetyfourvonemodifsteptwo{eigenvalue
\yprocessinmixedonedensitymatrixvaldiagdiffeigensubsetindexstd\
of
\yprocessoutmixedonedensitymatrix ,
with a multiplicity 
\yprocessinmixedonedensitymatrixvalmultiplicity ,}
corresponds to a
\yprocessinmixedonedensitymatrixvalmultiplicity -dimensional
eigenspace
1) that is orthogonal to the eigenspaces defined by all
other eigenvalues
and 2) that is
the
subspace defined by the columns of 
\yprocessmat\
that have the indices defined by
(\ref{eq-def-processinmixedoneeigensubsetindexstdvalindexall}).
But this eigenspace is nothing but
\yprocessinmixedoneeigensubsetindexstdspace ,
since
\yprocessoutmixedonedensitymatrix\
is equal to
\yprocessoutmixedonedensitymatrixestimhermunittrace\
(up to estimation errors).
This shows that this first part of our second 
type of methods
does not yet succeed in
separately identifying each column of
\yprocessmat , but it succeeds in 
finding a basis for
each of the
\yprocessinmixedonedensitymatrixvalnb\
subspaces 
\yprocessinmixedoneeigensubsetindexstdspace\
that 
corresponds to a value
\yprocessinmixedoneeigensubsetindexstd,
with
$
\Yprocessinmixedoneeigensubsetindexstd
\in
\{
1,
\dots
,
\Yprocessinmixedonedensitymatrixvalnb
\}
$,
and that
is
associated with the
\yprocessinmixedonedensitymatrixvalmultiplicity\
columns of
\yprocessmat\
that have the indices defined by
(\ref{eq-def-processinmixedoneeigensubsetindexstdvalindexall}).
We hereafter show how to extend that method so as to
identify each column of
\yprocessmat\
separately.
\subsection{%
Second part
of the methods}
\label{sec-two-stage-part-two}
The two important properties that we showed in Section
\ref{sec-two-stage-part-one} 
for the considered (possibly high)
dimension 
$
\Yketspacedim
=
\Yprocessinmixedonedensitymatrixvalmultiplicity
\Yprocessinmixedonedensitymatrixvalnb
$
are as follows.
First,
\ytextartitionehundredninetyfourvonemodifstepfour{we proved that}
we can ``partition''
the set of
\yketspacedim\
columns of
\yprocessmat\
in
\yprocessinmixedonedensitymatrixvalnb\
subsets,
in the sense that%
\ytextartitionehundredninetyfourvonemodifstepfour{,
in Section
\ref{sec-two-stage-part-one},}
we identified each
\yprocessinmixedonedensitymatrixvalmultiplicity -dimensional
subspace
\yprocessinmixedoneeigensubsetindexstdspace\
spanned by the columns of
\yprocessmat\
that compose the subset with index
\yprocessinmixedoneeigensubsetindexstd .
Second, 
\ytextartitionehundredninetyfourvonemodifstepfour{we proved that}
we can select which partition we perform:
the columns of
\yprocessmat\
that are gathered in the same subset are those
that correspond to the same eigenvalue of
\yprocessoutmixedonedensitymatrix\
and hence
\yprocessoutmixedonedensitymatrixestimhermunittrace .

In the present section, 
we take advantage of these
results 
to create another partition of
the columns of
\yprocessmat ,
in order to then jointly exploit these two partitions
\ytextartitionehundredninetyfourvonemodifsteptwo{as}
explained further in Section
\ref{sec-two-stage-part-three}.
We thus create a two-stage algorithm in the sense
that it successively uses two eigendecompositions
of density matrices estimated at the output of
the
process \yprocessmat ,
namely the eigendecompositions
respectively described in Section
\ref{sec-two-stage-part-one} 
and in the present subsection.
This should be contrasted with the algorithms of
Section
\ref{sec-single-stage},
that are single-stage methods in the sense that they
perform a single eigendecomposition.

More precisely, 
the input
density matrix
\ytwostageprocessinmixedtwodensitymatrix\
here applied 
to the considered unitary process
is different from
\yprocessinmixedonedensitymatrix\
but,
here again,
it 
is diagonal, it contains 
\yprocessinmixedonedensitymatrixvalnb\
different values
that each appear
\yprocessinmixedonedensitymatrixvalmultiplicity\
times
and these values are the same as in 
Section
\ref{sec-two-stage-part-one}.
However, they are here placed in a different
order:
the set of values on the main diagonal of
\ytwostageprocessinmixedtwodensitymatrix\
consists of a series of 
\yprocessinmixedonedensitymatrixvalmultiplicity\
identical 
subsets,
each of which 
successively
contains the
\yprocessinmixedonedensitymatrixvalnb\
values
$
\Yprocessinmixedonedensitymatrixvaldiagdifffirst
,
\dots
,
\Yprocessinmixedonedensitymatrixvaldiagdifflast
$.
The density matrix
\ytwostageprocessinmixedtwodensitymatrix\
thus reads
\begin{equation}
\Ytwostageprocessinmixedtwodensitymatrix
=
\left[
\begin{tabular}{lllllll}
\yprocessinmixedonedensitymatrixvaldiagdifffirst
&
0
&
$
\dots
$
&
&
&
&
0
\\
&
$
\ddots
$
\\
&
&
\yprocessinmixedonedensitymatrixvaldiagdifflast
\\
&
&
&
\yprocessinmixedonedensitymatrixvaldiagdifffirst
\\
&
&
&
&
$
\ddots
$
\\
&
&
&
&
&
\yprocessinmixedonedensitymatrixvaldiagdifflast
\\
0
&
&
&
&
&
&
$
\ddots
$
\end{tabular}
\right]
.
\label{eq-def-processinmixedonedensitymatrix-method-two-stage-part-two}
\end{equation}
\ytextartitionehundredninetyfourvfivemodifstepone{This
may also be expressed as
\begin{equation}
\Ytwostageprocessinmixedtwodensitymatrix
= 
I
_{\Yprocessinmixedonedensitymatrixvalmultiplicity}
\otimes
\mathrm{diag}
(
r
)
\label{eq-def-processinmixedonedensitymatrix-method-two-stage-part-two-compact}
\end{equation}
with the same 
notations as in
(\ref{eq-def-processinmixedonedensitymatrix-method-two-stage-compact}).}

This input density matrix
\ytwostageprocessinmixedtwodensitymatrix\
is used 
in the same way
as in
Section
\ref{sec-two-stage-part-one}.
So,
the density matrix
\ytwostageprocessoutmixedtwodensitymatrix\
of the corresponding process output
may be derived from
\ytwostageprocessinmixedtwodensitymatrix\
in the same way as in
(\ref{eq-processoutmixedonedensitymatrixvsinmixedonedensitymatrix}).
An estimate 
\yprocessoutmixedtwodensitymatrixestim\
of
\ytwostageprocessoutmixedtwodensitymatrix\
is derived by a QST algorithm and,
if required by the considered QST algorithm,
we preprocess
\yprocessoutmixedtwodensitymatrixestim\
in order to derive:
1)
its Hermitian version
\ytwostageprocessoutmixedtwodensitymatrixestimherm\
(in the same way as in
(\ref{eq-def-processoutmixedonedensitymatrixestimherm}))
and then
2)
its Hermitian and unit-trace version
\ytwostageprocessoutmixedtwodensitymatrixestimhermunittrace\
(in the same way as in
(\ref{eq-def-processoutmixedonedensitymatrixestimhermunittrace})).
We then perform an eigendecomposition of
\ytwostageprocessoutmixedtwodensitymatrixestimhermunittrace .
This yields
a diagonal matrix
\ytwostageprocessoutmixedtwodensitymatrixestimhermunittraceeigenmatval\
containing its eigenvalues
and a matrix
\ytwostageprocessoutmixedtwodensitymatrixestimhermunittraceeigenmatvec\
whose columns are its
associated
eigenvectors.

We then apply a joint permutation to the diagonal
elements of
\ytwostageprocessoutmixedtwodensitymatrixestimhermunittraceeigenmatval\
and to the columns of
\ytwostageprocessoutmixedtwodensitymatrixestimhermunittraceeigenmatvec,
so that the values corresponding to the
diagonal
elements of
\ytwostageprocessoutmixedtwodensitymatrixestimhermunittraceeigenmatval\
become in 
nonincreasing
order.
Here again,
each of the
\yprocessinmixedonedensitymatrixvalnb\
successive
subsets of 
\yprocessinmixedonedensitymatrixvalmultiplicity\
identical values
along the diagonal of the 
reordered version of
\ytwostageprocessoutmixedtwodensitymatrixestimhermunittraceeigenmatval\
(values equal to
eigenvalue
\yprocessinmixedtwodensitymatrixvaldiagdiffeigensubsetindexstd )
has an index
\yprocessinmixedtwoeigensubsetindexstd,
with
$
\Yprocessinmixedtwoeigensubsetindexstd
\in
\{
1,
\dots
,
\Yprocessinmixedonedensitymatrixvalnb
\}
$,
and
consists of elements with indices
\begin{equation}
(
\Yprocessinmixedtwoeigensubsetindexstd
-
1
)
\Yprocessinmixedonedensitymatrixvalmultiplicity
+
\Yprocessinmixedtwoeigensubsetindexstdvalindexstd,
\hspace{5mm}
\mathrm{with}
\hspace{5mm}
\Yprocessinmixedtwoeigensubsetindexstdvalindexstd
\in
\{
1,
\dots
,
\Yprocessinmixedonedensitymatrixvalmultiplicity
\}
.
\label{eq-def-processinmixedtwoeigensubsetindexstdvalindexall}
\end{equation}
Moreover,
the corresponding
matrix,
derived from
\ytwostageprocessoutmixedtwodensitymatrixestimhermunittraceeigenmatvec,
with reordered 
eigenvectors in its columns
is denoted as
\ytwostageprocessmatestimthree\
\footnote{In 
\ytextartitionehundredninetyfourvonemodifsteptwo{this whole}
paper, all the quantities
of importance related to estimates of the
considered process are denoted with the
same letter
and increasing indices 1, 2 and so on.
When describing our first 
type of methods
in Section
\ref{sec-single-stage},
for the sake of clarity we used detailed
notations for these estimates, 
\ytextartitionehundredninetyfourvfivemodifstepone{therefore}
with
a first notation (with index 1) for the
quantity before column reordering and a
second notation (with index 2) for the
quantity after column reordering.
However, we then did not use the first
notation.
We here want to make it clear that,
for the second 
type of methods
described in the
present section, we also consider
two such quantities, 
\ytextartitionehundredninetyfourvfivemodifstepone{namely}
before and
after reordering.
However, we do not introduce a notation
for the quantity before reordering
because, here again, we would not use
it afterwards.
Therefore, we here only introduce
a single quantity, with an
index equal to 3.}.
Its subset
of columns
that also have the indices defined by
(\ref{eq-def-processinmixedtwoeigensubsetindexstdvalindexall})
is a set of 
\yprocessinmixedonedensitymatrixvalmultiplicity\
eigenvectors that form a basis of the
\yprocessinmixedonedensitymatrixvalmultiplicity -dimensional
eigenspace of
\ytwostageprocessoutmixedtwodensitymatrixestimhermunittrace\
associated with its eigenvalue
\yprocessinmixedtwodensitymatrixvaldiagdiffeigensubsetindexstd .
This eigenspace, obtained in the second part of this method,
is hereafter denoted as
\yprocessinmixedtwoeigensubsetindexstdspace .

\yprocessinmixedtwoeigensubsetindexstdspace\
also has a relationship with
the actual matrix
\yprocessmat\
to be identified,
but it is different from the relationship disclosed for
\yprocessinmixedoneeigensubsetindexstdspace\
in Section
\ref{sec-two-stage-part-one}.
Here 
again,
\yprocessinmixedtwoeigensubsetindexstdspace\
is the 
\yprocessinmixedonedensitymatrixvalmultiplicity -dimensional
subspace spanned by the
\yprocessinmixedonedensitymatrixvalmultiplicity\
columns of 
\yprocessmat\
that correspond to the eigenvalue
\yprocessinmixedtwodensitymatrixvaldiagdiffeigensubsetindexstd .
However, here the indices of these columns are different from those
of Section
\ref{sec-two-stage-part-one}:
they are equal to the column indices of all occurrences of
\yprocessinmixedtwodensitymatrixvaldiagdiffeigensubsetindexstd\
in
(\ref{eq-def-processinmixedonedensitymatrix-method-two-stage-part-two}),
that is
\begin{equation}
\Yprocessinmixedtwoeigensubsetindexstd
+
\Yprocessinmixedtwoeigensubsetindexstdvalindexstd
\Yprocessinmixedonedensitymatrixvalnb
\hspace{5mm}
\mathrm{with}
\hspace{5mm}
\Yprocessinmixedtwoeigensubsetindexstdvalindexstd
\in
\{
0
\dots
,
\Yprocessinmixedonedensitymatrixvalmultiplicity
-
1
\}
.
\label{eq-def-processinmixedtwoprocessmatcolumnall}
\end{equation}

This shows that this second part of our second 
type of methods
succeeds in 
finding a basis for
each of the
\yprocessinmixedonedensitymatrixvalnb\
subspaces 
\yprocessinmixedtwoeigensubsetindexstdspace\
that:
1)
corresponds to a value
\yprocessinmixedtwoeigensubsetindexstd,
with
$
\Yprocessinmixedtwoeigensubsetindexstd
\in
\{
1,
\dots
,
\Yprocessinmixedonedensitymatrixvalnb
\}
$,
and 2)
is
associated with the
\yprocessinmixedonedensitymatrixvalmultiplicity\
columns of
\yprocessmat\
that have the indices defined by
(\ref{eq-def-processinmixedtwoprocessmatcolumnall}).
We hereafter show how to combine the properties of these subspaces
\yprocessinmixedtwoeigensubsetindexstdspace\
and of the subspaces
\yprocessinmixedoneeigensubsetindexstdspace\
derived in
the first part of this algorithm.
\subsection{%
Third part
of the methods}
\label{sec-two-stage-part-three}
We now consider all subspace intersections
\mbox{$
\Yprocessinmixedoneeigensubsetindexstdspace
\cap
\Yprocessinmixedtwoeigensubsetindexstdspace
$},
for
$
\Yprocessinmixedoneeigensubsetindexstd
\in
\{
1,
\dots
,
\Yprocessinmixedonedensitymatrixvalnb
\}
$
and
$
\Yprocessinmixedtwoeigensubsetindexstd
\in
\{
1,
\dots
,
\Yprocessinmixedonedensitymatrixvalnb
\}
$.
The subspaces
\yprocessinmixedoneeigensubsetindexstdspace\
and
\yprocessinmixedtwoeigensubsetindexstdspace\
are each defined by a set of column vectors of
\yprocessmat\
and we want 
each such pair of subspaces
to share 
{\it
only one} column vector
(or possibly no columns for some of these intersections).
When this condition is met,
\mbox{$
\Yprocessinmixedoneeigensubsetindexstdspace
\cap
\Yprocessinmixedtwoeigensubsetindexstdspace
$}
is equal to the one-dimensional subspace defined by this shared
column vector,
because
\yprocessmat\
is unitary and all its column vectors are therefore orthogonal
(see
\cite{book-horn-johnson} p. 67).
Moreover, we want the set of
$
\Yprocessinmixedonedensitymatrixvalnb
^2
$
intersection
column vectors thus created to yield
all
\yketspacedim\
columns of
\yprocessmat\
(up to 
scale
factors).
This requires
$
\Yprocessinmixedonedensitymatrixvalnb
^2
\geq
\Yketspacedim
$
and hence,
due to
(\ref{eq-def-ketspacedim-two-stage}):
\begin{equation}
\Yprocessinmixedonedensitymatrixvalmultiplicity
\leq
\Yprocessinmixedonedensitymatrixvalnb
.
\label{eq-def-processinmixedonedensitymatrixvalmultiplicity-leq-processinmixedonedensitymatrixvalnb}
\end{equation}
Besides,
\ytextartitionehundredninetyfourvfivemodifstepone{the number
\yprocessinmixedonedensitymatrixvalnb }
of different eigenvalues
of the input density matrix
(as defined in Section
\ytextartitionehundredninetyfourvfivemodifstepone{\ref{sec-two-stage-part-one})}
is upper bounded by the value
\yeigenvalnbmax
, as explained above.
Therefore,
due to
(\ref{eq-def-processinmixedonedensitymatrixvalmultiplicity-leq-processinmixedonedensitymatrixvalnb})
and
(\ref{eq-def-ketspacedim-two-stage}),
the 
state
space dimension
\yketspacedim\
that can be handled by the methods proposed in the present section
is upper bounded by:
\begin{equation}
\Yketspacedim
\leq
(
\Yeigenvalnbmax
)
^2
.
\label{eq-ketspacedim-bound-method-two-stage}
\end{equation}
This maximum space dimension is therefore much higher than the bound
(\ref{eq-ketspacedim-bound-method-one-stage})
faced with
the
methods of Section
\ref{sec-single-stage}%
, under the assumption that
\yeigenvalnbmax\
here remains the same as with single-stage methods,
as discussed above. 
In that case,
the methods of the present section are
much more attractive
for handling high-dimensional spaces
(but more complex).

In the remainder of this section, we only consider the case when
\begin{equation}
\Yprocessinmixedonedensitymatrixvalmultiplicity
=
\Yprocessinmixedonedensitymatrixvalnb
\end{equation}
which
makes the analysis 
simpler.
Combining this condition with
(\ref{eq-def-ketspacedim-two-stage})
yields
\begin{equation}
\Yketspacedim
=
\Yprocessinmixedonedensitymatrixvalmultiplicity
^2
=
\Yprocessinmixedonedensitymatrixvalnb
^2
.
\label{eq-ketspacedim-equal-square}
\end{equation}
When
\yprocessinmixedonedensitymatrixvalnb\
is moreover set to its maximum value
\yeigenvalnbmax ,
we obtain
$
\Yketspacedim
=
(
\Yeigenvalnbmax
)
^2
$,
\ytextartitionehundredninetyfourvfivemodifstepone{so that}
we reach the upper bound defined by
(\ref{eq-ketspacedim-bound-method-two-stage}).

One then analyzes how the indices of the columns of
\yprocessmat\
corresponding to
\yprocessinmixedoneeigensubsetindexstdspace\
(defined by
(\ref{eq-def-processinmixedoneeigensubsetindexstdvalindexall}))
are interleaved
with
the indices of the columns of
\yprocessmat\
corresponding to
\yprocessinmixedtwoeigensubsetindexstdspace\
(defined by
(\ref{eq-def-processinmixedtwoprocessmatcolumnall})),
for
any given couple of values
$
(
\Yprocessinmixedoneeigensubsetindexstd
,
\Yprocessinmixedtwoeigensubsetindexstd
)
$,
with
$
\Yprocessinmixedoneeigensubsetindexstd
\in
\{
1,
\dots
,
\Yprocessinmixedonedensitymatrixvalnb
\}
$
and
$
\Yprocessinmixedtwoeigensubsetindexstd
\in
\{
1,
\dots
,
\Yprocessinmixedonedensitymatrixvalnb
\}
$.
This 
\ytextartitionehundredninetyfourvonemodifsteptwo{shows}
that,
whatever
$
(
\Yprocessinmixedoneeigensubsetindexstd
,
\Yprocessinmixedtwoeigensubsetindexstd
)
$,
the sets of columns of
\yprocessmat\
respectively associated with
\yprocessinmixedoneeigensubsetindexstdspace\
and
\yprocessinmixedtwoeigensubsetindexstdspace\
share one and only one column,
and that its index is
\begin{equation}
(
\Yprocessinmixedoneeigensubsetindexstd
-
1
)
\Yprocessinmixedonedensitymatrixvalmultiplicity
+
\Yprocessinmixedtwoeigensubsetindexstd
.
\label{ed-share-column-index}
\end{equation}
Moreover, when
\yprocessinmixedoneeigensubsetindexstd\
and
\yprocessinmixedtwoeigensubsetindexstd\
are varied over all their ranges
$
\Yprocessinmixedoneeigensubsetindexstd
\in
\{
1,
\dots
,
\Yprocessinmixedonedensitymatrixvalnb
\}
$
and
$
\Yprocessinmixedtwoeigensubsetindexstd
\in
\{
1,
\dots
,
\Yprocessinmixedonedensitymatrixvalnb
\}
$,
the resulting set of 
$
\Yprocessinmixedonedensitymatrixvalnb
^2
$
indices
(\ref{ed-share-column-index})
spans all 
$
\Yketspacedim
=
\Yprocessinmixedonedensitymatrixvalmultiplicity
\Yprocessinmixedonedensitymatrixvalnb
=
\Yprocessinmixedonedensitymatrixvalnb
^2
$
columns of
\yprocessmat .
This approach therefore succeeds in identifying
all columns of
\yprocessmat ,
up to 
scale
factors 
at this stage.

The practical algorithm proposed for 
implementing the above approach operates as follows.
Successively for each couple
$
(
\Yprocessinmixedoneeigensubsetindexstd
,
\Yprocessinmixedtwoeigensubsetindexstd
)
$,
with
$
\Yprocessinmixedoneeigensubsetindexstd
\in
\{
1,
\dots
,
\Yprocessinmixedonedensitymatrixvalnb
\}
$
and
$
\Yprocessinmixedtwoeigensubsetindexstd
\in
\{
1,
\dots
,
\Yprocessinmixedonedensitymatrixvalnb
\}
$,
we need to determine a vector situated
in the 
1-dimensional intersection
of the subspaces
\yprocessinmixedoneeigensubsetindexstdspace\
and
\yprocessinmixedtwoeigensubsetindexstdspace .
The
data available in practice for
defining
\yprocessinmixedoneeigensubsetindexstdspace\
consist of the set of
columns of
\yprocessmatestimtwo\
that have the indices defined by
(\ref{eq-def-processinmixedoneeigensubsetindexstdvalindexall})
and
the data that
define
\yprocessinmixedtwoeigensubsetindexstdspace\
consist of the set of
columns of
\ytwostageprocessmatestimthree\
that have the indices defined by
(\ref{eq-def-processinmixedtwoeigensubsetindexstdvalindexall}).
To determine this 
1-dimensional
subspace intersection,
we here use 
the first direction provided by
the standard Canonical Correlation Analysis,
or CCA,
e.g. described in
\cite{bibref-ellipses-chap-mv-livresaporta}
and directly available in the Matlab scientific software
\cite{mathworks-cca}.
This approach is by the way closely related to the
eigendecomposition tool, that is the core of the methods
proposed
in the present paper%
\footnote{%
It should also be noted that an alternative to CCA,
based on
Singular Value Decomposition, 
and that allows one to determine
subspace intersections, 
is 
available in
\cite{amoi6-72}.
It is a
building block of the method called SIBIS,
for
``Subspace-Intersection Blind Identification
and Separation''.
Beyond its subspace-intersection
tool,
SIBIS has the following relationships with the present paper in terms of 
final
goal.
SIBIS
performs
1) Blind Mixture Identification, or BMI
\cite{amoi6-144,amoi6-48,ldelathauwer-2007-mixture-identif,paper-amuse},
which is a type of classical system
identification problem and is thus related
to (the classical counterpart of) QPT
and 2)
Blind Source Separation,
which is a problem closely connected to
classical BMI.}.

By calculating the intersection
\mbox{$
\Yprocessinmixedoneeigensubsetindexstdspace
\cap
\Yprocessinmixedtwoeigensubsetindexstdspace
$},
we obtain one
column vector
of
\yprocessmat ,
possibly
up to an arbitrary complex factor,
unless this vector is guaranteed to be normalized by the
considered CCA tool.
If it is not,
we then reduce
this factor to a phase factor, by
dividing this intersection vector 
by its norm,
so as to
enforce its norm 
to become equal to one
(again because
\yprocessmat\
is known to be unitary and thus to have unit-norm column vectors).

All these normalized 
results of CCA are used to create
an additional matrix related to the estimation
of
\yprocessmat\
and therefore denoted as
\ytwostageprocessmatestimfour
:
each application of CCA
to a given couple
$
(
\Yprocessinmixedoneeigensubsetindexstd
,
\Yprocessinmixedtwoeigensubsetindexstd
)
$
yields a normalized vector that is stored in the column
of
\ytwostageprocessmatestimfour\
whose index is defined by
(\ref{ed-share-column-index}).
The matrix
\ytwostageprocessmatestimfour\
obtained at this stage 
is therefore equal to
\yprocessmat\
up to an arbitrary
phase factor 
in each of its columns
(and up to estimation errors).
\subsection{%
Fourth part
of the methods}
\label{sec-two-stage-part-four}
The properties of
\ytwostageprocessmatestimfour\
defined at the end of the previous section are exactly the same as those
obtained in our single-stage methods
for
\yprocessmatestimtwo:
see
Section
\ref{sec-single-stage-part-one},
including
(\ref{eq-processmatestimtwo-vs-processmat-and-phases}).
Therefore, the final stage of the method considered here consists of
processing
\ytwostageprocessmatestimfour\
in the same way as we processed
\yprocessmatestimtwo\
in Section
\ref{sec-single-stage-part-two}.
This yields
a matrix
\yprocessmatestimthree\
that here again
succeeds in restoring
\yprocessmat ,
up to only a global phase factor.

A pseudo-code of the version of our 
algorithm described in this
Section
\ref{sec-two-stage}
is provided in
Algorithm
\ref{label-algodef-eqpttwo}.
This algorithm is called EQPT2.
Two variants (called EQPT3 and EQPT4)
of 
two-stage QPT methods
are moreover presented in Appendix~\ref{appendix-sec-two-stage-variants}.
\begin{algorithm}[htb!]
{%
{%
\SetKwInOut{Input}{Input}\SetKwInOut{Output}{Output}
\Input{%
a) 
Estimate 
\ytextartitionehundredninetyfourvfivemodifstepone{%
\yprocessoutmixedonedensitymatrixestimhermunittrace}
of output density matrix
\yprocessoutmixedonedensitymatrix\
(provided by 
QST
and obtained for
input density matrix
(\ref{eq-def-processinmixedonedensitymatrix-method-two-stage})
with the diagonal values defined in
(\ref{eq-opdensity-diag-val-uniform-two-stage})).
b) Estimate 
\ytextartitionehundredninetyfourvfivemodifstepone{%
\ytwostageprocessoutmixedtwodensitymatrixestimhermunittrace}
of output density matrix
\ytwostageprocessoutmixedtwodensitymatrix\
(provided by QST
and obtained for
input density matrix
(\ref{eq-def-processinmixedonedensitymatrix-method-two-stage-part-two})
with the diagonal values defined in
(\ref{eq-opdensity-diag-val-uniform-two-stage})).
c)~Estimate 
\yprocessoutpuretwoketestim\
of output ket
\yprocessoutpuretwoket\
(provided by QST
and obtained for all components 
of
input ket
\yprocessinpuretwoket\
equal to
$
1
/
\sqrt{
\Yketspacedim
}
$).%
}
\Output{%
Estimate
\yprocessmatestimthree\
of 
quantum process matrix
\yprocessmat .}
\BlankLine
\Begin{
\tcc{\ytextartitionehundredninetyfourvfivemodifstepone{Exploit}
\yprocessoutmixedonedensitymatrixestimhermunittrace :}
Eigendecomposition:
derive 
1) a diagonal matrix
\yprocessoutmixedonedensitymatrixestimhermunittraceeigenmatval\
that contains the eigenvalues of
\yprocessoutmixedonedensitymatrixestimhermunittrace\
in an arbitrary order
and 2) a matrix
\yprocessoutmixedonedensitymatrixestimhermunittraceeigenmatvec\
whose columns are 
eigenvectors of
\yprocessoutmixedonedensitymatrixestimhermunittrace\
in the same order as eigenvalues\;
Reorder the eigenvalues in
\yprocessoutmixedonedensitymatrixestimhermunittraceeigenmatval\
in 
nonincreasing
order
and 
apply the same permutation to the columns of
\yprocessoutmixedonedensitymatrixestimhermunittraceeigenmatvec\
to create the matrix
\yprocessmatestimtwo\;
\tcc{\ytextartitionehundredninetyfourvfivemodifstepone{Exploit}
\ytwostageprocessoutmixedtwodensitymatrixestimhermunittrace :}
Eigendecomposition:
derive 
1) a diagonal matrix
\ytwostageprocessoutmixedtwodensitymatrixestimhermunittraceeigenmatval\
that contains the eigenvalues of
\ytwostageprocessoutmixedtwodensitymatrixestimhermunittrace\
in an arbitrary order
and 2) a matrix
\ytwostageprocessoutmixedtwodensitymatrixestimhermunittraceeigenmatvec\
whose columns are 
eigenvectors of
\ytwostageprocessoutmixedtwodensitymatrixestimhermunittrace\
in the same order as eigenvalues\;
Reorder the eigenvalues in
\ytwostageprocessoutmixedtwodensitymatrixestimhermunittraceeigenmatval\
in 
nonincreasing
order
and 
apply the same permutation to the columns of
\ytwostageprocessoutmixedtwodensitymatrixestimhermunittraceeigenmatvec\
to create the matrix
\ytwostageprocessmatestimthree\;
\tcc{%
\ytextartitionehundredninetyfourvfivemodifstepone{Combine}
\yprocessmatestimtwo\
and
\ytwostageprocessmatestimthree
\ytextartitionehundredninetyfourvfivemodifstepone{:}%
}
\For{$
\Yprocessinmixedoneeigensubsetindexstd
=
1
$
to
\yprocessinmixedonedensitymatrixvalnb}
{
\For{$
\Yprocessinmixedtwoeigensubsetindexstd
=
1
$
to
\yprocessinmixedonedensitymatrixvalnb}
{
Set column of \ytwostageprocessmatestimfour\
\ytextartitionehundredninetyfourvfivemodifstepone{with
index equal to
$
(
(
\Yprocessinmixedoneeigensubsetindexstd
-
1
)
\Yprocessinmixedonedensitymatrixvalmultiplicity
+
\Yprocessinmixedtwoeigensubsetindexstd
)
$: Set}
it to
first vector provided by CCA of
1) the set of
columns of
\yprocessmatestimtwo\
with
indices 
\ytextartitionehundredninetyfourvfivemodifstepone{$
(
(
\Yprocessinmixedoneeigensubsetindexstd
-
1
)
\Yprocessinmixedonedensitymatrixvalmultiplicity
+
\Yprocessinmixedoneeigensubsetindexstdvalindexstd
)
$
with
$
\Yprocessinmixedoneeigensubsetindexstdvalindexstd
\in
\{
1,
\dots
,
\Yprocessinmixedonedensitymatrixvalmultiplicity
\}$}
and 
2)
the set of
columns of
\ytwostageprocessmatestimthree\
with indices 
\ytextartitionehundredninetyfourvfivemodifstepone{$
(
(
\Yprocessinmixedtwoeigensubsetindexstd
-
1
)
\Yprocessinmixedonedensitymatrixvalmultiplicity
+
\Yprocessinmixedtwoeigensubsetindexstdvalindexstd
)
$
with
$
\Yprocessinmixedtwoeigensubsetindexstdvalindexstd
\in
\{
1,
\dots
,
\Yprocessinmixedonedensitymatrixvalmultiplicity
\}
$}%
\;
\tcc{If the
CCA tool
does not provide unit-norm vectors, then do 
it:}
Divide 
above
column of
\ytwostageprocessmatestimfour\ by its norm%
\;
}
}
\tcc{\ytextartitionehundredninetyfourvfivemodifstepone{Exploit}
\yprocessoutpuretwoketestim :}
\ytextartitionehundredninetyfourvfivemodifstepone{$
\Yprocessmatestimthreeketinterm
=
\Ytwostageprocessmatestimfour
\Ytransconjug
\Yprocessoutpuretwoketestim
$\;}
\ytextartitionehundredninetyfourvfivemodifstepone{$
\Yprocessmatestimthree
=
\Ytwostageprocessmatestimfour
\Ymathspace
\mathrm{diag}
(
\Yprocessmatestimthreeketinterm
\oslash
\Yprocessinpuretwoket
)$\;}
}
\caption{EQPT2: 
second
Eigenanalysis-based 
QPT
algorithm (composed of two stages).
\ytextartitionehundredninetyfourvfivemodifstepone{This version
is for the case when the QST algorithm provides
estimates that meet the known properties of the
actual quantities. Otherwise, see text.}%
}
}
\label{label-algodef-eqpttwo}
}
\end{algorithm}
\section{Proposed multi-stage QPT methods}
\label{sec-multi-stage}
The QPT methods proposed in the previous sections
can be 
extended as follows,
by 
\ytextartitionehundredninetyfourvonemodifstepthree{%
selecting}
density matrices
\yprocessinmixedonedensitymatrix\
of input mixed states 
that
are diagonal%
\ytextartitionehundredninetyfourvonemodifstepthree{, that}
contain 
{\it
several occurrences} of each value
that appears on 
their diagonal
\ytextartitionehundredninetyfourvonemodifstepthree{and
that have additional properties.}
More precisely,
here again,
each
value that appears on the main diagonal of
\yprocessinmixedonedensitymatrix\
appears
\yprocessinmixedonedensitymatrixvalmultiplicity\
times%
\ytextartitionehundredninetyfourvonemodifstepthree{%
, with
$
\Yprocessinmixedonedensitymatrixvalmultiplicity
>
1
$,}
and
that diagonal contains
\yprocessinmixedonedensitymatrixvalnb\
different values,
here
possibly with
$
\Yprocessinmixedonedensitymatrixvalnb
\neq
\Yprocessinmixedonedensitymatrixvalmultiplicity
$.
The number of rows of
\yprocessinmixedonedensitymatrix ,
equal to the dimension of the considered 
state
space,
is thus here again defined by
(\ref{eq-def-ketspacedim-two-stage}).
The new feature introduced here consists of
storing
all 
\yprocessinmixedonedensitymatrixvalmultiplicity\
occurrences of any given diagonal value of
\yprocessinmixedonedensitymatrix\
in
\ymultistageyprocessinmixeddensitymatrixblocknb\
blocks
of adjacent
rows,
with each block consisting of
\ymultistageyprocessinmixeddensitymatrixblocksize\
rows,
and hence with
\ytextartitionehundredninetyfourvonemodifstepone{%
\begin{equation}
\Ymultistageyprocessinmixeddensitymatrixblocknb
\Ymultistageyprocessinmixeddensitymatrixblocksize
=
\Yprocessinmixedonedensitymatrixvalmultiplicity
.
\label{eq-processinmixedonedensitymatrixvalmultiplicity-product}
\end{equation}%
}
Besides, we consider the case when
\ymultistageyprocessinmixeddensitymatrixblocknb ,
\ymultistageyprocessinmixeddensitymatrixblocksize\
and
\yprocessinmixedonedensitymatrixvalmultiplicity\
are powers of 2.
We denote as
\ymultistageyprocessinmixeddensitymatrixblocknblogbasetwo\
the 
integer-valued
logarithm 
of
\ymultistageyprocessinmixeddensitymatrixblocknb\
in base 2:
we have
\ytextartitionehundredninetyfourvonemodifstepone{%
\begin{equation}
\Ymultistageyprocessinmixeddensitymatrixblocknb
=
2^
\Ymultistageyprocessinmixeddensitymatrixblocknblogbasetwo
.
\label{eq-multistageyprocessinmixeddensitymatrixblocknb-power-two}
\end{equation}%
}
We hereafter consider all possible integer values of
\ymultistageyprocessinmixeddensitymatrixblocknblogbasetwo,
for a fixed value
\yprocessinmixedonedensitymatrixvalmultiplicity\
and varying values of
\ymultistageyprocessinmixeddensitymatrixblocknb\
and
\ymultistageyprocessinmixeddensitymatrixblocksize .
In other words, we consider all 
linked
powers of 2 
for
\ymultistageyprocessinmixeddensitymatrixblocknb\
and
\ymultistageyprocessinmixeddensitymatrixblocksize,
in their acceptable range.
The minimum of
\ymultistageyprocessinmixeddensitymatrixblocknblogbasetwo\
is 0,
corresponding to the minimum of
\ymultistageyprocessinmixeddensitymatrixblocknb ,
equal to
1,
and to the maximum of
\ymultistageyprocessinmixeddensitymatrixblocksize ,
equal to
\yprocessinmixedonedensitymatrixvalmultiplicity.
Similarly,
the maximum of
\ymultistageyprocessinmixeddensitymatrixblocknblogbasetwo\
is equal to
$
\log_2
\Yprocessinmixedonedensitymatrixvalmultiplicity
$,
corresponding to 
the maximum of
\ymultistageyprocessinmixeddensitymatrixblocknb, that is
equal to
\yprocessinmixedonedensitymatrixvalmultiplicity,
while
\ymultistageyprocessinmixeddensitymatrixblocksize\
reaches
its minimum, 
equal to 1.

Moreover, these blocks are organized as follows:
from the top to the bottom of the
diagonal of
\yprocessinmixedonedensitymatrix,
we have
a block of values equal to
\yprocessinmixedonedensitymatrixvaldiagdifffirst,
then a block of values
\yprocessinmixedonedensitymatrixvaldiagdiffsecond,
and so on until
a block of values
\yprocessinmixedonedensitymatrixvaldiagdifflast,
and then the same structure again and again,
\ytextartitionehundredninetyfourvfivemodifstepone{that is,}
blocks of values
$
\Yprocessinmixedonedensitymatrixvaldiagdifffirst
,
\Yprocessinmixedonedensitymatrixvaldiagdiffsecond
,
\dots
,
\Yprocessinmixedonedensitymatrixvaldiagdifflast
$
(when there exist several blocks containing the same value,
i.e. when
$
\Ymultistageyprocessinmixeddensitymatrixblocknb
>
1
$).
Here again, we request these 
possible
values 
to be defined in decreasing order, i.e.
$
\Yprocessinmixedonedensitymatrixvaldiagdifffirst
>
\Yprocessinmixedonedensitymatrixvaldiagdiffsecond
>
\dots
>
\Yprocessinmixedonedensitymatrixvaldiagdifflast
$,
and to take the same values as in
(\ref{eq-opdensity-diag-val-uniform-two-stage}).
The two extreme cases of this general framework
are nothing but
the two
specific cases that we already used for the
above-defined two-stage methods:
when setting
$
\Ymultistageyprocessinmixeddensitymatrixblocknblogbasetwo
=
0
$
and hence
$
\Ymultistageyprocessinmixeddensitymatrixblocknb
=
1
$
and
$
\Ymultistageyprocessinmixeddensitymatrixblocksize
=
\Yprocessinmixedonedensitymatrixvalmultiplicity
$,
we obtain the density matrix of
(\ref{eq-def-processinmixedonedensitymatrix-method-two-stage})
whereas with
$
\Ymultistageyprocessinmixeddensitymatrixblocknblogbasetwo
=
\log_2
\Yprocessinmixedonedensitymatrixvalmultiplicity
$
and hence
$
\Ymultistageyprocessinmixeddensitymatrixblocknb
=
\Yprocessinmixedonedensitymatrixvalmultiplicity
$
and
$
\Ymultistageyprocessinmixeddensitymatrixblocksize
=
1
$,
we get the density matrix of
(\ref{eq-def-processinmixedonedensitymatrix-method-two-stage-part-two}).

The main idea of the above-defined two-stage methods was
to exploit intersections of two subspaces of eigendecompositions
resulting from the above density matrices
(\ref{eq-def-processinmixedonedensitymatrix-method-two-stage})
and
(\ref{eq-def-processinmixedonedensitymatrix-method-two-stage-part-two}).
Here, we extend this idea to general multi-stage methods,
with
$
(
\log_2
\Yprocessinmixedonedensitymatrixvalmultiplicity
+
1
)
$
stages.
To this end,
we consider all
$
(
\log_2
\Yprocessinmixedonedensitymatrixvalmultiplicity
+
1
)
$
input density matrices that belong to the
above-defined general framework, 
by considering all
$
(
\log_2
\Yprocessinmixedonedensitymatrixvalmultiplicity
+
1
)
$
possible values of
\ymultistageyprocessinmixeddensitymatrixblocknblogbasetwo ,
i.e. all integers from
0
to
$
\log_2
\Yprocessinmixedonedensitymatrixvalmultiplicity
$.
Using the same approach as in Section
\ref{sec-two-stage}
\ytextartitionehundredninetyfourvfivemodifstepone{(that is,}
performing the eigendecomposition and
reordering for the estimated output
density matrix),
for each value of
\ymultistageyprocessinmixeddensitymatrixblocknblogbasetwo\
and hence of
\ymultistageyprocessinmixeddensitymatrixblocknb\
and
\ymultistageyprocessinmixeddensitymatrixblocksize ,
we obtain a set of 
\yprocessinmixedonedensitymatrixvalnb\
subspaces
denoted as
\ymultistageblocknblogbasetwoeigensubsetindexstdspace ,
with
$
\Ymultistageblocknblogbasetwoeigensubsetindexstd
\in
\{
1,
\dots
,
\Yprocessinmixedonedensitymatrixvalnb
\}
$
(each index
\ymultistageblocknblogbasetwoeigensubsetindexstd\
is used to define each subspace
in the eigendecomposition of the stage
\ymultistageyprocessinmixeddensitymatrixblocknblogbasetwo\
of the algorithm
proposed here;
its correspondence 
with the structure of the input density matrix 
is not the same as
when we used the indices
\yprocessinmixedoneeigensubsetindexstd\
and
\yprocessinmixedtwoeigensubsetindexstd\
in Section
\ref{sec-two-stage}).
\ytextartitionehundredninetyfourvonemodifstepone{These subspaces
\ymultistageblocknblogbasetwoeigensubsetindexstdspace\
are defined in more detail in Appendix
\ref{sec-appendix-relevance-multi-stage}.}
We then extend the approach of
Section
\ref{sec-two-stage} by performing subspace intersections as follows.
We successively consider all
$
\Yprocessinmixedonedensitymatrixvalnb
^
{
(
\log_2
\Yprocessinmixedonedensitymatrixvalmultiplicity
+
1
)
}
$
possible values of the set of
$
(
\log_2
\Yprocessinmixedonedensitymatrixvalmultiplicity
+
1
)
$
subspace indices
\ymultistageblocknblogbasetwoeigensubsetindexstd\
(with
$
\Ymultistageblocknblogbasetwoeigensubsetindexstd
\in
\{
1,
\dots
,
\Yprocessinmixedonedensitymatrixvalnb
\}
$),
i.e.
we consider all possible values of the set of indices
$
\{
m_0
,
\dots
,
m_
{%
\ytextartitionehundredninetyfourvonemodifstepone{%
\log_2
\Yprocessinmixedonedensitymatrixvalmultiplicity}
}
\}
$.
For each 
\ytextartitionehundredninetyfourvfivemodifstepone{such}
set of indices, we derive the intersection of
the
corresponding
$
(
\log_2
\Yprocessinmixedonedensitymatrixvalmultiplicity
+
1
)
$
subspaces
\ymultistageblocknblogbasetwoeigensubsetindexstdspace .
We thus form
$
\Yprocessinmixedonedensitymatrixvalnb
^
{
(
\log_2
\Yprocessinmixedonedensitymatrixvalmultiplicity
+
1
)
}
$
subspace intersections.
As in Section
\ref{sec-two-stage-part-three},
we want each of these intersections to be one-dimensional
(or possibly to be restricted to the null vector
for some of these intersections).
In order to identify all
\yketspacedim\
columns of
\yprocessmat,
the number of subspace intersections should be at least equal to
\yketspacedim
\ytextartitionehundredninetyfourvfivemodifstepone{.
Using
(\ref{eq-def-ketspacedim-two-stage}),
this condition reads
}
\begin{equation}
\Yprocessinmixedonedensitymatrixvalnb
^
{
\log_2
\Yprocessinmixedonedensitymatrixvalmultiplicity
}
\Yprocessinmixedonedensitymatrixvalnb
\geq
\Yprocessinmixedonedensitymatrixvalmultiplicity
\Yprocessinmixedonedensitymatrixvalnb
\ytextartitionehundredninetyfourvfivemodifstepone{.}
\label{eq-multistage-bound-columns-begin}
\end{equation}
\ytextartitionehundredninetyfourvfivemodifstepone{Taking}
the logarithm, in base 2, of that 
\ytextartitionehundredninetyfourvfivemodifstepone{equation
yields}
\begin{equation}
\log_2
\Yprocessinmixedonedensitymatrixvalnb
\geq
1
\end{equation}
that is
\begin{equation}
\Yprocessinmixedonedensitymatrixvalnb
\geq
2
.
\label{eq-multistage-bound-columns-end}
\end{equation}
In particular, for
$
\Yprocessinmixedonedensitymatrixvalnb
=
2
$,
we have equalities in
(\ref{eq-multistage-bound-columns-end})
back to
(\ref{eq-multistage-bound-columns-begin})
and
the
number
$
\Yprocessinmixedonedensitymatrixvalnb
^
{
(
\log_2
\Yprocessinmixedonedensitymatrixvalmultiplicity
+
1
)
}
$
of intersections is exactly equal to
the dimension
(\ref{eq-def-ketspacedim-two-stage}).
It should be noted that, unlike in the two-stage methods,
we here do not face the constraints
(\ref{eq-def-processinmixedonedensitymatrixvalmultiplicity-leq-processinmixedonedensitymatrixvalnb})
and hence
(\ref{eq-ketspacedim-bound-method-two-stage}).
This is because we pay the price by 
using a multi-stage,
hence more complex,
approach:
the method proposed here 
potentially
applies to any 
state
space dimension
\yketspacedim\
and any number
\yprocessinmixedonedensitymatrixvalnb\
of different eigenvalues
(with
(\ref{eq-multistage-bound-columns-end})),
provided we accept to set accordingly the value of
\yprocessinmixedonedensitymatrixvalmultiplicity,
defined by
(\ref{eq-def-ketspacedim-two-stage}),
and hence the number
$
(
\log_2
\Yprocessinmixedonedensitymatrixvalmultiplicity
+
1
)
$
of
stages
of the method and of eigendecompositions to be performed
(and, again, provided this method remains numerically accurate enough,
despite its higher computational complexity).

Moreover, 
also in this multi-stage
extension, we aim at reducing the number 
\yprocessinmixedonedensitymatrixvalnb\
of different eigenvalues,
because we thus hope to better
succeed in assigning each estimated eigenvalue to 
the correct value among the possible ones, despite estimation errors.
A very attractive feature of the 
approach
proposed here it that it
allows one to decrease
\yprocessinmixedonedensitymatrixvalnb\
to 2, thus leading to
Dichotomic 
multi-stage Eigenanalysis-based QPT%
, or DEQPT,
methods
(where ``dichotomic'' refers to the fact that the complete
set of eigenvalues
is split into two subsets, with identical values in each subset).

\ytextartitionehundredninetyfourvonemodifstepone{In Appendix
\ytextartitionehundredninetyfourvfivemodifstepone{%
\ref{sec-appendix-relevance-multi-stage-relevance}}
we prove that,
for
$
\Yprocessinmixedonedensitymatrixvalnb
= 2
$,}
the above-defined
subspace intersections are 
\ytextartitionehundredninetyfourvonemodifstepone{each}
indeed restricted 
\ytextartitionehundredninetyfourvonemodifstepone{to one}
dimension and
yield all columns of
\yprocessmat\ (up to scale or phase factors).
\ytextartitionehundredninetyfourvonemodifstepone{Moreover, 
\ytextartitionehundredninetyfourvfivemodifstepone{we
illustrate}
this property and thus the proposed complete
multi-stage methods 
\ytextartitionehundredninetyfourvfivemodifstepone{in
Appendix
\ref{sec-appendix-relevance-multi-stage-example}.}%
}
\ytextartitionehundredninetyfourvonemodifstepthree{
\ytextartitionehundredninetyfourvfivemodifstepone{Finally,
Appendix
\ref{sec-appendix-relevance-multi-stage-code}
contains a pseudo-code for
that multi-stage QPT method
applicable to}
an arbitrary state space dimension
\ytextartitionehundredninetyfourvfivemodifstepone{and
called EQPT5.}%
}
\section{Test results}
\label{sec-test-results}
To validate the 
EQPT1,
\ytextartitionehundredninetyfourvonemodifstepthree{EQPT2,
EQPT3 and EQPT5}
methods
detailed
in the previous sections,
and to evaluate 
their
accuracy,
we performed numerical tests
with
data derived from
a software simulation of the 
considered configuration.
As mentioned in Section 
\ref{sec-intro},
all the methods proposed in the present paper belong to the class of QPT
methods that first perform a complete QST for each considered process
output state and then exploit the results of these QST to achieve the
core of QPT.
The input of our QPT methods thus consists of estimates 
(more precisely, models of such estimates, as detailed below)
of pure or mixed
states provided by QST.
Various QST methods were
reported in
the literature
(see e.g.
\cite{booknielsen,amq28officiel,amq51officiel,amq41,amq52-physical-review,
amq49officiel,
amq43,amq112officiel,amq120,amq121officiel,amq122officiel,amq29officiel,
amq-sflammia-newjphys-2012,amq-ddong-IFAC2017,amq179,amq180})
and one may hereafter e.g.
use any of the available QST methods
compatible with the types of quantum
states encountered in each of the
QPT methods proposed in the present paper.
Therefore,
to here provide
numerical
results that define the accuracy of our
QPT methods, we do not restrict this
investigation to the use of a single
QST method, from which the reader could not easily
extrapolate to his preferred QST approach,
with his preferred parameter values (especially, the number of 
\ytextartitionehundredninetyfourvonemodifstepfour{times
he prepares a copy of a considered input state
and he performs a measurement for the corresponding
output state), whereas these parameter
values}
may have a strong influence on QPT
performance.
Instead,
we 
create 
a numerical model
of
quantum state estimates with a
given 
accuracy
and we focus on 
the resulting
accuracy of our QPT methods,
not only to estimate these absolute accuracies, but mainly to
determine which of these methods yields the best performance
when they use the 
{\it
same} accuracy for QST.
This moreover
allows us to investigate the numerical
performance of our QPT methods over a wide range of
conditions, by varying the accuracy of the
modeled
quantum state estimates.

It should also be noted that some papers from the literature focus on
reducing the complexity of QST algorithms
(see e.g.
\cite{amq179,amq180}).
The overall complexity of a QST-based QPT algorithm therefore does not
necessarily come from its QST part, as opposed to its subsequent part
that consists of using QST results for the core of QPT.
This also motivates us to perform tests as explained above, in order
to uncouple the investigation of the complexity of the core of QPT
(e.g. expressed in terms of required CPU time) from that of QST.

More precisely, to 
model
a ket estimate
(e.g. for the ket
\yprocessoutpuretwoket\
of Section
\ref{sec-single-stage-part-two}),
we compute its theoretical value in the considered conditions
(e.g. using
(\ref{eq-processoutpuretwoket})
with the input ket defined before
(\ref{eq-def-matrix-diag})
and a known process matrix
\yprocessmat )
and we add random, independent, complex-valued fluctuations
to all components of this ket.
Their real and imaginary parts are separately
created with a random generator that has
a probability density function that is uniform over the range
$
[
-
\Yketestimflucwidth
/ 2
,
\Yketestimflucwidth
/ 2
]
$,
where the parameter 
\yketestimflucwidth\
defines the magnitude of the modeled
estimation fluctuations
(their variance is thus equal to
$
\Yketestimflucwidth
^2
/
12
$).
This parameter
is hereafter varied so as to investigate the sensitivity of the
proposed QPT methods to the magnitude of QST estimation errors.
This simple error 
model has the advantage of being easily
interpretable.
The error model for density matrix
estimation
is defined accordingly
and is presented in Appendix
\ref{sec-appendix-model-fluctuations}.

Each elementary test uses a randomly drawn
$
\Yketspacedim
\times
\Yketspacedim
$
process matrix
\yprocessmat ,
obtained as follows.
We first create a  
$
\Yketspacedim
\times
\Yketspacedim
$
real-valued
matrix with each element randomly, uniformly, 
drawn in the interval
$
[
0,
1
]
$.
We then 
\ytextartitionehundredninetyfourvfivemodifstepone{perform
a QR decomposition of this matrix
(e.g. using one of the algorithms defined in
\cite{book-horn-johnson} p. 112,
\cite{book-golub-cite-chap-adap} p. 223,
\cite{mathworks-qr-decomposition}).}
This especially yields the so-called ``Q factor'' of that 
QR
decomposition
\ytextartitionehundredninetyfourvfivemodifstepone{(see e.g.
the software package
\cite{mathworks-qr-decomposition})}%
.
This factor is guaranteed to be a unitary matrix
(see
\cite{book-horn-johnson} p. 112%
\ytextartitionehundredninetyfourvfivemodifstepone{,
\cite{book-golub-cite-chap-adap} p. 223,
\cite{mathworks-qr-decomposition}}%
)
and we therefore use it as our
matrix
\yprocessmat .

Each above-defined matrix
\yprocessmat\
is used by feeding it with the process input states defined in the
previous sections,
computing the corresponding exact output states
(e.g. 
using
(\ref{eq-processoutmixedonedensitymatrixvsinmixedonedensitymatrix})
and
(\ref{eq-processoutpuretwoket})),
deriving their above-defined modeled estimates associated with QST
and
applying the considered QPT method to those estimates,
so as to derive the process matrix estimate
\yprocessmatestimthree .

The criterion used to evaluate the performance obtained in such a test
aims at measuring 
how close we get to the case when
\yprocessmatestimthree\ is equal to
\yprocessmat\
up to a phase factor.
The criterion used to this end is 
defined by first introducing
\begin{equation}
NMSE
(
\Yprocessmat
,
\Yprocessmatestimthree
)
=
\min
_{\theta}
\left[
\frac{1}
{
2
\Yketspacedim
}
||
\Yprocessmat
-
e^{
\Ysqrtminusone
\theta
}
\Yprocessmatestimthree
||
_F
^2
\right]
\end{equation}
where
$
||
.
||
_F
$
stands for the Frobenius norm
and where the phase
$
\theta
$
is chosen so as to minimize this
NMSE.
The NMSE value resulting from this minimization
may be shown to be equal to
\begin{equation}
NMSE
(
\Yprocessmat
,
\Yprocessmatestimthree
)
=
\frac{1}
{
2
\Yketspacedim
}
\left[
||
\Yprocessmat
||
_F
^2
+
||
\Yprocessmatestimthree
||
_F
^2
-
2
\
|
\mathrm{Tr}
(
\Yprocessmat
\Ytransconjug
\Yprocessmatestimthree
)
|
\right]
.
\end{equation}
This criterion is a Normalized Mean squared Error (NMSE)
since it is based on the 
sum of the squared moduli of the elements of the error matrix
$
(
\Yprocessmat
-
e^{
\Ysqrtminusone
\theta
}
\Yprocessmatestimthree
)
$,
then normalized by a scale factor 
chosen so that the maximum
value
of this parameter is equal to one
when
\yprocessmat\
and
\yprocessmatestimthree\
are unitary
(because their squared norms are then equal to
\yketspacedim ).
The criterion that we eventually consider is the
Normalized Root Mean Square Error (NRMSE) associated with the above
NMSE:
\begin{equation}
NRMSE
(
\Yprocessmat
,
\Yprocessmatestimthree
)
=
\sqrt{
NMSE
(
\Yprocessmat
,
\Yprocessmatestimthree
)
}
.
\end{equation}

For each dimension
\yketspacedim\
of the 
state
space,
the results reported below were obtained by running 100
above-defined elementary tests,
each with a different, randomly drawn,
process matrix
\yprocessmat\
\ytextartitionehundredninetyfourvonemodifstepfive{(except that we
only performed 10 tests when applying the EQPT5 method to 11 qubits,
as explained below).}
The performance figure then provided is the mean of
$
NRMSE
(
\Yprocessmat
,
\Yprocessmatestimthree
)
$
over all these tests.
This protocol is repeated for various values of
\yketspacedim .

\Figure[t!](topskip=0pt, botskip=0pt, midskip=0pt)[width=\linewidth]{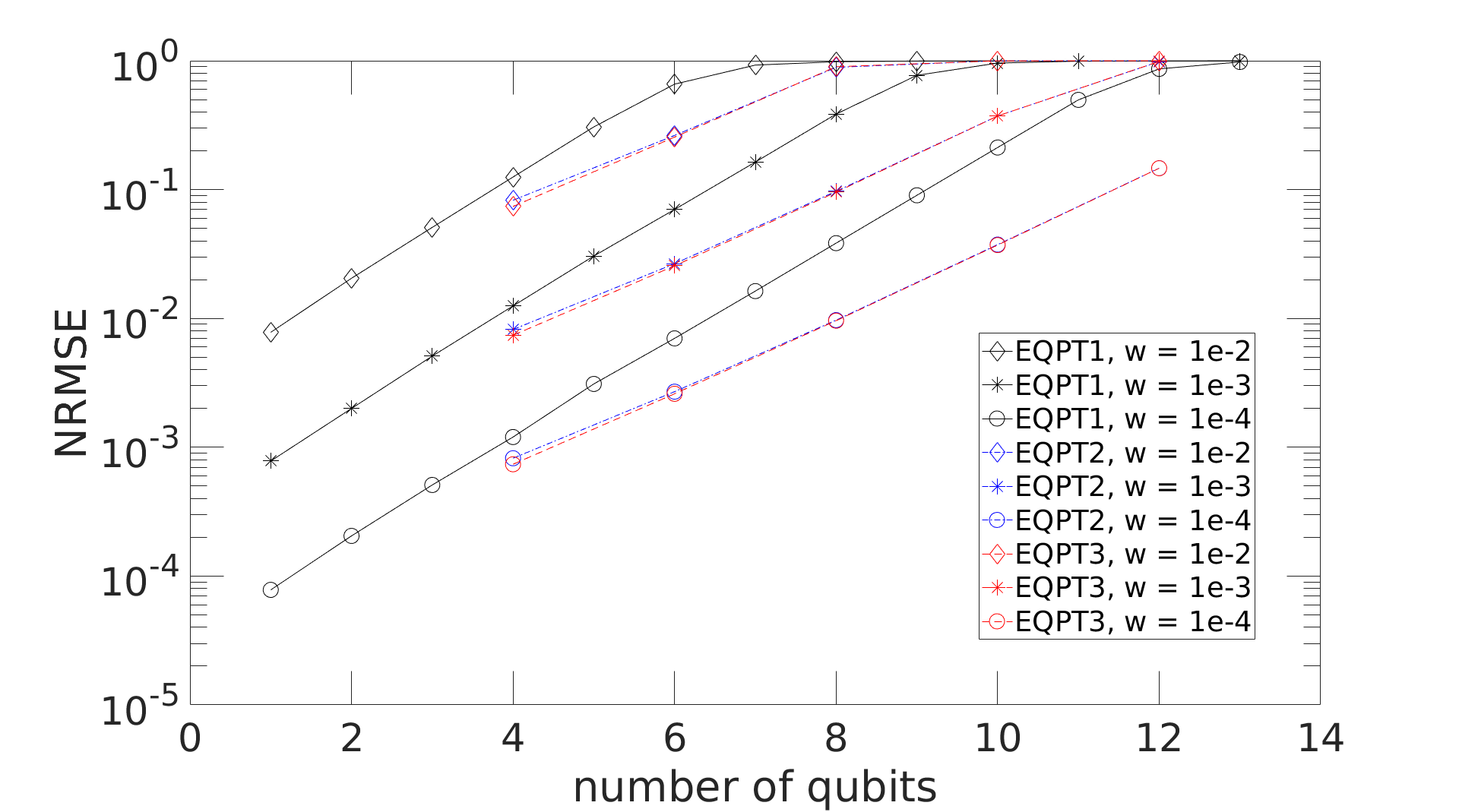}
{%
Normalized Root Mean Square Error (NRMSE) of estimation of process 
matrix \yprocessmat\ 
of proposed Quantum Process Tomography (QPT)
methods EQPT1
\ytextartitionehundredninetyfourvfivemodifstepone{(black solid lines)}%
, EQPT2
\ytextartitionehundredninetyfourvfivemodifstepone{(blue
dash doted lines)}
and EQPT3
\ytextartitionehundredninetyfourvfivemodifstepone{(red
dashed lines)}%
,
vs. 
\ytextartitionehundredninetyfourvfivemodifstepone{number
\yqubitnb\
of
qubits,}
for several values of the
parameter
\yketestimflucwidth\
that defines the magnitude of the estimation errors of 
Quantum State Tomography (QST).%
\label{fig-NRMSE}}

The reported tests were
carried out with a
standard PC
(Intel(R) Core(TM) i5-7500 CPU running at 3.40 GHz with 16 
GB 
of RAM),
using the Matlab
software environment.
\ytextartitionehundredninetyfourvonemodifstepthree{As a first step,}
Fig.
\ref{fig-NRMSE}
shows the results thus obtained with our
\ytextartitionehundredninetyfourvonemodifstepthree{single-stage
and two-stage methods, namely
EQPT1
(black solid lines),
EQPT2
(blue dash doted lines)
and EQPT3
(red dashed lines).}
The results for EQPT1 are provided for all values of the 
state
space
dimension
\yketspacedim\
that take the form
$
\Yketspacedim
=
2^{\Yqubitnb}
$,
where
\yqubitnb\
is the equivalent number of qubits,
that is here varied from
2 to 13.
$
\Yqubitnb
=
13
$
is the maximum number that was accepted in the considered 
hardware and software environment
without running out of memory
(for non-optimized code).
The same approach applies to
EQPT2 and EQPT3
except that,
due to
(\ref{eq-ketspacedim-equal-square}),
only even values of
\yqubitnb\
were considered.
Moreover, for EQPT2 and EQPT3,
performance turned out to often be somewhat degraded
for the lowest possible 
state
space dimension, namely
$
\Yketspacedim
=
4
$,
that corresponds to
$
\Yqubitnb
=
2
$
qubits.
This is not an issue, since EQPT2 and EQPT3 were developed to address
higher values of
\yketspacedim\
(EQPT1 can be used instead for lower values of
\yketspacedim ).
Therefore, for better readability, the results of
EQPT2 and EQPT3 for
\ytextartitionehundredninetyfourvfivemodifstepone{%
$
\Yqubitnb
=
2
$
}
are skipped
in
Fig.
\ref{fig-NRMSE}.
It should also be noted that the complexity of the configurations that
could be tested was limited by memory requirements, as stated above,
\ytextartitionehundredninetyfourvfivemodifstepone{hence}
not by the computational load of the proposed algorithms,
that remained reasonable up to 
$
\Yqubitnb
=
13
$.
For instance,
for each of the 100 process matrix estimations,
the average execution time
used
for
a) creating data
(thus only ``modeling''
QST instead of running a QST algorithm: see above),
b)
running the QPT algorithm and 
c) displaying results,
was as follows,
respectively for EQPT1, EQPT2 and EQPT3:
1) 
\ytextartitionehundredninetyfourvfivemodifstepone{$
\simeq
1
$~ms,}
2.6~ms
and
2.8~ms
for
$
\Yqubitnb
=
4
$
qubits,
2) 
22~ms,
170~ms
and
200~ms
for
$
\Yqubitnb
=
8
$
qubits
and
3) 
23~s,
180~s
and
230~s
for
$
\Yqubitnb
=
12
$
qubits.
\ytextartitionehundredninetyfourvfivemodifstepone{The execution times
for the other numbers of qubits are provided in
Table
\ref{tab-execution-time}.}

\begin{table*}[t]
\ytextartitionehundredninetyfourvfivemodifstepone{%
\begin{center}
\begin{tabular}{|l||r|r|r|r|r|r|r|r|r|r|r|r|r|}
\hline
Method
&
\multicolumn{13}{c|}{number of qubits}
\\
\cline{2-14}
&
1
&
2
&
3
&
4
&
5
&
6
&
7
&
8
&
9
&
10
&
11
&
12
&
13
\\
\hline
\hline
EQPT1
&
$
\simeq 1$ ms
&
$
\simeq 1$ ms
&
$
\simeq 1$ ms
&
$
\simeq 1$ ms
&
2.2 ms
&
3.9 ms
&
7.2 ms
&
22~ms
&
68 ms
&
400 ms
&
3.1 s
&
23~s
&
170 s
\\
\hline
EQPT2
&
&
&
&
2.6~ms
&
&
15 ms
&
&
170~ms
&
&
4.3 s
&
&
180~s
&
\\
\hline
EQPT3
&
&
&
&
2.8~ms
&
&
18 ms
&
&
200~ms
&
&
5.0 s
&
&
230~s
&
\\
\hline
EQPT5
&
&
&
2.3 ms
&
5.4 ms
&
27 ms
&
100 ms
&
580 ms
&
4.4 s
&
38 s
&
\ytextartitionehundredninetyfourvonemodifstepfive{%
490~s
}
&
7600~s
&
&
\\
&
&
&
&
&
&
&
&
&
&
$
\simeq
$
8 mn
&
$
\simeq
$
2 h 07 mn
&
&
\\
\hline
\end{tabular}
\end{center}
\caption{Execution times (averaged over tests with all values of
\yketestimflucwidth )
of the EQPT1, EQPT2, EQPT3
and EQPT5 methods vs. 
number of qubits.}
\label{tab-execution-time}
}
\end{table*}

The following conlusions can be drawn from
Fig.
\ref{fig-NRMSE},
when comparing the proposed methods
for any given QST quality, defined by the considered value of
\yketestimflucwidth.
The two-stage methods EQPT2 and EQPT3 
yield almost the same performance, with a slight advantage for
EQPT3
(as expected from Appendix
\ref{appendix-sec-two-stage-variants})%
,
that is more significant for lower values of the space dimension
\yketspacedim .
Moreover, these two-stage methods
always yield better and often
much better performance
than the single-stage method EQPT1,
as expected from the beginning of Section
\ref{sec-two-stage}.
Up to a certain
space dimension
\yketspacedim ,
the gap between these two types of methods
increases with the space dimension
\ytextartitionehundredninetyfourvfivemodifstepone{and
hence with the
associated number \yqubitnb\ of qubits:}
whereas all methods yield linear variations of the 
logarithm of NRMSE with respect to 
\ytextartitionehundredninetyfourvfivemodifstepone{%
\yqubitnb\
(i.e.
logarithm of
\yketspacedim),}
the slope of these variations is significantly lower for
EQPT2 and EQPT3, and these methods already start with a 
lower NRMSE than EQPT1 for the lowest values of
\yketspacedim .
Depending on the considered conditions,
the EQPT2 and EQPT3 methods decrease the NRMSE by
a factor up to 6, 
and hence the NMSE by a factor up to 36,
as compared with EQPT1.
Moreover, for the lowest QST error magnitude
considered
in these tests, for all analyzed 
state
space
dimensions,
EQPT2 and EQPT3
yield an NRMSE limited to about
$
10^{-1}
$
and
hence an NMSE 
limited to about
$
10^{-2}
$.

\ytextartitionehundredninetyfourvonemodifstepthree{%

\Figure[t!](topskip=0pt, botskip=0pt, midskip=0pt)[width=\linewidth]{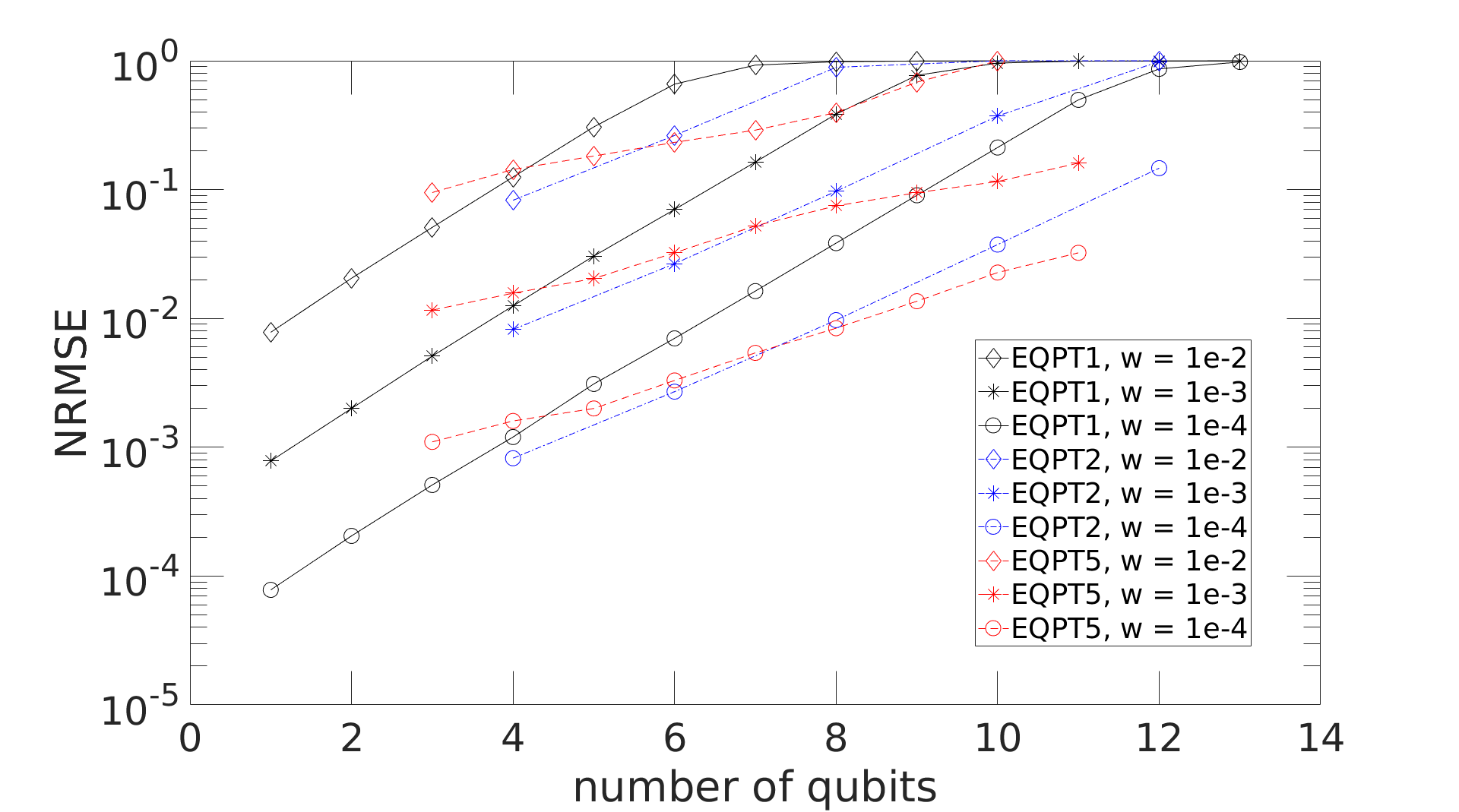}
{%
Normalized 
Root Mean Square Error (NRMSE) of estimation of process 
matrix \yprocessmat\ of proposed Quantum Process Tomography (QPT)
methods EQPT1
\ytextartitionehundredninetyfourvfivemodifstepone{(black
solid lines)}%
, EQPT2
\ytextartitionehundredninetyfourvfivemodifstepone{(blue
dash doted lines)}
and EQPT5
\ytextartitionehundredninetyfourvfivemodifstepone{(red
dashed lines)}%
,
vs. 
\ytextartitionehundredninetyfourvfivemodifstepone{(number
\yqubitnb\
of qubits)}
for several values of the
parameter
\yketestimflucwidth\
that defines the magnitude of the estimation errors of 
Quantum State Tomography (QST).%
\label{fig-NRMSE-eqpt125}}

We now move to the multi-stage method
EQPT5, whose performance is shown in
Fig.
\ref{fig-NRMSE-eqpt125}
(red dashed lines),
together with the results already provided above for
EQPT1
(black solid lines)
and
EQPT2
(blue dash doted lines).
The results for EQPT5 are reported for 
values of the 
state 
space
dimension
\yketspacedim\
that take the form
$
\Yketspacedim
=
2^{\Yqubitnb}
$,
where
\yqubitnb\
is the equivalent number of qubits,
varied as follows.
For a large \yqubitnb,
EQPT5 yields a 
significantly higher computational complexity than
the single-stage and two-stage methods, due to
the higher number of eigendecompositions it performs
and to the associated program structure
(recursive function with a binary tree structure,
as explained in Appendix
\ref{sec-appendix-relevance-multi-stage}).
Unlike for the previous proposed methods,
the applicability 
of EQPT5 to a large number of qubits
was therefore more limited
by its computational load than by its
memory requirements,
when using the considered basic hardware 
and software means.
Therefore, to keep a reasonable simulation duration
(for 
{\it
all} estimations of the process
matrix),
EQPT5 was only tested up to
$
\Yqubitnb
=
10
$
qubits
\ytextartitionehundredninetyfourvonemodifstepfive{when performing
100 above-defined elementary tests, and also for
$
\Yqubitnb
=
11
$
qubits
but then with only
10 
elementary tests.}
Moreover, as with
EQPT2,
we hereafter skip 
the results for a low dimension, that is for
$
\Yketspacedim
\leq
4
$,
which is not the case of interest for a multi-stage
method.
In the considered range of space dimensions,
the above-defined
average execution time per process
matrix estimation was e.g. as follows for EQPT5:
\ytextartitionehundredninetyfourvfivemodifstepone{5.4~ms}
for
$
\Yqubitnb
=
4
$
qubits,
\ytextartitionehundredninetyfourvfivemodifstepone{4.4~s}
for
$
\Yqubitnb
=
8
$
\ytextartitionehundredninetyfourvonemodifstepfive{qubits,}
\ytextartitionehundredninetyfourvonemodifstepfive{%
490~s
$
\simeq
$
8 mn}
for
$
\Yqubitnb
=
10
$
qubits
\ytextartitionehundredninetyfourvonemodifstepfive{and
7600~s
$
\simeq
$
2 h 07 mn
for
$
\Yqubitnb
=
11
$
qubits.}
\ytextartitionehundredninetyfourvfivemodifstepone{The execution times
for the other numbers of qubits are provided in
Table
\ref{tab-execution-time}.}

Fig.
\ref{fig-NRMSE-eqpt125}
yields the following conclusions
for EQPT5.
For the lowest considered values of the 
space dimension, EQPT5 starts with a somewhat
higher NRMSE than EQPT1 and EQPT2.
But, an attractive feature of EQPT5 is that its
NRMSE then increases more slowly that those of
the other methods,
with respect to the space
dimension%
\ytextartitionehundredninetyfourvfivemodifstepone{. This}
again yields 
{\it
linear} variations
\ytextartitionehundredninetyfourvfivemodifstepone{
of the logarithm of
NRMSE
with
respect to the number of qubits.}
EQPT5 thus outperforms the other two methods for higher
space dimensions, and the gap between
them increases with that dimension, up to a
certain dimension.
EQPT5 is thus attractive for high space dimensions,
so that it
actually
reaches the goal for which it was designed.
Depending on the considered conditions,
EQPT5 decreases the NRMSE by
a factor up to 3.2, 
and hence the NMSE by a factor up to more than 
10,
as compared with EQPT2.
Moreover, for the lowest QST error magnitude
considered
in these tests, for all analyzed 
state
space
dimensions
(which are 
\ytextartitionehundredninetyfourvonemodifstepfive{slightly}
lower than those
considered above for
EQPT1 and EQPT2, however),
EQPT5
yield an NRMSE limited to around
\ytextartitionehundredninetyfourvonemodifstepfive{$
3.2
\times
10^{-2}
$}
and
hence an NMSE 
limited to around
\ytextartitionehundredninetyfourvonemodifstepfive{$
10^{-3}
$.}%
}
\section{%
Conclusion
and future works}
\label{sec-concl}
Quantum Process Tomography
(QPT) is a research topic of major interest, due to its
importance for experimentally characterizing the actual behavior of
quantum gates.
In the present paper, we addressed QPT configurations that have
two features:
first, we consider unitary processes,
and second, we aim at developing QPT methods that are applicable to
a high state space dimension, corresponding to
a significant number of 
\ytextartitionehundredninetyfourvonemodifsteptwo{qubits,
as explained in Section
\ref{sec-intro}.}

Since the
considered process is unitary, its
input-output relationship takes a specific
form, that suggests that eigenanalysis may
be used to estimate at least part of the
parameters of the process matrix.
We first built upon that idea, and we then
extended it so as to also estimate the
other process parameters that 
a plain eigenanalysis 
cannot
estimate.
This resulted in a first class of QPT
methods (including EQPT1), that are stated to
be single-stage since they only require
one eigendecomposition.

The above approach may however be expected
to yield limited performance when considering
high-dimensional state spaces, due to the
estimation errors 
of
the 
Quantum State Tomography (QST) that it uses
as a building block.
We 
\ytextartitionehundredninetyfourvonemodifstepthree{very
significantly reduced}
this problem by introducing
multi-stage eigenanalysis-based extensions
of our approach, 
\ytextartitionehundredninetyfourvonemodifstepthree{where}
the term ``multi-stage''
refers to the fact these methods successively
perform several eigendecompositions.
We explained in detail how two-stage methods
(especially EQPT2 and EQPT3) operate, and 
\ytextartitionehundredninetyfourvfivemodifstepone{then}
how to extend them to an arbitrary
number of stages in order to minimize their
sensitivity to QST errors
\ytextartitionehundredninetyfourvonemodifstepthree{%
(especially with the dichotomic EQPT5 method).}

The relevance of this approach was validated by means
of numerical tests performed with a simulated
version of this QPT problem.
This especially showed that, 
depending on the considered conditions,
the tested two-stage methods (EQPT2 and EQPT3) decrease the 
NMSE by
a factor up to 36, 
as compared with 
the tested single-stage method (EQPT1).
Moreover, for the lowest QST error magnitude
considered
in these tests, for all analyzed 
state
space
dimensions,
the two-stage methods
yield 
an NMSE 
limited to about
$
10^{-2}
$.
\ytextartitionehundredninetyfourvonemodifstepthree{%
Besides, the
multi-stage dichotomic EQPT5 method 
may
yield a
higher accuracy
(NMSE decreased by a factor up to 10 as compared
with the two-stage EQPT2 method and
NMSE 
limited to about
\ytextartitionehundredninetyfourvonemodifstepfive{$
10^{-3}
$
in almost the same conditions as above)}
but at the expense of a higher computational
load.}

\ytextartitionehundredninetyfourvonemodifstepthree{We}
here 
had to restrict
ourselves to tests performed on a
standard PC.
\ytextartitionehundredninetyfourvonemodifstepthree{Our future
works will therefore especially include additional tests,
performed}
on a much more powerful platform, to derive the
performance of our methods for a significantly higher number of
qubits.
Anyway,
the configurations reported in the present paper
already involve a quite significant number of qubits
\ytextartitionehundredninetyfourvonemodifstepthree{(up
to 
\ytextartitionehundredninetyfourvonemodifstepfive{11 to}
13, depending on the considered method)}
and hence a 
high state space dimension
\ytextartitionehundredninetyfourvonemodifstepone{\yketspacedim}
(up to a few thousands)
as compared 
with
various investigations from the literature%
,
in particular with only
$
\Yketspacedim
=
5
$
in
\cite{amq-baldwin-physreva-2014}
that also deals with unitary processes.
Our tests also 
showed that the applicability of our
\ytextartitionehundredninetyfourvonemodifstepthree{%
single-stage and two-stage}
methods 
was 
not
limited by their complexity
in terms of computational cost, unlike various alternative QPT
methods, but only by the memory size available 
on the platform that was used for running them.
\ytextartitionehundredninetyfourvonemodifstepthree{Our
dichotomic multi-stage methods have the opposite
features.}
\appendices
\section{%
\ytextartitionehundredninetyfourvfivemodifstepone{Connections between
our QPT methods and classical signal processing}}
\label{sec-appendix-connections-clasical-processing}
\ytextartitionehundredninetyfourvfivemodifstepone{As stated in Section
\ref{sec-intro},
the type of QPT approaches proposed in this paper
has 
connections with
works that were previously reported
in the classical domain.
These connections may be defined as follows.}
QPT may be considered as the quantum version of so-called
(classical)
system identification
\cite{booknielsen}.
The latter 
classical
problem
consists of estimating
the parameter values of
a transform from the input to the output of
a system.
That system may have multiple inputs,
so that it then performs a 
\ytextartitionehundredninetyfourvonemodifsteptwo{combination
of them.
In the classical framework, such a combination is
called a mixture.}
The 
\ytextartitionehundredninetyfourvonemodifsteptwo{classical
problem then considered in the literature}
is therefore also referred to as
mixture identification
and it
especially includes blind mixture identification, or BMI
\cite{ldelathauwer-2007-mixture-identif,amoi6-48,amoi6-144,paper-amuse}.
\ytextartitionehundredninetyfourvonemodifsteptwo{This
terminology should be used with care when 
also considering the
quantum domain, because the above classical ``mixtures''
have nothing to do with the ``statistical mixtures'',
i.e. ``mixed states'', faced in the quantum domain
and used further in this paper.
Only considering previous works in the 
{\it
classical}
domain at this stage of our discussion, the above-mentioned BMI}
has a close relationship with source separation,
especially including blind source separation,
or BSS
\ytextartitionehundredninetyfourvfourmodifstepone{%
\cite{icabook-cichocki,
book-comon-jutten-ap,
amoi6-48,
amoi6-144,
icabook-oja,
book-makino-bss-speech,
book-multi-way-smiled-bro-geladi}.}
BSS
does not focus on
identifying the
transform performed by the above, i.e.
direct, system itself, but 
mainly
on deriving an
estimate of the inverse of that transform,
so as to apply it to the outputs of
the above direct system and to thus restore
the unknown source signals that form the input of the direct system.
This relationship 
between BMI and BSS
even appears in the name of the well-known SOBI method
\cite{a320belouch},
which stands for ``Second-Order Blind Identification'', 
whereas it is claimed to be a BSS method.
\ytextartitionehundredninetyfourvonemodifsteptwo{The
term ``Second-Order'' here refers to the fact
that this method applies to random signals and
uses the statistical properties of the measured
signals, more precisely their usual 
second-order statistical parameters that consist of
their covariance matrices involving different
time delays between signals.}

\ytextartitionehundredninetyfourvfivemodifstepone{So
far, we especially discussed how
SOBI-like methods and the 
methods proposed in the present paper 
are connected in terms
of the nature of data processing problems
that they address.
These two
types of methods 
also have relationships in terms of the
approaches that they use to solve the
considered problems. }
More precisely,
in
classical BSS/BMI,
when the transform performed by the 
direct 
system
is linear (and memoryless) and its identification
is performed by using the 
\ytextartitionehundredninetyfourvonemodifsteptwo{above-mentioned}
second-order statistics
of the output signals of the direct system, 
an equation
similar to 
(\ref{eq-processoutmixedonedensitymatrixvsinmixedonedensitymatrix})
is obtained
\cite{thesefety,a320belouch,paper-amuse,amoi5-50,paper-molgedey-schuster}.
\ytextartitionehundredninetyfourvonemodifsteptwo{That}
classical counterpart of the structure of
(\ref{eq-processoutmixedonedensitymatrixvsinmixedonedensitymatrix})
was already exploited 
to perform classical BSS/BMI
by means of several
methods based on eigendecomposition.
The real-valued version of the basic structure of
(\ref{eq-processoutmixedonedensitymatrixvsinmixedonedensitymatrix})
was 
first 
used in
\cite{thesefety,paper-amuse}
together with a 
required 
modified form of this equation
(a related approach also appears in
\cite{paper-molgedey-schuster}).
It was then
extended in
\cite{a320belouch},
by combining it with
more than one modified form of this equation,
to obtain a more robust method.
Another 
version
\cite{amoi5-50}
aims at 
extending that type of approach from the standard case of stationary
time-domain
source signals
to nonstationary time-domain sources.
That version 
eventually
uses complex-valued signals
(whereas the mixing transform remains real-valued)
because one moves
to the Fourier domain.
\section{%
\ytextartitionehundredninetyfourvfivemodifstepone{Creating a mixed state
that has a diagonal density matrix}}
\label{sec-appendix-diag-density-matrix}
\ytextartitionehundredninetyfourvfivemodifstepone{The algorithm proposed
in Section
\ref{sec-single-stage-part-one}
uses an input density matrix
\yprocessinmixedonedensitymatrix\
that is diagonal.}
It should be noted that,
when one considers that one is able to create
a pure state,
then,
creating the type of state considered here
is not a problem, as will now be shown.
We here consider the 
approach initially
used by
von Neumann
in
\cite{bookvonneumannfoundationsqm} to
define a statistical mixture:
such a
state is expressed 
as being composed
of a set of
pure states, each of them being associated
with a probability of occurrence.
Then, to 
\ytextartitionehundredninetyfourvonemodifsteptwo{create experimental
data corresponding to}
the state defined
by the diagonal density matrix 
\yprocessinmixedonedensitymatrix,
one just has to repeatedly, randomly, select
one of the pure states that form the considered basis
of the state space, moreover selecting
the 
$
k
$th
basis state
with a probability that is equal to the
$
k
$%
th
value on the diagonal of
\yprocessinmixedonedensitymatrix.
This simplicity of implementation is worth
noting because, in constrast, the authors of
\cite{amq153,amq-baldwin-physreva-2014}
consider that using a mixed input state is too
complex
(maybe because they have non-diagonal density matrices in mind),
so that they choose to restrict themselves
to pure states, 
which has the 
drawback of
requesting 
many such 
\ytextartitionehundredninetyfourvonemodifsteptwo{multiqubit input states,}
as opposed to 
only two 
\ytextartitionehundredninetyfourvonemodifsteptwo{multiqubit input}
states in the complete
approach that we here start to describe
\ytextartitionehundredninetyfourvonemodifsteptwo{(namely the mixed
state with a density matrix
\yprocessinmixedonedensitymatrix\
considered here,
and then the pure state
\yprocessinpuretwoket\
in Section
\ref{sec-single-stage-part-two}).}
Beyond their relationship based on considering
(\ref{eq-processoutmixedonedensitymatrixvsinmixedonedensitymatrix}),
our approach and the one in 
\cite{amq153}
are therefore quite different.
\section{%
\ytextartitionehundredninetyfourvfivemodifstepone{Preprocessing steps
of the single-stage QPT algorithm}}
\label{sec-appendix-single-stage-algo}
\ytextartitionehundredninetyfourvfivemodifstepone{The 
first part of our single-stage algorithm, described
in Section
\ref{sec-single-stage-part-one},
is based on an estimated density matrix
\yprocessoutmixedonedensitymatrixestim\
provided by a QST algorithm. If}
this estimate 
\yprocessoutmixedonedensitymatrixestim\
is not guaranteed to be Hermitian
due to 
\ytextartitionehundredninetyfourvthreemodifstepone{the
estimation errors of the considered ``external''
(i.e. 
external to 
{\it
our} QPT)
QST algorithm,}
a first recommended preprocessing step of our
method consists of 
deriving a Hermitian
estimate
\yprocessoutmixedonedensitymatrixestimherm\
from
\yprocessoutmixedonedensitymatrixestim .
This may be achieved by computing the
Hermitian part of
\yprocessoutmixedonedensitymatrixestim ,
that is
(see
\cite{book-horn-johnson} p. 109
or
\cite{icabook-oja} p. 343
for 
\ytextartitionehundredninetyfourvfourmodifstepone{its}
real-valued counterpart):
\begin{equation}
\Yprocessoutmixedonedensitymatrixestimherm
=
\frac{1}{2}
\left(
\Yprocessoutmixedonedensitymatrixestim
+
\Yprocessoutmixedonedensitymatrixestim
\Ytransconjug
\right).
\label{eq-def-processoutmixedonedensitymatrixestimherm}
\end{equation}
Similarly,
the actual matrix
\yprocessoutmixedonedensitymatrix\
is known to have a trace equal to 1, since it is a density matrix
(see e.g. 
\cite{booknielsen} p. 101,
\cite{livreperes1995} p. 73,
\cite{livremessiahtomeun}
p. 283).
Therefore,
if its estimate
(\yprocessoutmixedonedensitymatrixestim\
or)
\yprocessoutmixedonedensitymatrixestimherm\
obtained 
so far
is not guaranteed to have a unit trace
due to QST estimation errors,
a second recommended preprocessing step of our
method consists of
deriving a
(still Hermitian and moreover)
unit-trace
estimate
\yprocessoutmixedonedensitymatrixestimhermunittrace\
from
(\yprocessoutmixedonedensitymatrixestim\
or)
\yprocessoutmixedonedensitymatrixestimherm,
by dividing all its elements by its trace:
\begin{equation}
\Yprocessoutmixedonedensitymatrixestimhermunittrace
=
\frac{
\Yprocessoutmixedonedensitymatrixestimherm
}
{
\mathrm{Trace}
(
\Yprocessoutmixedonedensitymatrixestimherm
)
}
.
\label{eq-def-processoutmixedonedensitymatrixestimhermunittrace}
\end{equation}

\ytextartitionehundredninetyfourvfivemodifstepone{Similarly,
the 
second part of our single-stage algorithm, described
in Section
\ref{sec-single-stage-part-two},
uses
an estimated 
ket 
\yprocessoutpuretwoketestim,
provided by a QST algorithm.}
If the considered QST algorithm does not guarantee that the norm of
\yprocessoutpuretwoketestim\
is equal to 1,
a recommended 
pre-processing step
\ytextartitionehundredninetyfourvfivemodifstepone{of our
QPT method}
consists of
dividing
\yprocessoutpuretwoketestim\
by its norm.
\section{%
\ytextartitionehundredninetyfourvfivemodifstepone{Preferred input density
matrix for 
the first part of the single-stage QPT algorithm}}
\label{sec-appendix-input-density-matrix}
\ytextartitionehundredninetyfourvonemodifsteptwo{In the first step of
\ytextartitionehundredninetyfourvfivemodifstepone{our 
first
method defined in 
Section
\ref{sec-single-stage-part-one},}
we
preferably
use the value of
\yprocessinmixedonedensitymatrix\
defined as follows.
The}
values on its diagonal are the eigenvalues
of the considered density operator 
\yprocessoutmixedonedensitymatrix ,
they are real and non-negative
and their sum is
equal to one
(see e.g.
\cite{booknielsen}
p. 101).
As shown 
\ytextartitionehundredninetyfourvfivemodifstepone{in
Section
\ref{sec-single-stage-part-one},}
the proposed method 
assumes that 
these eigenvalues
are different and
ordered, and it
is based on
reordering the eigenvalues in the eigendecomposition
of the estimated output
density matrix and on permuting 
the
eigenvectors
accordingly.
Therefore, if
using quite close 
actual
eigenvalues
when choosing
\yprocessinmixedonedensitymatrix ,
one may fear that
the corresponding 
estimated
eigenvalues obtained for the output
density matrix be permuted due to
estimation errors, so that the corresponding
eigenvectors may then be permuted accordingly.
This is a major problem, because the
intermediate
estimate
\yprocessmatestimtwo\
related to
the considered process, obtained at the
output of this first step of our QPT method,
is primarily based on
these eigenvectors.
In this paper, we handle this problem as 
follows: 
first, for the method imposed in
\ytextartitionehundredninetyfourvfivemodifstepone{Section
\ref{sec-single-stage-part-one},}
we select the value of 
\yprocessinmixedonedensitymatrix\
in order to limit that problem 
to the greatest extent that is possible
within the frame of that
method;
then, in Sections
\ref{sec-two-stage}
and
\ref{sec-multi-stage},
we propose more advanced methods that allow
one to further reduce that problem.
So, for the method considered here, 
we choose eigenvalues that are uniformly
distributed
\ytextartitionehundredninetyfourvonemodifsteptwo{(i.e. multiples of the
same step, with zero excluded)}%
, 
so as to maximize the gap between
any two adjacent values.
Moreover constraining them to be 
positive, with a sum equal to one and in
decreasing order as explained above,
it is easily shown that one obtains the
following values for the diagonal elements 
$
\Yopdensity
_{
1,
\Yprocessmatcolindexstd
\Yprocessmatcolindexstd
}
$
of
\yprocessinmixedonedensitymatrix :
\begin{equation}
\Yopdensity
_{
1,
\Yprocessmatcolindexstd
\Yprocessmatcolindexstd
}
=
\frac{
2
(
\Yketspacedim
-
\Yprocessmatcolindexstd
+
1
)
}
{
\Yketspacedim
(
\Yketspacedim
+
1
)
}
\hspace{5mm}
\Yprocessmatcolindexstd
\in
\{
1
,
\dots
,
\Yketspacedim
\}
.
\label{eq-opdensity-diag-val-uniform}
\end{equation}

\section{%
\ytextartitionehundredninetyfourvfivemodifstepone{Preferred input ket for 
the second part of the single-stage QPT algorithm}}
\label{sec-appendix-single-stage-algo-input-ket}
\ytextartitionehundredninetyfourvfivemodifstepone{In the second step of
first
method defined in 
Section
\ref{sec-single-stage-part-two},
we select the input ket 
\yprocessinpuretwoket\
as follows.
Based on the condition on nonzero components
provided 
in 
Section
\ref{sec-single-stage-part-two}}
for
\yprocessinpuretwoket, we 
preferably
select a state
\yprocessinpuretwoket\
with all component moduli as high as possible.
Considering positive real-valued components and
taking into account that
\yprocessinpuretwoket\
has unit norm, this yields a single
solution:
all components of
\yprocessinpuretwoket\
in the considered basis
are equal to
$
1
/
\sqrt{
\Yketspacedim
}
$.
\section{Mathematical
\ytextartitionehundredninetyfourvonemodifstepfour{definitions and}
properties}
\label{sec-math-appendix}
\ytextartitionehundredninetyfourvonemodifstepfour{
\begin{definition}
The element-wise division
$
\oslash
$
for vectors is defined as follows.
Considering two vectors
$
u
$
and
$
v
$,
respectively with
$
k
$%
th
elements
$
u_k
$
and
$
v_k
$,
the quantity
$
u
\oslash
v
$
is the vector whose
$
k
$%
th
element
is equal to
$
u_k
/
v_k
$.
\end{definition}}
\begin{theorem}
\label{theorem-normalmatrix-eigendecomposition}
Every normal matrix 
$
A
\in
M_n
$
can be written as
\begin{equation}
\label{eq-theorem-normalmatrix-eigendecomposition}
A
=
U
\Lambda
U
\Ytransconjug
,
\end{equation}
where
$
U
\in
M_n
$
is unitary and
$
\Lambda
\in
M_n
$
is a diagonal matrix whose main diagonal entries are the eigenvalues of
$
A
$
(%
$
M_n
$
is the set of 
$
n
$-by-%
$
n
$
complex matrices%
).
\end{theorem}
\begin{proof}
See \cite{book-horn-johnson} p. 157 (e).
See also \cite{book-horn-johnson} 
p. 101
for 
unitarily diagonalizable matrices
(every normal matrix is unitarily diagonalizable)
and the spectral theorem for normal matrices.
In particular, this implies the
spectral theorem for Hermitian matrices:
see \cite{book-horn-johnson} p. 104.%
\qed
\end{proof}
\begin{theorem}
\label{theorem-eigendecomposition-eigenvectors-eigenvalues}
If
$
A
\in
M_n
$
reads
\begin{equation}
A
=
P
D
P^{-1}
\label{eq-eigendecomposition-eigenvectors-eigenvalues}
\end{equation}
where 
$
D
\in
M_n
$
is a diagonal matrix and
$
P
\in
M_n
$
is invertible
then,
for any index
$
k
\in
\{
1,
\dots
,
n
\}
$,
the
$k$th
column of
$
P
$
is an eigenvector of
$
A
$
and the associated eigenvalue
is
the
$k$th
value on the diagonal of
$
D
$.
\end{theorem}
\begin{proof}
Right-multiplying
(\ref{eq-eigendecomposition-eigenvectors-eigenvalues})
by
$
P
$
yields
\begin{equation}
A
P
=
P
D
.
\label{eq-eigendecomposition-eigenvectors-eigenvalues-version-two}
\end{equation}
For any index
$
k
\in
\{
1,
\dots
,
n
\}
$,
the
$k$th column of the matrix equation
(\ref{eq-eigendecomposition-eigenvectors-eigenvalues-version-two})
reads
\begin{equation}
A
v_k
=
\lambda_k
v_k
\label{eq-eigendecomposition-eigenvectors-eigenvalues-version-three}
\end{equation}
where
$
v_k
$
is the 
$k$th column of
$
P
$
and
$
\lambda_k
$
is the
$k$th
element on the diagonal of 
$
D
$.
Eq.
(\ref{eq-eigendecomposition-eigenvectors-eigenvalues-version-three})
shows that
$
v_k
$
is an eigenvector of
$
A
$
and that the associated eigenvalue is
$
\lambda_k
$.\qed
\end{proof}
\begin{theorem}
\label{theorem-normal-matrix-eigenvectors-orthogonal}
If
$
A
\in
M_n
$
is normal,
and if
$x$
and
$y$
are eigenvectors corresponding to distinct eigenvalues,
$x$
and
$y$
are orthogonal.
\end{theorem}
\begin{proof}
See \cite{book-horn-johnson} p. 103.\qed
\end{proof}
\begin{theorem}
\label{theorem-closest-unitary-matrix}
A best least-squares approximation of a given matrix
$
A
\in
M_n
$
by a unitary matrix
$
U
_A
\in
M_n
$
is given by
$
U
_A
=
V
W
\Ytransconjug
$,
where
$
A
=
V
\Sigma
W
\Ytransconjug
$
is
a
singular value decomposition 
\ytextartitionehundredninetyfourvfivemodifstepone{(SVD)}
of
$
A
$.
\end{theorem}
\begin{proof}
See \cite{book-horn-johnson} p. 431.\qed
\end{proof}
\section{A variant 
for the first part of
single-stage QPT methods}
\label{appendix-sec-single-stage-part-one}
We here again consider
the first part of
single-stage QPT methods.
As in
the variant that we proposed
in Section
\ref{sec-single-stage-part-one}
for that first part, we here use
a mixed input state, 
represented by the density matrix
\yprocessinmixedonedensitymatrix .
However, for the sake of generality,
this matrix is here not any more 
constrained to be diagonal.
It is still Hermitian and therefore 
unitarily
diagonalizable.
In the version considered here, 
as
in Section
\ref{sec-single-stage-part-one},
we 
request 
all its eigenvalues to be different.
That input density matrix
\yprocessinmixedonedensitymatrix\
is here requested to be known
(%
i.e. fixed a priori or estimated from copies of that state,
so that this variant
is supervised
or semi-supervised%
).

Using the same properties of eigendecomposition as in Section
\ref{sec-single-stage-part-one},
we consider all unitarily diagonalized forms of
\yprocessinmixedonedensitymatrix\
that have the following properties:
\begin{enumerate}
\item The eigenvalues are in a known order along the diagonal of
\yprocessinmixedonedensitymatrix .
Without loss of generality, we hereafter consider the case when they are
in decreasing order.
\item All eigenvectors have unit norm.
\end{enumerate}
All these eigendecompositions of
\yprocessinmixedonedensitymatrix\
may be 
expressed as
\begin{equation}
\Yprocessinmixedonedensitymatrix
=
\Yprocessinmixedonedensitymatrixeigenunit
\Ymathspace
\Yprocessinmixedonedensitymatrixeigendiag
\Ymathspace
\Yprocessinmixedonedensitymatrixeigenunit
\Ytransconjug
\label{eq-processinmixedonedensitymatrix-eigendecompos}
\end{equation}
where
\yprocessinmixedonedensitymatrixeigenunit\
and
\yprocessinmixedonedensitymatrixeigendiag\
are respectively a
unitary matrix
and a diagonal matrix
associated with that first input of the considered process.

Starting from a given
\yprocessinmixedonedensitymatrix ,
the first step of the algorithm proposed in the present section
consists of determining one (arbitrary) of these 
decompositions
(\ref{eq-processinmixedonedensitymatrix-eigendecompos})
meeting the above-defined conditions.
This is performed
by 1) running an eigendecomposition algorithm, then 2) reordering the
values in
\ytextartitionehundredninetyfourvonemodifstepfour{the matrix obtained
for}
\yprocessinmixedonedensitymatrixeigendiag\
if they are not directly provided in decreasing order by the considered
eigendecomposition algorithm, 3) reordering the columns of
\ytextartitionehundredninetyfourvonemodifstepfour{the matrix obtained
for}
\yprocessinmixedonedensitymatrixeigenunit\
accordingly and 4) dividing these columns by their norms if 
\ytextartitionehundredninetyfourvonemodifstepfour{the considered
practical eigendecomposition algorithm is not guaranteed
to yield unit-norm vectors.}
Thus, we especially get a known value
\yprocessinmixedonedensitymatrixeigenunit\
(which contains indeterminacies consisting of
a phase factor separately for each of its columns).
\yprocessinmixedonedensitymatrixeigenunit\
is used in the second part of this version of our algorithm:
see
Appendix
\ref{appendix-sec-single-stage-nondiag-inputone-part-two}.

Besides, using
(\ref{eq-processinmixedonedensitymatrix-eigendecompos}),
\ytextartitionehundredninetyfourvonemodifstepfour{Equation}
(\ref{eq-processoutmixedonedensitymatrixvsinmixedonedensitymatrix})
may be expressed as
\begin{equation}
\Yprocessoutmixedonedensitymatrix
=
\Yprocessmattimesyprocessinmixedonedensitymatrixeigenunit
\Ymathspace
\Yprocessinmixedonedensitymatrixeigendiag
\Ymathspace
\Yprocessmattimesyprocessinmixedonedensitymatrixeigenunit
\Ytransconjug
\label{eq-processoutmixedonedensitymatrix-vs-processinmixedonedensitymatrixeigendiag}
\end{equation}
where
the global unitary matrix
\begin{equation}
\Yprocessmattimesyprocessinmixedonedensitymatrixeigenunit
=
\Yprocessmat
\Yprocessinmixedonedensitymatrixeigenunit
\label{eq-def-processmattimesyprocessinmixedonedensitymatrixeigenunit}
\end{equation}
includes the effect of the eigendecomposition of
\yprocessinmixedonedensitymatrix,
in addition to the considered process
\yprocessmat .

Formally,
(\ref{eq-processoutmixedonedensitymatrix-vs-processinmixedonedensitymatrixeigendiag})
adresses exactly the same problem as in
Section
\ref{sec-single-stage-part-one}
(see
(\ref{eq-processoutmixedonedensitymatrixvsinmixedonedensitymatrix})
with a diagonal \yprocessinmixedonedensitymatrix ),
but now with a diagonal input density matrix
\yprocessinmixedonedensitymatrixeigendiag\
and a process
\yprocessmattimesyprocessinmixedonedensitymatrixeigenunit\
(that is also unitary).
We can therefore apply the algorithm and results of
Section
\ref{sec-single-stage-part-one}
to this reformulated problem,
i.e.,
briefly,
we diagonalize
\yprocessoutmixedonedensitymatrixestimhermunittrace\
and we then reorder the eigenvalues 
(in the same order as in
\yprocessinmixedonedensitymatrixeigendiag )
and associated
normalized eigenvectors.
This guarantees that the 
algorithm 
of
Section
\ref{sec-single-stage-part-one}
here
provides an estimate
\yprocessmatestimtwo\
that
restores
\yprocessmattimesyprocessinmixedonedensitymatrixeigenunit\
up to a phase factor for each of its columns,
with
(\ref{eq-processmatestimtwo-vs-processmatestimtwophaseindetermmatrixdiag})
here replaced by
\begin{eqnarray}
\Yprocessmatestimtwo
&
=
&
\Yprocessmattimesyprocessinmixedonedensitymatrixeigenunit
\Yprocessmatestimtwophaseindetermmatrixdiag
\label{eq-processmatestimtwo-vs-processmattimesyprocessinmixedonedensitymatrixeigenunit}
\\
&
=
&
\Yprocessmat
\Ymathspace
\Yprocessinmixedonedensitymatrixeigenunit
\Ymathspace
\Yprocessmatestimtwophaseindetermmatrixdiag
\label{eq-processmatestimtwo-vs-processmattimesyprocessinmixedonedensitymatrixeigenunit-explicit}
\end{eqnarray}
where
\yprocessmatestimtwophaseindetermmatrixdiag\
is again a
diagonal matrix that contains 
unknown phase
factors
$
e^
{
\Yprocessmatestimtwophaseindetermfirst
}
,
\dots
,
e^
{
\Yprocessmatestimtwophaseindetermlast
}
$
on its main diagonal.

As opposed to the basic version of the first step defined in
Section \ref{sec-single-stage-part-one},
the estimate
\yprocessmatestimtwo\
here
contains the effect of
\yprocessinmixedonedensitymatrixeigenunit\
in addition.
The second part of the method must then be modified
accordingly, as compared with its basic version defined
in Section
\ref{sec-single-stage-part-two}.
A solution to this problem is provided in Appendix
\ref{appendix-sec-single-stage-nondiag-inputone-part-two}.
\section{A variant 
for the second part of
single-stage QPT methods}
\label{appendix-sec-single-stage-part-two}
In the framework of 
single-stage QPT methods
that include a first part as defined in
Section
\ref{sec-single-stage-part-one},
we here again consider
the associated second part,
that aims at estimating
the phase factors 
$
e^
{
\Ysqrtminusone
\Yprocessmatestimtwophaseindetermfirst
}
,
\dots
,
e^
{
\Ysqrtminusone
\Yprocessmatestimtwophaseindetermlast
}
$,
by using a second input state of the considered process.
The method that we proposed
in Section
\ref{sec-single-stage-part-two}
for that second part uses
a pure input state
defined by a ket
\yprocessinpuretwoket .
However, a mixed input
state 
having a density matrix
\yprocessinmixedtwodensitymatrix\
may be used instead, as will now be shown.
As
\yprocessinpuretwoket\
in 
Section
\ref{sec-single-stage-part-two},
the input
density matrix
\yprocessinmixedtwodensitymatrix\
is here supposed to be initially known (supervised case)
or estimated by using part of the available copies of that
state
(semi-supervised case).

The input state
\yprocessinmixedtwodensitymatrix\
yields a state
\yprocessoutmixedtwodensitymatrix\
at the output of the considered process,
and these two states have the same relationship as in
(\ref{eq-processoutmixedonedensitymatrixvsinmixedonedensitymatrix}).
In practice, what is available, thanks to QST, is an estimate
\yprocessoutmixedtwodensitymatrixestimhermunittrace\
of
\yprocessoutmixedtwodensitymatrix,
here again after it has been
pre-processed, if required, so as to be Hermitian and to have unit
trace.
Thus, up to estimation errors,
\yprocessoutmixedtwodensitymatrixestimhermunittrace\
meets the condition
\begin{equation}
\Yprocessoutmixedtwodensitymatrixestimhermunittrace
=
\Yprocessmat
\Ymathspace
\Yprocessinmixedtwodensitymatrix
\Ymathspace
\Yprocessmat
\Ytransconjug
.
\label{eq-processoutmixedtwodensitymatrixestimhermunittrace-vs-processinmixedtwodensitymatrix}
\end{equation}

Then, knowing 
\yprocessoutmixedtwodensitymatrixestimhermunittrace\
and
\yprocessmatestimtwo\
obtained in the first part of our algorithm
(see Section
\ref{sec-single-stage-part-one}),
the next step of our algorithm here consists of combining
them 
to form
\begin{equation}
\Yprocessmatestimthreedensitymatrixinterm
=
\Yprocessmatestimtwo
\Ytransconjug
\Ymathspace
\Yprocessoutmixedtwodensitymatrixestimhermunittrace
\Ymathspace
\Yprocessmatestimtwo
.
\label{eq-def-processmatestimthreedensopinterm}
\end{equation}
This may be seen as the extension, to
a mixed state, of the quantity
(\ref{eq-def-processmatestimthreeketinterm})
in the variant of the present method that uses a pure 
input state instead:
for the latter type of state,
\yprocessoutmixedtwodensitymatrixestimhermunittrace\
reduces to
\begin{equation}
\Yprocessoutmixedtwodensitymatrixestimhermunittrace\
=
\Yprocessoutpuretwoketestim
\langle
\widehat{\Yketnotstdone}
_{2}
\lvert
\label{eq-projector}
\end{equation}
and
the state
\yprocessmatestimthreeketinterm\
considered in Section
\ref{sec-single-stage-part-one}
corresponds to
$
\Yprocessmatestimthreedensitymatrixinterm
=
\Yprocessmatestimthreeketinterm
\langle
\Yketnotstdone
_{3}
\lvert
$.
Combining the latter expression with
(\ref{eq-def-processmatestimthreeketinterm})
and using
(\ref{eq-projector}),
the resulting expression of
\yprocessmatestimthreedensitymatrixinterm\
is fully compatible with its extension
(\ref{eq-def-processmatestimthreedensopinterm})
to an arbitrary mixed input state
considered here.

Combining
(\ref{eq-processoutmixedtwodensitymatrixestimhermunittrace-vs-processinmixedtwodensitymatrix}),
(\ref{eq-def-processmatestimthreedensopinterm})
and
(\ref{eq-processmatestimtwo-vs-processmatestimtwophaseindetermmatrixdiag})%
, with
\yprocessmat\ unitary,
yields
\begin{equation}
\Yprocessmatestimthreedensitymatrixinterm
=
\Yprocessmatestimtwophaseindetermmatrixdiag
\Yconjug
\Ymathspace
\Yprocessinmixedtwodensitymatrix
\Ymathspace
\Yprocessmatestimtwophaseindetermmatrixdiag
.
\label{eq-processmatestimthreedensopinterm-express-two}
\end{equation}

This shows that a diagonal matrix
\yprocessinmixedtwodensitymatrix\
cannot be used in this approach, because all three matrices in the
right-hand term of
(\ref{eq-processmatestimthreedensopinterm-express-two})
are then diagonal, so that
$
\Yprocessmatestimtwophaseindetermmatrixdiag
\Yconjug
$
and
\yprocessmatestimtwophaseindetermmatrixdiag\
cancel out
and
(\ref{eq-processmatestimthreedensopinterm-express-two})
reduces to
\begin{equation}
\Yprocessmatestimthreedensitymatrixinterm
=
\Yprocessinmixedtwodensitymatrix
\end{equation}
and
\yprocessmatestimthreedensitymatrixinterm\
does not yield any information about the phase factors
in
\yprocessmatestimtwophaseindetermmatrixdiag .
This is not surprising because all the information that can be extracted
with a diagonal input density matrix was already extracted by using such a
matrix in the first part of our algorithm
(see Section
\ref{sec-single-stage-part-one}).
We therefore consider a non-diagonal
matrix
\yprocessinmixedtwodensitymatrix\
hereafter.

Whatever the input density matrix
\yprocessinmixedtwodensitymatrix,
denoting
$
\Yopdensity
_{
k
l
}
$
its elements,
(\ref{eq-processmatestimthreedensopinterm-express-two})
yields
(again up to estimation errors)
\begin{equation}
\Yprocessmatestimthreedensitymatrixinterm
=
\left[
\begin{tabular}{llll}
$
\Yopdensity
_{
1
1
}
$
&
$
\Yopdensity
_{
1
2
}
e^
{
\Ysqrtminusone
(
\Yprocessmatestimtwophaseindetermsecond
-
\Yprocessmatestimtwophaseindetermfirst
)
}
$
&
$
\dots
$
&
$
\Yopdensity
_{
1
\Yketspacedim
}
e^
{
\Ysqrtminusone
(
\Yprocessmatestimtwophaseindetermlast
-
\Yprocessmatestimtwophaseindetermfirst
)
}
$
\\
$
\dots
$
&
&
&
$
\dots
$
\end{tabular}
\right]
.
\label{eq-def-processmatestimthreedensitymatrixinterm}
\end{equation}
Therefore, we consider a value of
\yprocessinmixedtwodensitymatrix\
that is arbitrary except that all the elements of its first row are
nonzero 
(this method may be adapted to the case when
another row only has nonzero elements).
The next step of our algorithm consists of computing the quantity
\begin{equation}
\left[
\Yprocessmatestimthreedensitymatrixinterm
\right]
_
{1
\bullet
}
\oslash
\left[
\Yprocessinmixedtwodensitymatrix
\right]
_
{1
\bullet
}
\label{eq-processmatestimthreedensitymatrixinterm-row-one-rescaled}
\end{equation}
where
$
\left[
.
\right]
_
{1
\bullet
}
$
represents the row vector equal to
the first row of the considered matrix (i.e.
row 1 and all columns,
or ``columns $\bullet$'').
Thanks to
(\ref{eq-def-processmatestimthreedensitymatrixinterm}),
we have
\begin{equation}
\left[
\Yprocessmatestimthreedensitymatrixinterm
\right]
_
{1
\bullet
}
\oslash
\left[
\Yprocessinmixedtwodensitymatrix
\right]
_
{1
\bullet
}
=
\left[
1
,
e^
{
\Ysqrtminusone
(
\Yprocessmatestimtwophaseindetermsecond
-
\Yprocessmatestimtwophaseindetermfirst
)
}
,
\dots
,
e^
{
\Ysqrtminusone
(
\Yprocessmatestimtwophaseindetermlast
-
\Yprocessmatestimtwophaseindetermfirst
)
}
\right]
\end{equation}
and hence
\begin{equation}
\mathrm{diag}
(
(
\left[
\Yprocessmatestimthreedensitymatrixinterm
\right]
_
{1
\bullet
}
\oslash
\left[
\Yprocessinmixedtwodensitymatrix
\right]
_
{1
\bullet
}
)
\Yconjug
)
=
e^
{
\Ysqrtminusone
\Yprocessmatestimtwophaseindetermfirst
}
\Yprocessmatestimtwophaseindetermmatrixdiag
\Yconjug
.
\label{eq-algo-single-stage-part-two-version-opdens-diag}
\end{equation}

The next step of our algorithm therefore consists of computing
\begin{eqnarray}
\Yprocessmatestimthree
&
=
&
\Yprocessmatestimtwo
\Ymathspace
\mathrm{diag}
(
(
\left[
\Yprocessmatestimthreedensitymatrixinterm
\right]
_
{1
\bullet
}
\oslash
\left[
\Yprocessinmixedtwodensitymatrix
\right]
_
{1
\bullet
}
)
\Yconjug
)
\label{eq-processmatestimthree-implicit}
\\
&
=
&
\Yprocessmatestimtwo
\Ymathspace
\mathrm{diag}
(
(
\left[
\Yprocessmatestimtwo
\Ytransconjug
\Ymathspace
\Yprocessoutmixedtwodensitymatrixestimhermunittrace
\Ymathspace
\Yprocessmatestimtwo
\right]
_
{1
\bullet
}
\oslash
\left[
\Yprocessinmixedtwodensitymatrix
\right]
_
{1
\bullet
}
)
\Yconjug
)
\label{eq-processmatestimthree-explicit}
\end{eqnarray}
where
(\ref{eq-processmatestimthree-explicit})
results from
(\ref{eq-def-processmatestimthreedensopinterm})
and defines the complete processing performed in this second part of
our algorithm, starting from the estimate
\yprocessmatestimtwo\
of
\yprocessmat\
obtained in the first part of the algorithm and
combining it with
the density matrices
\yprocessinmixedtwodensitymatrix\
and
\yprocessoutmixedtwodensitymatrixestimhermunittrace\
used in that second part.

Combining
(\ref{eq-processmatestimthree-implicit}),
(\ref{eq-algo-single-stage-part-two-version-opdens-diag})
and
(\ref{eq-processmatestimtwo-vs-processmatestimtwophaseindetermmatrixdiag})
yields
\begin{equation}
\Yprocessmatestimthree
=
e^
{
\Ysqrtminusone
\Yprocessmatestimtwophaseindetermfirst
}
\Yprocessmat
\end{equation}
i.e.
\yprocessmatestimthree\
succeeds in restoring
\yprocessmat\
(once again,
up to a global phase factor
that has no physical consequences
and up to the estimation errors that were mentioned above but implicit
in the above equations).
\section{Another variant 
for the second part of
single-stage QPT methods}
\label{appendix-sec-single-stage-nondiag-inputone-part-two}
In the framework of 
single-stage QPT methods,
we here again consider the case when the
first part is defined as in
Appendix
\ref{appendix-sec-single-stage-part-one}.
We present an associated solution for the
second part of the method,
using a pure input state
defined by a ket
\yprocessinpuretwoket .
\yprocessinpuretwoket\
is requested to be known,
i.e. fixed a priori
(supervised case)
or estimated from a subset of its copies
(semi-supervised case).

Starting with
the
same approach as in Section
\ref{sec-single-stage-part-two},
this here again yields the output ket
estimate
(\ref{eq-processoutpuretwoketestim-vs-processinpuretwoket}).
Our algorithm then computes
\yprocessmatestimthreeketinterm\
defined by
(\ref{eq-def-processmatestimthreeketinterm}),
as in
Section
\ref{sec-single-stage-part-two},
but now with
\yprocessmatestimtwo\
that meets
(\ref{eq-processmatestimtwo-vs-processmattimesyprocessinmixedonedensitymatrixeigenunit-explicit}).
Combining
(\ref{eq-processmatestimtwo-vs-processmattimesyprocessinmixedonedensitymatrixeigenunit-explicit})
with
(\ref{eq-def-processmatestimthreeketinterm})
and
(\ref{eq-processoutpuretwoketestim-vs-processinpuretwoket})
and considering that
\yprocessmat\ is unitary
(hence
$
\Yprocessmat
\Ytransconjug
\Yprocessmat
=
I
$)
,
we
get
\begin{equation}
\Yprocessmatestimthreeketinterm
=
e
^{
\Ysqrtminusone
\Yprocessoutpuretwoketestimphase
}
\Yprocessmatestimtwophaseindetermmatrixdiag
\Yconjug
\Yprocessmatestimthreeketintermtwo
\label{eq-processmatestimthreeketinterm-vs-processmatestimthreeketintermtwo}
\end{equation}
where we 
introduce
\begin{equation}
\Yprocessmatestimthreeketintermtwo
=
\Yprocessinmixedonedensitymatrixeigenunit
\Ytransconjug
\Yprocessinpuretwoket
\end{equation}
which is a known quantity.
Comparing
(\ref{eq-processmatestimthreeketinterm-vs-processmatestimthreeketintermtwo})
with
(\ref{eq-processmatestimthreeketinterm-vs-processinpuretwoket})
shows that
\yprocessmatestimthreeketintermtwo\
here replaces the ket
\yprocessinpuretwoket\
used in the method of 
Section
\ref{sec-single-stage-part-two}
(which is coherent with the fact that the latter method is a
reduced form of the present one, for the case when
\yprocessinmixedonedensitymatrix\
is diagonal and hence
\yprocessinmixedonedensitymatrixeigenunit\
is the identity matrix).
The remainder of the present method is then derived
by modifying the end of the method of Section
\ref{sec-single-stage-part-two}
accordingly:
here using
\begin{equation}
\Yprocessmatestimthree
=
\Yprocessmatestimtwo
\Ymathspace
\mathrm{diag}
(
\Yprocessmatestimthreeketinterm
\oslash
\Yprocessmatestimthreeketintermtwo
)
\Yprocessinmixedonedensitymatrixeigenunit
\Ytransconjug
\end{equation}
it may be shown in the same way as in
Section
\ref{sec-single-stage-part-two}
that we here again get
(\ref{eq-processmatestimthree-vs-processmat}),
i.e.
\yprocessmatestimthree\ here again
succeeds in restoring
\yprocessmat ,
up to a global phase factor
(and again up to estimation errors).
\section{Variants
of two-stage QPT methods}
\label{appendix-sec-two-stage-variants}
Since the columns of 
\ytwostageprocessmatestimfour\ 
are estimated independently,
one may fear that they are not exactly
orthogonal, due to estimation errors.
Then,
\ytwostageprocessmatestimfour\ 
and
hence
\yprocessmatestimthree\
are not unitary,
although they should be because they
aim at estimating
\yprocessmat,
that is constrained to be unitary.
A first solution to this problem
consists of postprocessing
\ytwostageprocessmatestimfour,
so as to determine
a best least-squares unitary approximation of
\ytwostageprocessmatestimfour .
An algorithm that may be used to this end is
defined by
Theorem
\ref{theorem-closest-unitary-matrix}.
The resulting modified QPT method is called EQPT3
hereafter.
\ytextartitionehundredninetyfourvfivemodifstepone{Its
pseudo-code is provided in
Algorithm
\ref{label-algodef-eqptthree}.}
A slightly different version consists of
finding a unitary approximation of
\yprocessmatestimthree\
instead of
\ytwostageprocessmatestimfour .
The resulting QPT method is called EQPT4.
\ytextartitionehundredninetyfourvfivemodifstepone{Its
pseudo-code is provided in
Algorithm
\ref{label-algodef-eqptfour}.}

\begin{algorithm}[t]
{%
{%
\small
\ytextartitionehundredninetyfourvfivemodifstepone{%
\SetKwInOut{Input}{Input}\SetKwInOut{Output}{Output}
\Input{%
a) 
Estimate 
\ytextartitionehundredninetyfourvfivemodifstepone{%
\yprocessoutmixedonedensitymatrixestimhermunittrace}
of output density matrix
\yprocessoutmixedonedensitymatrix\
(provided by 
QST
and obtained for
input density matrix
(\ref{eq-def-processinmixedonedensitymatrix-method-two-stage})
with the diagonal values defined in
(\ref{eq-opdensity-diag-val-uniform-two-stage})).
b) Estimate 
\ytextartitionehundredninetyfourvfivemodifstepone{%
\ytwostageprocessoutmixedtwodensitymatrixestimhermunittrace}
of output density matrix
\ytwostageprocessoutmixedtwodensitymatrix\
(provided by QST
and obtained for
input density matrix
(\ref{eq-def-processinmixedonedensitymatrix-method-two-stage-part-two})
with the diagonal values defined in
(\ref{eq-opdensity-diag-val-uniform-two-stage})).
c)~Estimate 
\yprocessoutpuretwoketestim\
of output ket
\yprocessoutpuretwoket\
(provided by QST
and obtained for all components 
of
input ket
\yprocessinpuretwoket\
equal to
$
1
/
\sqrt{
\Yketspacedim
}
$).%
}
\Output{%
Estimate
\yprocessmatestimthree\
of 
quantum process matrix
\yprocessmat .}
\BlankLine
\Begin{
\tcc{\ytextartitionehundredninetyfourvfivemodifstepone{Exploit}
\yprocessoutmixedonedensitymatrixestimhermunittrace :}
Eigendecomposition:
derive 
1) a diagonal matrix
\yprocessoutmixedonedensitymatrixestimhermunittraceeigenmatval\
that contains the eigenvalues of
\yprocessoutmixedonedensitymatrixestimhermunittrace\
in an arbitrary order
and 2) a matrix
\yprocessoutmixedonedensitymatrixestimhermunittraceeigenmatvec\
whose columns are 
eigenvectors of
\yprocessoutmixedonedensitymatrixestimhermunittrace\
in the same order as eigenvalues\;
Reorder the eigenvalues in
\yprocessoutmixedonedensitymatrixestimhermunittraceeigenmatval\
in 
nonincreasing
order
and 
apply the same permutation to the columns of
\yprocessoutmixedonedensitymatrixestimhermunittraceeigenmatvec\
to create the matrix
\yprocessmatestimtwo\;
\tcc{\ytextartitionehundredninetyfourvfivemodifstepone{Exploit}
\ytwostageprocessoutmixedtwodensitymatrixestimhermunittrace :}
Eigendecomposition:
derive 
1) a diagonal matrix
\ytwostageprocessoutmixedtwodensitymatrixestimhermunittraceeigenmatval\
that contains the eigenvalues of
\ytwostageprocessoutmixedtwodensitymatrixestimhermunittrace\
in an arbitrary order
and 2) a matrix
\ytwostageprocessoutmixedtwodensitymatrixestimhermunittraceeigenmatvec\
whose columns are 
eigenvectors of
\ytwostageprocessoutmixedtwodensitymatrixestimhermunittrace\
in the same order as eigenvalues\;
Reorder the eigenvalues in
\ytwostageprocessoutmixedtwodensitymatrixestimhermunittraceeigenmatval\
in 
nonincreasing
order
and 
apply the same permutation to the columns of
\ytwostageprocessoutmixedtwodensitymatrixestimhermunittraceeigenmatvec\
to create the matrix
\ytwostageprocessmatestimthree\;
\tcc{%
\ytextartitionehundredninetyfourvfivemodifstepone{Combine}
\yprocessmatestimtwo\
and
\ytwostageprocessmatestimthree
\ytextartitionehundredninetyfourvfivemodifstepone{:}%
}
\For{$
\Yprocessinmixedoneeigensubsetindexstd
=
1
$
to
\yprocessinmixedonedensitymatrixvalnb}
{
\For{$
\Yprocessinmixedtwoeigensubsetindexstd
=
1
$
to
\yprocessinmixedonedensitymatrixvalnb}
{
Set column of \ytwostageprocessmatestimfour\
\ytextartitionehundredninetyfourvfivemodifstepone{with
index equal to
$
(
(
\Yprocessinmixedoneeigensubsetindexstd
-
1
)
\Yprocessinmixedonedensitymatrixvalmultiplicity
+
\Yprocessinmixedtwoeigensubsetindexstd
)
$: Set}
it to
first vector provided by CCA of
1) the set of
columns of
\yprocessmatestimtwo\
with
indices 
\ytextartitionehundredninetyfourvfivemodifstepone{$
(
(
\Yprocessinmixedoneeigensubsetindexstd
-
1
)
\Yprocessinmixedonedensitymatrixvalmultiplicity
+
\Yprocessinmixedoneeigensubsetindexstdvalindexstd
)
$
with
$
\Yprocessinmixedoneeigensubsetindexstdvalindexstd
\in
\{
1,
\dots
,
\Yprocessinmixedonedensitymatrixvalmultiplicity
\}$}
and 
2)
the set of
columns of
\ytwostageprocessmatestimthree\
with indices 
\ytextartitionehundredninetyfourvfivemodifstepone{$
(
(
\Yprocessinmixedtwoeigensubsetindexstd
-
1
)
\Yprocessinmixedonedensitymatrixvalmultiplicity
+
\Yprocessinmixedtwoeigensubsetindexstdvalindexstd
)
$
with
$
\Yprocessinmixedtwoeigensubsetindexstdvalindexstd
\in
\{
1,
\dots
,
\Yprocessinmixedonedensitymatrixvalmultiplicity
\}
$}%
\;
\tcc{If the
CCA tool
does not provide unit-norm vectors, then do 
it:}
Divide 
above
column of
\ytwostageprocessmatestimfour\ by its norm%
\;
}
}
Find decomposition of
\ytwostageprocessmatestimfour\
as
$
\Ytwostageprocessmatestimfour
=
V
\Sigma
W
\Ytransconjug
$
with SVD\;
$
\Ytwostageprocessmatestimfour 
\hspace{0mm} 
_{u}
=
V
W
\Ytransconjug
$%
\;
\tcc{\ytextartitionehundredninetyfourvfivemodifstepone{Exploit}
\yprocessoutpuretwoketestim :}
\ytextartitionehundredninetyfourvfivemodifstepone{$
\Yprocessmatestimthreeketinterm
=
\Ytwostageprocessmatestimfour
\hspace{0mm} 
_{u}
\hspace{0.1mm}
\Ytransconjug
\Yprocessoutpuretwoketestim
$\;}
\ytextartitionehundredninetyfourvfivemodifstepone{$
\Yprocessmatestimthree
=
\Ytwostageprocessmatestimfour
\hspace{0mm} 
_{u}
\Ymathspace
\mathrm{diag}
(
\Yprocessmatestimthreeketinterm
\oslash
\Yprocessinpuretwoket
)$\;}
}
\caption{%
EQPT3: third
Eigenanalysis-based 
QPT
algorithm (composed of two stages).
\ytextartitionehundredninetyfourvfivemodifstepone{This version
is for the case when the QST algorithm provides
estimates that meet the known properties of the
actual quantities. Otherwise, see text.}%
}
}
\label{label-algodef-eqptthree}
}
}
\end{algorithm}

\begin{algorithm}[t]
{%
{%
\small
\ytextartitionehundredninetyfourvfivemodifstepone{%
\SetKwInOut{Input}{Input}\SetKwInOut{Output}{Output}
\Input{%
a) 
Estimate 
\ytextartitionehundredninetyfourvfivemodifstepone{%
\yprocessoutmixedonedensitymatrixestimhermunittrace}
of output density matrix
\yprocessoutmixedonedensitymatrix\
(provided by 
QST
and obtained for
input density matrix
(\ref{eq-def-processinmixedonedensitymatrix-method-two-stage})
with the diagonal values defined in
(\ref{eq-opdensity-diag-val-uniform-two-stage})).
b) Estimate 
\ytextartitionehundredninetyfourvfivemodifstepone{%
\ytwostageprocessoutmixedtwodensitymatrixestimhermunittrace}
of output density matrix
\ytwostageprocessoutmixedtwodensitymatrix\
(provided by QST
and obtained for
input density matrix
(\ref{eq-def-processinmixedonedensitymatrix-method-two-stage-part-two})
with the diagonal values defined in
(\ref{eq-opdensity-diag-val-uniform-two-stage})).
c)~Estimate 
\yprocessoutpuretwoketestim\
of output ket
\yprocessoutpuretwoket\
(provided by QST
and obtained for all components 
of
input ket
\yprocessinpuretwoket\
equal to
$
1
/
\sqrt{
\Yketspacedim
}
$).%
}
\Output{%
Estimate
\yprocessmatestimthree\
of 
quantum process matrix
\yprocessmat .}
\BlankLine
\Begin{
\tcc{\ytextartitionehundredninetyfourvfivemodifstepone{Exploit}
\yprocessoutmixedonedensitymatrixestimhermunittrace :}
Eigendecomposition:
derive 
1) a diagonal matrix
\yprocessoutmixedonedensitymatrixestimhermunittraceeigenmatval\
that contains the eigenvalues of
\yprocessoutmixedonedensitymatrixestimhermunittrace\
in an arbitrary order
and 2) a matrix
\yprocessoutmixedonedensitymatrixestimhermunittraceeigenmatvec\
whose columns are 
eigenvectors of
\yprocessoutmixedonedensitymatrixestimhermunittrace\
in the same order as eigenvalues\;
Reorder the eigenvalues in
\yprocessoutmixedonedensitymatrixestimhermunittraceeigenmatval\
in 
nonincreasing
order
and 
apply the same permutation to the columns of
\yprocessoutmixedonedensitymatrixestimhermunittraceeigenmatvec\
to create the matrix
\yprocessmatestimtwo\;
\tcc{\ytextartitionehundredninetyfourvfivemodifstepone{Exploit}
\ytwostageprocessoutmixedtwodensitymatrixestimhermunittrace :}
Eigendecomposition:
derive 
1) a diagonal matrix
\ytwostageprocessoutmixedtwodensitymatrixestimhermunittraceeigenmatval\
that contains the eigenvalues of
\ytwostageprocessoutmixedtwodensitymatrixestimhermunittrace\
in an arbitrary order
and 2) a matrix
\ytwostageprocessoutmixedtwodensitymatrixestimhermunittraceeigenmatvec\
whose columns are 
eigenvectors of
\ytwostageprocessoutmixedtwodensitymatrixestimhermunittrace\
in the same order as eigenvalues\;
Reorder the eigenvalues in
\ytwostageprocessoutmixedtwodensitymatrixestimhermunittraceeigenmatval\
in 
nonincreasing
order
and 
apply the same permutation to the columns of
\ytwostageprocessoutmixedtwodensitymatrixestimhermunittraceeigenmatvec\
to create the matrix
\ytwostageprocessmatestimthree\;
\tcc{%
\ytextartitionehundredninetyfourvfivemodifstepone{Combine}
\yprocessmatestimtwo\
and
\ytwostageprocessmatestimthree
\ytextartitionehundredninetyfourvfivemodifstepone{:}%
}
\For{$
\Yprocessinmixedoneeigensubsetindexstd
=
1
$
to
\yprocessinmixedonedensitymatrixvalnb}
{
\For{$
\Yprocessinmixedtwoeigensubsetindexstd
=
1
$
to
\yprocessinmixedonedensitymatrixvalnb}
{
Set column of \ytwostageprocessmatestimfour\
\ytextartitionehundredninetyfourvfivemodifstepone{with
index equal to
$
(
(
\Yprocessinmixedoneeigensubsetindexstd
-
1
)
\Yprocessinmixedonedensitymatrixvalmultiplicity
+
\Yprocessinmixedtwoeigensubsetindexstd
)
$: Set}
it to
first vector provided by CCA of
1) the set of
columns of
\yprocessmatestimtwo\
with
indices 
\ytextartitionehundredninetyfourvfivemodifstepone{$
(
(
\Yprocessinmixedoneeigensubsetindexstd
-
1
)
\Yprocessinmixedonedensitymatrixvalmultiplicity
+
\Yprocessinmixedoneeigensubsetindexstdvalindexstd
)
$
with
$
\Yprocessinmixedoneeigensubsetindexstdvalindexstd
\in
\{
1,
\dots
,
\Yprocessinmixedonedensitymatrixvalmultiplicity
\}$}
and 
2)
the set of
columns of
\ytwostageprocessmatestimthree\
with indices 
\ytextartitionehundredninetyfourvfivemodifstepone{$
(
(
\Yprocessinmixedtwoeigensubsetindexstd
-
1
)
\Yprocessinmixedonedensitymatrixvalmultiplicity
+
\Yprocessinmixedtwoeigensubsetindexstdvalindexstd
)
$
with
$
\Yprocessinmixedtwoeigensubsetindexstdvalindexstd
\in
\{
1,
\dots
,
\Yprocessinmixedonedensitymatrixvalmultiplicity
\}
$}%
\;
\tcc{If the
CCA tool
does not provide unit-norm vectors, then do 
it:}
Divide 
above
column of
\ytwostageprocessmatestimfour\ by its norm%
\;
}
}
\tcc{\ytextartitionehundredninetyfourvfivemodifstepone{Exploit}
\yprocessoutpuretwoketestim :}
\ytextartitionehundredninetyfourvfivemodifstepone{$
\Yprocessmatestimthreeketinterm
=
\Ytwostageprocessmatestimfour
\Ytransconjug
\Yprocessoutpuretwoketestim
$\;}
\ytextartitionehundredninetyfourvfivemodifstepone{$
\Yprocessmatestimthree
\hspace{0mm} 
_{p}
\hspace{0.1mm}
=
\Ytwostageprocessmatestimfour
\Ymathspace
\mathrm{diag}
(
\Yprocessmatestimthreeketinterm
\oslash
\Yprocessinpuretwoket
)$\;}
Find decomposition of
$
\Yprocessmatestimthree
\hspace{0mm} 
_{p}
$
as
$
\Yprocessmatestimthree
\hspace{0mm} 
_{p}
=
V
\Sigma
W
\Ytransconjug
$
with SVD\;
$
\Yprocessmatestimthree
=
V
W
\Ytransconjug
$%
\;
}
\caption{%
EQPT4: fourth
Eigenanalysis-based 
QPT
algorithm (composed of two stages).
\ytextartitionehundredninetyfourvfivemodifstepone{This version
is for the case when the QST algorithm provides
estimates that meet the known properties of the
actual quantities. Otherwise, see text.}%
}
}
\label{label-algodef-eqptfour}
}
}
\end{algorithm}
\section{\ytextartitionehundredninetyfourvfivemodifstepone{Additional
information for the proposed multi-stage methods}}
\label{sec-appendix-relevance-multi-stage}
\subsection{%
\ytextartitionehundredninetyfourvfivemodifstepone{Relevance 
of the 
methods}}
\label{sec-appendix-relevance-multi-stage-relevance}
\ytextartitionehundredninetyfourvonemodifstepone{%
In this appendix, we 
\ytextartitionehundredninetyfourvonemodifstepthree{first}
prove that the approach 
with
\mbox{$
\Yprocessinmixedonedensitymatrixvalnb
=
2
$}
that we proposed in
Section
\ref{sec-multi-stage},
based on subspace intersections, is relevant.

Each subspace
\ymultistageblocknblogbasetwoeigensubsetindexstdspace\
defined in
Section 
\ref{sec-multi-stage}
is associated with a set of columns of the process matrix
\yprocessmat\ to be identified, in the following sense.
We consider a given arbitrary
stage 
of our algorithm,
defined by the index
\ymultistageyprocessinmixeddensitymatrixblocknblogbasetwo,
and 
the
subset of columns of
\yprocessmat\
that 
correspond to
(i.e. that
have the same column indices as)
a 
given arbitrary
diagonal value
\ymultistageblocknblogbasetwoeigensubsetindexstdinvaldiagdiffeigensubsetindexstd\
of the input density matrix 
\yprocessinmixedonedensitymatrix\
when forming 
(\ref{eq-processoutmixedonedensitymatrixvsinmixedonedensitymatrix})%
.
This
subset of columns of
\yprocessmat\
and
\ymultistageblocknblogbasetwoeigensubsetindexstdinvaldiagdiffeigensubsetindexstd\
are
defined by the
index
$
\Ymultistageblocknblogbasetwoeigensubsetindexstd
\in
\{
1
,
2
\}
$%
.
We then
have the following property:
the eigendecomposition 
plus reordering algorithm
proposed in
Section
\ref{sec-multi-stage}
yields a set of column vectors 
associated with
\ymultistageyprocessinmixeddensitymatrixblocknblogbasetwo\
and
\ymultistageblocknblogbasetwoeigensubsetindexstd ,
that form a basis of a subspace
denoted as
\ymultistageblocknblogbasetwoeigensubsetindexstdspace ,
and that subspace is also the subspace spanned by the
columns of
\yprocessmat\
associated with the
index
\ymultistageblocknblogbasetwoeigensubsetindexstd\
in the above-defined sense.

The analysis of the indices of these columns presented in
the present appendix is made easier
by introducing the following modified notations,
related to the expression of these indices in base 2,
which is motivated by the structure of the matrix
\yprocessinmixedonedensitymatrix\
based on powers of 2 that was introduced 
in
Section
\ref{sec-multi-stage}:
\begin{itemize}
\item The index of the first row and first column of
\yprocessinmixedonedensitymatrix\ 
and
\yprocessmat\
is here equal to 0
(instead of 1 in the other parts of this paper).
\item
Each variable
\ymultistageblocknblogbasetwoeigensubsetindexstd ,
used to split in two subsets the columns of
\yprocessinmixedonedensitymatrix ,
takes the values 1 and 2
(because here
$
\Yprocessinmixedonedensitymatrixvalnb
=
2
$).
With this, we here associate another binary variable, that takes the
values 0 and 1,
namely
\mbox{$
\Ymultistageblocknblogbasetwoeigensubsetindexshiftedstd
=
\Ymultistageblocknblogbasetwoeigensubsetindexstd
-
1
$.}
\item
When using the variable
\ymultistageblocknblogbasetwoeigensubsetindexshiftedstd ,
the subspaces are denoted with the letter
\ymultistageblocknblogbasetwoeigensubsetindexshiftedanyspacenot ,
to avoid any misunderstanding with the notation
\ymultistageblocknblogbasetwoeigensubsetindexstdspace\
used above, especially when considering explicit numerical values
of their indices: for instance,
$
\Ymultistageblocknblogbasetwoeigensubsetindexshiftedanyspacenot
_{
0,
0}
$
is nothing but
$
{\cal S}_{
0
,
1}
$.
More generally speaking, we have
$
\Ymultistageblocknblogbasetwoeigensubsetindexshiftedstdspace
=
\Ymultistageblocknblogbasetwoeigensubsetindexstdspace
$,
whatever the values of the indices.
\end{itemize}

With these notations, using the description provided in
Section
\ref{sec-multi-stage},
the indices of the columns of 
\yprocessmat\
corresponding to
\ymultistageblocknblogbasetwoeigensubsetindexshiftedstdspace\
may 
be shown to be
\begin{eqnarray}
&
&
2
\Ymultistageyprocessinmixeddensitymatrixblocksize
\Ymultistageyprocessinmixeddensitymatrixblockindex
+
\Ymultistageblocknblogbasetwoeigensubsetindexshiftedstd
\Ymultistageyprocessinmixeddensitymatrixblocksize
+
\{
0,
\dots
,
(
\Ymultistageyprocessinmixeddensitymatrixblocksize
-
1
)
\}
\nonumber
\\
&
&
\hspace{20mm}
\mathrm{with}
\hspace{5mm}
\Ymultistageyprocessinmixeddensitymatrixblockindex
\in
\{
0,
\dots
,
(
\Ymultistageyprocessinmixeddensitymatrixblocknb
-
1
)
\}
.
\label{eq-processmat-column-index-version-one}
\end{eqnarray}
This is more easily interpreted by 
indexing
the stages of the algorithm in the reverse order, thus introducing
\begin{equation}
\Ymultistageyprocessinmixeddensitymatrixblocknblogbasetworevers
=
\Ymultistageyprocessinmixeddensitymatrixblocknblogbasetwomax
-
\Ymultistageyprocessinmixeddensitymatrixblocknblogbasetwo
\label{eq-def-multistageyprocessinmixeddensitymatrixblocknblogbasetworevers}
\end{equation}
where
$
\Ymultistageyprocessinmixeddensitymatrixblocknblogbasetwomax
=
\log_2
\Yprocessinmixedonedensitymatrixvalmultiplicity
$
is the maximum value of
\ymultistageyprocessinmixeddensitymatrixblocknblogbasetwo\
and therefore
\begin{equation}
\Yprocessinmixedonedensitymatrixvalmultiplicity
=
2
^
\Ymultistageyprocessinmixeddensitymatrixblocknblogbasetwomax
.
\label{eq-Yprocessinmixedonedensitymatrixvalmultiplicity-vs-Ymultistageyprocessinmixeddensitymatrixblocknblogbasetwomax}
\end{equation}
Using
(\ref{eq-processinmixedonedensitymatrixvalmultiplicity-product}),
(\ref{eq-multistageyprocessinmixeddensitymatrixblocknb-power-two}),
(\ref{eq-Yprocessinmixedonedensitymatrixvalmultiplicity-vs-Ymultistageyprocessinmixeddensitymatrixblocknblogbasetwomax})
and
(\ref{eq-def-multistageyprocessinmixeddensitymatrixblocknblogbasetworevers}),
we obtain
\begin{equation}
\Ymultistageyprocessinmixeddensitymatrixblocksize
=
2
^
\Ymultistageyprocessinmixeddensitymatrixblocknblogbasetworevers
.
\end{equation}
Also using
(\ref{eq-multistageyprocessinmixeddensitymatrixblocknb-power-two}),
the column indices in
(\ref{eq-processmat-column-index-version-one})
may therefore
be rewritten as
\begin{eqnarray}
&
&
\Ymultistageyprocessinmixeddensitymatrixblockindex
2
^
{
(
\Ymultistageyprocessinmixeddensitymatrixblocknblogbasetworevers
+
1
)
}
+
\Ymultistageblocknblogbasetwoeigensubsetindexshiftedstd
2
^
\Ymultistageyprocessinmixeddensitymatrixblocknblogbasetworevers
+
\{
0,
\dots
,
(
2
^
\Ymultistageyprocessinmixeddensitymatrixblocknblogbasetworevers
-
1
)
\}
\nonumber
\\
&
&
\hspace{20mm}
\mathrm{with}
\hspace{5mm}
\Ymultistageyprocessinmixeddensitymatrixblockindex
\in
\{
0,
\dots
,
(
2^
\Ymultistageyprocessinmixeddensitymatrixblocknblogbasetwo
-
1
)
\}
.
\label{eq-processmat-column-index-version-two}
\end{eqnarray}
This should be compared to the representation in base 2 of the index
of any column of
\yprocessmat,
that may be expressed as
\begin{equation}
\sum
_
{
j
=
0
}
^
{
\Ymultistageyprocessinmixeddensitymatrixblocknblogbasetwomax
}
c
_
j
2
^
j
\label{eq-number-in-base-2}
\end{equation}
with
each bit
$
c
_
j
$
equal to 0 or 1.
Eq.
(\ref{eq-number-in-base-2})
allows one to interpret the set of indices
in
(\ref{eq-processmat-column-index-version-two}),
i.e. 
the set of indices of
the columns of
\yprocessmat\
that correspond to
$
\Ymultistageblocknblogbasetwoeigensubsetindexstdspace
=
\Ymultistageblocknblogbasetwoeigensubsetindexshiftedstdspace
$,
for given values of
\ymultistageblocknblogbasetwoeigensubsetindexstd ,
\ymultistageyprocessinmixeddensitymatrixblocknblogbasetwo\
and hence
\ymultistageyprocessinmixeddensitymatrixblocknblogbasetworevers,
due to
(\ref{eq-def-multistageyprocessinmixeddensitymatrixblocknblogbasetworevers}):
these column indices are all the indices 
whose representation in base 2 is such that
their bit
$
c
_
j
$
with index
$
j
=
\Ymultistageyprocessinmixeddensitymatrixblocknblogbasetworevers
$
is constrained to be equal to
\ymultistageblocknblogbasetwoeigensubsetindexshiftedstd ,
whereas all possible combinations are gathered in this set of
index values, for all the bits
$
c
_
j
$
with indices
$
j
\neq
\Ymultistageyprocessinmixeddensitymatrixblocknblogbasetworevers
$.
To summarize,
in
\ymultistageblocknblogbasetwoeigensubsetindexstdspace ,
only
the bit
with index
$
j
=
\Ymultistageyprocessinmixeddensitymatrixblocknblogbasetworevers
$
of the column index is imposed.

Then, we eventually aim at analyzing the intersections of
$
(
\Ymultistageyprocessinmixeddensitymatrixblocknblogbasetwomax
+
1
)
$
subspaces
introduced
in
Section
\ref{sec-multi-stage}.
This may be rephrased as follows,
using the notations introduced in the present appendix.
Each considered intersection of
$
(
\Ymultistageyprocessinmixeddensitymatrixblocknblogbasetwomax
+
1
)
$
subspaces
$
\Ymultistageblocknblogbasetwoeigensubsetindexstdspace
=
\Ymultistageblocknblogbasetwoeigensubsetindexshiftedstdspace
$
corresponds to one
value of the set of indices
$
\{
m_0
,
\dots
,
m_
{
\Ymultistageyprocessinmixeddensitymatrixblocknblogbasetwomax
}
\}
$
and therefore
$
\{
\mu _0
,
\dots
,
\mu _
{
\Ymultistageyprocessinmixeddensitymatrixblocknblogbasetwomax
}
\}
$.
As explained above, separately for each value
of 
\ymultistageyprocessinmixeddensitymatrixblocknblogbasetwo\
and hence of
\ymultistageyprocessinmixeddensitymatrixblocknblogbasetworevers,
the columns of \yprocessmat\ that are kept in this
intersection are those whose index represented in base 2
has a
bit
with index
$
j
=
\Ymultistageyprocessinmixeddensitymatrixblocknblogbasetworevers
$
that is fixed
and equal to
\ymultistageblocknblogbasetwoeigensubsetindexshiftedstd .
The intersection of the
complete set of
$
(
\Ymultistageyprocessinmixeddensitymatrixblocknblogbasetwomax
+
1
)
$
subspaces
obtained when varying
\ymultistageyprocessinmixeddensitymatrixblocknblogbasetwo\
thus yields one and only one column of
\yprocessmat ,
namely the one whose bits are all defined by the 
selected complete set of values
$
\{
\mu _0
,
\dots
,
\mu _
{
\Ymultistageyprocessinmixeddensitymatrixblocknblogbasetwomax
}
\}
$.
This also entails that, when successively
considering the subspace intersections corresponding to
all values of
$
\{
\mu _0
,
\dots
,
\mu _
{
\Ymultistageyprocessinmixeddensitymatrixblocknblogbasetwomax
}
\}
$,
all columns of \yprocessmat\ are successively
obtained.
This proves that the proposed method based on
subspace intersections is guaranteed to
completely identify
\yprocessmat\
(here again up to one phase factor per column,
these phase factors being then handled as in the
other proposed methods).

The successive subspace intersections may be
considered in such an order that the
sets of indices
$
\{
\mu _0
,
\dots
,
\mu _
{
\Ymultistageyprocessinmixeddensitymatrixblocknblogbasetwomax
}
\}
$
correspond to the representation in base 2 of
the integers in increasing order.
Then, the columns of
\yprocessmat\ are obtained with indices in increasing
order.
This is exactly what is obtained in the successive
rows of Table~\ref{tab-multistage-QPT_ketspacedim-eight}
\ytextartitionehundredninetyfourvfivemodifstepone{in
the example provided below.}%
}
\subsection{%
\ytextartitionehundredninetyfourvfivemodifstepone{Example}}
\label{sec-appendix-relevance-multi-stage-example}
\ytextartitionehundredninetyfourvonemodifstepone{%
\ytextartitionehundredninetyfourvfivemodifstepone{We here
illustrate the above}
property and thus the proposed complete
multi-stage methods by considering}
the case
$
\Yketspacedim
=
8
$,
$
\Yprocessinmixedonedensitymatrixvalnb
=
2
$
and
hence
$
\Yprocessinmixedonedensitymatrixvalmultiplicity
=
4
$
and
$
(
\log_2
\Yprocessinmixedonedensitymatrixvalmultiplicity
+
1
)
=
3
$
stages.
During the first stage, that corresponds to
$
\Ymultistageyprocessinmixeddensitymatrixblocknblogbasetwo
=
0
$,
the eigendecomposition allows one to identify two
subspaces
\ymultistageblocknblogbasetwoeigensubsetindexzerospace,
respectively with
$
\Ymultistageblocknblogbasetwoeigensubsetindexzero
=
1
$
and
$
\Ymultistageblocknblogbasetwoeigensubsetindexzero
=
2
$,
and that respectively correspond to the following sets of
4
indices of columns of
\yprocessmat :
$
\{
1, 2, 3, 4
\}$
and
$
\{
5, 6, 7, 8
\}$.
Then, during the second stage, that corresponds to
$
\Ymultistageyprocessinmixeddensitymatrixblocknblogbasetwo
=
1
$,
the eigendecomposition allows one to identify two
subspaces
\ymultistageblocknblogbasetwoeigensubsetindexonespace ,
respectively with
$
\Ymultistageblocknblogbasetwoeigensubsetindexone
=
1
$
and
$
\Ymultistageblocknblogbasetwoeigensubsetindexone
=
2
$,
and that respectively correspond to the following sets of
4
indices of columns of
\yprocessmat :
$
\{
1, 2, 5, 6
\}$
and
$
\{
3, 4, 7, 8
\}$.
Finally, the third stage, that corresponds to
$
\Ymultistageyprocessinmixeddensitymatrixblocknblogbasetwo
=
2
$,
yields
the following sets of
4
indices of columns of
\yprocessmat :
$
\{
1, 3, 5, 7
\}$
and
$
\{
2, 4, 6, 8
\}$.
We then derive the corresponding
subspace intersections,
first considering
$
\Ymultistageblocknblogbasetwoeigensubsetindexzerospace
\cap
\Ymultistageblocknblogbasetwoeigensubsetindexonespace
$
as intermediate quantities,
and finally deriving
$
\Ymultistageblocknblogbasetwoeigensubsetindexzerospace
\cap
\Ymultistageblocknblogbasetwoeigensubsetindexonespace
\cap
\Ymultistageblocknblogbasetwoeigensubsetindextwospace
$.
All these intersections
are listed in Table
\ref{tab-multistage-QPT_ketspacedim-eight}.
This proves that the proposed approach yields 8 intersections that are
each restricted to one column of
\yprocessmat\ and that this set of intersections yields all the
individual columns of
\yprocessmat ,
as required.

\begin{table*}[t!]
\begin{center}
\begin{tabular}{|r|r|r||r|r|}
\hline
\ymultistageblocknblogbasetwoeigensubsetindexzerospace
&
\ymultistageblocknblogbasetwoeigensubsetindexonespace
&
\ymultistageblocknblogbasetwoeigensubsetindextwospace
&
$
\Ymultistageblocknblogbasetwoeigensubsetindexzerospace
\cap
\Ymultistageblocknblogbasetwoeigensubsetindexonespace
$
&
$
\Ymultistageblocknblogbasetwoeigensubsetindexzerospace
\cap
\Ymultistageblocknblogbasetwoeigensubsetindexonespace
\cap
\Ymultistageblocknblogbasetwoeigensubsetindextwospace
$
\\
\hline
$
\{
1, 2, 3, 4
\}
$
&
$
\{
1, 2, 5, 6
\}$
&
$
\{
1, 3, 5, 7
\}$
&
$
\{
1, 2
\}$
&
$
\{
1
\}$
\\
\hline
$
\{
1, 2, 3, 4
\}
$
&
$
\{
1, 2, 5, 6
\}$
&
$
\{
2, 4, 6, 8
\}$
&
$
\{
1, 2
\}$
&
$
\{
2
\}$
\\
\hline
$
\{
1, 2, 3, 4
\}
$
&
$
\{
3, 4, 7, 8
\}$
&
$
\{
1, 3, 5, 7
\}$
&
$
\{
3, 4
\}$
&
$
\{
3
\}$
\\
\hline
$
\{
1, 2, 3, 4
\}
$
&
$
\{
3, 4, 7, 8
\}$
&
$
\{
2, 4, 6, 8
\}$
&
$
\{
3, 4
\}$
&
$
\{
4
\}$
\\
\hline
$
\{
5, 6, 7, 8
\}
$
&
$
\{
1, 2, 5, 6
\}$
&
$
\{
1, 3, 5, 7
\}$
&
$
\{
5, 6
\}$
&
$
\{
5
\}$
\\
\hline
$
\{
5, 6, 7, 8
\}
$
&
$
\{
1, 2, 5, 6
\}$
&
$
\{
2, 4, 6, 8
\}$
&
$
\{
5, 6
\}$
&
$
\{
6
\}$
\\
\hline
$
\{
5, 6, 7, 8
\}
$
&
$
\{
3, 4, 7, 8
\}$
&
$
\{
1, 3, 5, 7
\}$
&
$
\{
7, 8
\}$
&
$
\{
7
\}$
\\
\hline
$
\{
5, 6, 7, 8
\}
$
&
$
\{
3, 4, 7, 8
\}$
&
$
\{
2, 4, 6, 8
\}$
&
$
\{
7, 8
\}$
&
$
\{
8
\}$
\\
\hline
\end{tabular}
\end{center}
\caption{Multi-stage QPT method.
Each column of the table contains the indices of the column vectors
that correspond
to the considered subspace or subspace intersection
defined in the first row of the table.}
\label{tab-multistage-QPT_ketspacedim-eight}
\end{table*}
\subsection{%
\ytextartitionehundredninetyfourvfivemodifstepone{Pseudo-code}}
\label{sec-appendix-relevance-multi-stage-code}
\ytextartitionehundredninetyfourvonemodifstepthree{The above
analysis also has the advantage of suggesting a possible
structure for the software program used to implement this
multi-stage QPT method.
This program has a recursive structure, combined with
a binary tree as follows.
It is based on a recursive function, 
with each call of this function corresponding to one
value of the stage index
\ymultistageyprocessinmixeddensitymatrixblocknblogbasetwo\
(and hence of
\ymultistageyprocessinmixeddensitymatrixblocknblogbasetworevers )
and to one bit of the binary representation
(\ref{eq-number-in-base-2}) of the
index
of a column of
\yprocessmat.
In each such call, this function successively
considers the two possible values of that bit,
that is 0 and 1.
For each of them, it determines the 
intersection 
of the subspaces with indices 0 to
\ymultistageyprocessinmixeddensitymatrixblocknblogbasetwo\
(by determining the subspace with index
\ymultistageyprocessinmixeddensitymatrixblocknblogbasetwo\
and
combining 
it with the previously determined intersection of the
subspaces 
with indices 0 to
$
(
\Ymultistageyprocessinmixeddensitymatrixblocknblogbasetwo
-
1
)
$)
and it recursively calls itself with this bit value.
Successively performing this
for each of the two values of the considered bit
creates two branches in the tree structure.
In each final call of this recursive function,
i.e. each leave of the tree,
all bits of the column index
are assigned,
the function determines the corresponding complete subspace
intersection and sends it back as one column of the
estimate of
\yprocessmat.
By gathering these columns at each level back
from the recursion, the complete matrix
that estimates
\yprocessmat\
(up to phase factors at this stage)
is obtained.}

\ytextartitionehundredninetyfourvfivemodifstepone{The
corresponding pseudo-code is provided in
Algorithm~\ref{label-algodef-eqptfive-main}
for the main part of the program
and
Algorithm
\ref{label-algodef-eqptfive-sub}
for the recursive function.}
\begin{algorithm}[htb!]
\ytextartitionehundredninetyfourvfivemodifstepone{%
\SetKwInOut{Input}{Input}\SetKwInOut{Output}{Output}
\Input{a) Set 
$
\{
\widehat{\Yopdensity}_{
out,
\Ymultistageyprocessinmixeddensitymatrixblocknblogbasetwo
}
\}
$
of
estimates 
of output density matrices
of process
(provided by QST
and obtained for
input density matrices
defined in Section
\ref{sec-multi-stage}):
one estimate
for each stage
\ymultistageyprocessinmixeddensitymatrixblocknblogbasetwo\
of recursion in
Algorithm
\ref{label-algodef-eqptfive-sub},
b)~Estimate 
\yprocessoutpuretwoketestim\
of output ket
\yprocessoutpuretwoket\
(provided by QST
and obtained for
all components 
of
input ket
\yprocessinpuretwoket\
equal to
$
1
/
\sqrt{
\Yketspacedim
}
$).%
}
\Output{%
Estimate
\yprocessmatestimthree\
of 
quantum process matrix
\yprocessmat .}
\BlankLine
\Begin{
\tcc{%
Exploit
$
\{
\widehat{\Yopdensity}_{
out,
\Ymultistageyprocessinmixeddensitymatrixblocknblogbasetwo
}
\}
$:}
\yprocessmatestimtwo
=
EQPT5-sub(0,
$
\{
\widehat{\Yopdensity}_{
out,
\Ymultistageyprocessinmixeddensitymatrixblocknblogbasetwo
}
\}
$,0)\;
\tcc{Exploit
\yprocessoutpuretwoketestim :}
$
\Yprocessmatestimthreeketinterm
=
\Yprocessmatestimtwo
\Ytransconjug
\Yprocessoutpuretwoketestim
$\;
$
\Yprocessmatestimthree
=
\Yprocessmatestimtwo
\Ymathspace
\mathrm{diag}
(
\Yprocessmatestimthreeketinterm
\oslash
\Yprocessinpuretwoket
)$\;
\caption{EQPT5-main: 
main part of
fifth
Eigenanalysis-based 
QPT
algorithm (composed of multiple stages).
This version
is for the case when the QST algorithm provides
estimates that meet the known properties of the
actual quantities. Otherwise, see text.}
}
\label{label-algodef-eqptfive-main}
}
\end{algorithm}
\begin{algorithm}[htb!]
{\small
\ytextartitionehundredninetyfourvfivemodifstepone{%
\SetKwInOut{Input}{Input}\SetKwInOut{Output}{Output}
\Input{a) Index
\ymultistageyprocessinmixeddensitymatrixblocknblogbasetwo\
of recursion stage,
b) Set 
$
\{
\widehat{\Yopdensity}_{
out,
\Ymultistageyprocessinmixeddensitymatrixblocknblogbasetwo
}
\}
$
of
estimates 
of output density matrices
of process
(see Algorithm
\ref{label-algodef-eqptfive-main}),
c) Matrix
\ysubspaceintersecrecurs\
whose column vectors form a basis of subspace intersection
determined in previous stages of recursion
(none in first stage).}
\Output{%
matrix
$
{\widehat{\Yprocessmat}_{part}}
$
related to estimation of part
of 
quantum process matrix
\yprocessmat .}
\BlankLine
\Begin{%
\tcc{Set auxiliary variables
(using
\yketspacedim
:
state space dimension;
$
\Yprocessinmixedonedensitymatrixvalmultiplicity
=
\Yketspacedim
/
2
$):}
$
\Ymultistageyprocessinmixeddensitymatrixblocknblogbasetwomax
=
\log_2
\Yprocessinmixedonedensitymatrixvalmultiplicity
$\;
$
\Ymultistageyprocessinmixeddensitymatrixblocknb
=
2^
\Ymultistageyprocessinmixeddensitymatrixblocknblogbasetwo
$\;
$
\Ymultistageyprocessinmixeddensitymatrixblocksize
=
\Yprocessinmixedonedensitymatrixvalmultiplicity
/
\Ymultistageyprocessinmixeddensitymatrixblocknb
$\;
\tcc{Exploit single estimated output density matrix
$
\widehat{\Yopdensity}_{
out,
\Ymultistageyprocessinmixeddensitymatrixblocknblogbasetwo
}
$
corresponding to
current stage
\ymultistageyprocessinmixeddensitymatrixblocknblogbasetwo\
of recursion:}
Eigendecomposition:
derive 
1) a diagonal matrix
\yprocessoutmixedonedensitymatrixestimhermunittraceeigenmatval\
that contains the eigenvalues of
$
\widehat{\Yopdensity}_{
out,
\Ymultistageyprocessinmixeddensitymatrixblocknblogbasetwo
}
$
in an arbitrary order
and 2) a matrix
\yprocessoutmixedonedensitymatrixestimhermunittraceeigenmatvec\
whose columns are 
eigenvectors of
$
\widehat{\Yopdensity}_{
out,
\Ymultistageyprocessinmixeddensitymatrixblocknblogbasetwo
}
$
in the same order as eigenvalues\;
Reorder the eigenvalues in
\yprocessoutmixedonedensitymatrixestimhermunittraceeigenmatval\
in nonincreasing order
and 
apply the same permutation to the columns of
\yprocessoutmixedonedensitymatrixestimhermunittraceeigenmatvec\
to create the matrix
$
\widehat{\Yprocessmat}_{
\Ymultistageyprocessinmixeddensitymatrixblocknblogbasetwo
}
$\;
Store the 
first
\yprocessinmixedonedensitymatrixvalmultiplicity\
columns
and last
\yprocessinmixedonedensitymatrixvalmultiplicity\
columns of
$
\widehat{\Yprocessmat}_{
\Ymultistageyprocessinmixeddensitymatrixblocknblogbasetwo
}
$
respectively in matrices
$
\widehat{\Yprocessmat}_{
\Ymultistageyprocessinmixeddensitymatrixblocknblogbasetwo
}
(1)
$
and
$
\widehat{\Yprocessmat}_{
\Ymultistageyprocessinmixeddensitymatrixblocknblogbasetwo
}
(2)
$\;
\tcc{Derive intersections 
$
\Ysubspaceintersecrecurs
(1)
$
and
$
\Ysubspaceintersecrecurs
(2)
$
of subspaces defined by
$
\widehat{\Yprocessmat}_{
\Ymultistageyprocessinmixeddensitymatrixblocknblogbasetwo
}
(1)
$
and
$
\widehat{\Yprocessmat}_{
\Ymultistageyprocessinmixeddensitymatrixblocknblogbasetwo
}
(2)
$
with subspace defined by
\ysubspaceintersecrecurs\
(if any):}
\If{$
\Ymultistageyprocessinmixeddensitymatrixblocknblogbasetwo
=
0
$}
{
$
\Ysubspaceintersecrecurs
(1)
=
\widehat{\Yprocessmat}_{
\Ymultistageyprocessinmixeddensitymatrixblocknblogbasetwo
}
(1)
$\;
$
\Ysubspaceintersecrecurs
(2)
=
\widehat{\Yprocessmat}_{
\Ymultistageyprocessinmixeddensitymatrixblocknblogbasetwo
}
(2)
$\;
}
\Else{%
$
\Ysubspaceintersecrecurs
(1)
=
$
first
\ymultistageyprocessinmixeddensitymatrixblocksize\
column
vectors provided by CCA of
\ysubspaceintersecrecurs\
and
$
\widehat{\Yprocessmat}_{
\Ymultistageyprocessinmixeddensitymatrixblocknblogbasetwo
}
(1)
$\;
$
\Ysubspaceintersecrecurs
(2)
=
$
first
\ymultistageyprocessinmixeddensitymatrixblocksize\
column
vectors provided by CCA of
\ysubspaceintersecrecurs\
and
$
\widehat{\Yprocessmat}_{
\Ymultistageyprocessinmixeddensitymatrixblocknblogbasetwo
}
(2)
$\;
}
\If{%
$
\Ymultistageyprocessinmixeddensitymatrixblocknblogbasetwo
<
\Ymultistageyprocessinmixeddensitymatrixblocknblogbasetwomax
$}
{
$
\widehat{\Yprocessmat}_{part}
(1)
$
=
EQPT5-sub
$
(
\Ymultistageyprocessinmixeddensitymatrixblocknblogbasetwo
+1,
\{
\widehat{\Yopdensity}_{
out,
\Ymultistageyprocessinmixeddensitymatrixblocknblogbasetwo
}
\},
\Ysubspaceintersecrecurs
(1)
)
$
\;
$
\widehat{\Yprocessmat}_{part}
(2)
$
=
EQPT5-sub
$
(
\Ymultistageyprocessinmixeddensitymatrixblocknblogbasetwo
+1,
\{
\widehat{\Yopdensity}_{
out,
\Ymultistageyprocessinmixeddensitymatrixblocknblogbasetwo
}
\},
\Ysubspaceintersecrecurs
(2)
)
$
\;
}
\Else{
\tcc{If the
CCA tool
does not provide unit-norm vectors, then do 
it:} 
$
\widehat{\Yprocessmat}_{part}
(1)
=
\Ysubspaceintersecrecurs
(1)
/
\mathrm{norm}(
\Ysubspaceintersecrecurs
(1)
)
$\;
$
\widehat{\Yprocessmat}_{part}
(2)
=
\Ysubspaceintersecrecurs
(2)
/
\mathrm{norm}(
\Ysubspaceintersecrecurs
(2)
)
$\;
}
$
\widehat{\Yprocessmat}_{part}
$
=
concatenation of columns of
$
\widehat{\Yprocessmat}_{part}
(1)
$
and
$
\widehat{\Yprocessmat}_{part}
(2)
$\;
\caption{EQPT5-sub:
recursively called sub-program of
fifth
Eigenanalysis-based 
QPT
algorithm (composed of multiple stages).
This version
is for the case when the QST algorithm provides
estimates that meet the known properties of the
actual quantities. Otherwise, see text.}
}
\label{label-algodef-eqptfive-sub}
}
}
\end{algorithm}
\section{Fluctuation model for density matrix
estimation}
\label{sec-appendix-model-fluctuations}
In Section 
\ref{sec-test-results},
we presented the statistical model that we used in our tests to represent
the fluctuations
that may be
faced
when
estimating a ket with a QST algorithm.
We hereafter describe a corresponding model for the estimation of
density matrices.
Whereas we eventually apply it to mixed states,
we first
build it by considering
a pure state, 
to make it compatible with the model used for ket estimation in Section
\ref{sec-test-results}.
We therefore consider a pure state
represented by the density
matrix
\begin{equation}
\Yopdensity
=
\Yketdeterm
\Ybradeterm
.
\end{equation}
Denoting
$
c_k
$
each actual component
of
\yketdeterm\
and
$
\epsilon _k
$
the 
additive
random 
complex-valued
error made when estimating it,
each actual element of the density matrix
\yopdensity\
reads
\begin{equation}
\Yopdensity
_{
k
\ell
}
=
c_k
c_{\ell}
^*
\label{eq-opdensity-noiseless-one-el}
\end{equation}
and its estimated version may be expressed as
\begin{eqnarray}
\widehat{\Yopdensity}
_{
k
\ell
}
&
=
&
(
c_k
+
\epsilon _k
)
(
c_{\ell}
+
\epsilon _{\ell}
)
^*
\label{eq-opdensity-noisy-one-el-factored}
\\
&
=
&
c_k
c_{\ell}
^*
+
c_k
\epsilon _{\ell}
^*
+
\epsilon _k
c_{\ell}
^*
+
\epsilon _k
\epsilon _{\ell}
^*
.
\label{eq-opdensity-noisy-one-el-developed}
\end{eqnarray}
We aim at building an \emph{approximation} of that quantity that,
as with the ``noisy ket model'' of 
Section 
\ref{sec-test-results},
yields a noisy model of the density matrix that is expressed
as the sum of the actual density matrix and of a term that represents
fluctuations that are consistent with those of the random model used
in Section
\ref{sec-test-results}
for kets.
To this end, we first note that, in
(\ref{eq-opdensity-noisy-one-el-developed}),
the first term,
namely
$
c_k
c_{\ell}
^*
$,
is nothing but the actual value
(\ref{eq-opdensity-noiseless-one-el}),
whereas all other terms are 
the random fluctuations.
We aim at deriving an approximation of these terms, that only depends on
1) the actual value
(\ref{eq-opdensity-noiseless-one-el}),
2)
a single random variable
$
\epsilon
_R
$
for the real part of the fluctuations
and
3)
a single random variable
$
\epsilon
_I
$
for the imaginary part of the fluctuations.

As a simple approximation, we separately consider the real and
imaginary parts of the fluctuations and then just add them in our
final approximate model.
The real part is handled as follows.
The random variable 
$
\epsilon
_R
$
takes both
$
\epsilon _k
$
and
$
\epsilon _{\ell}
$
of
(\ref{eq-opdensity-noisy-one-el-developed})
into account.
Therefore,
it represents an error for 
a ket component and it is
drawn with the same statistics as the fluctuations of a
ket component of 
Section
\ref{sec-test-results},
i.e.
it is real-valued and
uniformly drawn over the range
$
[
-
\Yketestimflucwidth
/ 2
,
\Yketestimflucwidth
/ 2
]
$.
In other words,
to obtain a simple model, both
$
\epsilon _k
$
and
$
\epsilon _{\ell}
$
are here
replaced by
$
\epsilon
_R
$
in
(\ref{eq-opdensity-noisy-one-el-developed})
(without taking their possible statistical independence into account).
Then, to express this real-valued part of the 
fluctuation
model only with respect to
$
\Yopdensity
_{
k
\ell
}
$
in addition
to
$
\epsilon
_R
$,
as explained above,
we ignore the phases of
$
c_k
$
and
$
c_{\ell}
^*
$
in the second and third terms of
(\ref{eq-opdensity-noisy-one-el-developed})
and we
replace both of them with
$
\sqrt{
|
\Yopdensity
_{
k
\ell
}
|
}
$,
based on
(\ref{eq-opdensity-noiseless-one-el}).
The resulting approximate model 
for the real part of fluctuations, corresponding to the
real part of the last three terms of
(\ref{eq-opdensity-noisy-one-el-developed}),
is thus equal to
$
2
\sqrt{
|
\Yopdensity
_{
k
\ell
}
|
}
\epsilon
_R
+
\epsilon
_R
^2
$.

As explained above,
the imaginary part of fluctuations is handled in the same way 
and just added to the above real part.
As an overall result, the complete approximate 
noisy
model 
for any density matrix element reads
\begin{equation}
\widehat{\Yopdensity}
_{
k
\ell
}
=
\Yopdensity
_{
k
\ell
}
+
2
\sqrt{
|
\Yopdensity
_{
k
\ell
}
|
}
\epsilon
_R
+
\epsilon
_R
^2
+
\Ysqrtminusone
\left(
2
\sqrt{
|
\Yopdensity
_{
k
\ell
}
|
}
\epsilon
_I
+
\epsilon
_I
^2
\right)
.
\end{equation}

We again stress that this model contains approximations but
meets the required contraints:
it allows one to create a ``noisy'' version
$
\widehat{\Yopdensity}
_{
k
\ell
}
$
of a given ``noiseless''
density matrix element
$
\Yopdensity
_{
k
\ell
}
$
by just drawing a sample of each of the random variables
$
\epsilon
_R
$
and
$
\epsilon
_I
$,
whose statistics are defined in a way which is 
qualitatively
consistent with those
of the random variable used in 
Section
\ref{sec-test-results}
for deriving a ``noisy'' version of a ket component.
This then mainly allows us to compare the performance of several
QPT methods with the \emph{same} (relevant) fluctuation
statistics.
This qualitative relevance of our simplified model is thus sufficient
for our needs. 
This is to be contrasted with, e.g., the situation when one
would consider a given physical system and one would have to develop
a model of it that is as accurate as possible.

\bibliography{biblio_yd}

\EOD

\end{document}